\documentclass[prd,twocolumn,superscriptaddress,floatfix,amsmath,amssymb,amsfonts,nofootinbib]{revtex4-1}

\usepackage{float} 

\usepackage[normalem]{ulem}
\usepackage{graphicx}
\usepackage{dcolumn}
\usepackage{bm}
\usepackage{blindtext}
\usepackage{verbatim}
\usepackage{relsize}
\usepackage{mathrsfs}
\usepackage{musicography}
\usepackage{amsmath}
\usepackage{cancel}
\usepackage{physics}
\usepackage{epstopdf}
\usepackage{mathtools}
\usepackage{tensor}
\usepackage{color}
\usepackage[usenames,dvipsnames]{pstricks}
\usepackage{epsfig}
\usepackage{pst-grad} 
\usepackage{pst-plot} 
\usepackage{hyperref}
\usepackage{verbatim}
\usepackage{slashed}

\newcommand{\mf}{\mathsf}
\newcommand{\tb}[1]{\textcolor{black}{#1}}
\newcommand{\tc}[1]{\textcolor{black}{#1}}
\newcommand{\tm}[1]{\textcolor{black}{#1}}
\newcommand{\ii}{\mathrm{i}}

\DeclareMathOperator*{\sumint}{%
\mathchoice%
  {\ooalign{$\displaystyle\sum$\cr\hidewidth$\displaystyle\int$\hidewidth\cr}}
  {\ooalign{\raisebox{.14\height}{\scalebox{.7}{$\textstyle\sum$}}\cr\hidewidth$\textstyle\int$\hidewidth\cr}}
  {\ooalign{\raisebox{.2\height}{\scalebox{.6}{$\scriptstyle\sum$}}\cr$\scriptstyle\int$\cr}}
  {\ooalign{\raisebox{.2\height}{\scalebox{.6}{$\scriptstyle\sum$}}\cr$\scriptstyle\int$\cr}}
}

\allowdisplaybreaks[1] 

\begin{document}

\title{Anti-particle detector models in QFT}

\author{T. Rick Perche}
\email{trickperche@perimeterinstitute.ca}
\affiliation{Department of Applied Mathematics, University of Waterloo, Waterloo, Ontario, N2L 3G1, Canada}
\affiliation{Perimeter Institute for Theoretical Physics, Waterloo, Ontario, N2L 2Y5, Canada}

\author{Eduardo Mart\'{i}n-Mart\'{i}nez}
\email{emartinmartinez@uwaterloo.ca}
\affiliation{Institute for Quantum Computing, University of Waterloo, Waterloo, Ontario, N2L 3G1, Canada}
\affiliation{Department of Applied Mathematics, University of Waterloo, Waterloo, Ontario, N2L 3G1, Canada}
\affiliation{Perimeter Institute for Theoretical Physics, Waterloo, Ontario, N2L 2Y5, Canada}

\begin{abstract}
    We analyze families of particle detector models that linearly couple to different kinds of fermionic and bosonic fields. We also study the response of these detectors to particle and anti-particle excitations of the field. We propose a simple linear complex scalar particle detector model that captures the fundamental features of fermionic field detectors similarly to how the Unruh-DeWitt model captures the features of the light matter interaction.  We also discuss why we do not need to limit ourselves to quadratic models commonly employed in past literature. \tm{Namely, we provide a physically motivated mechanism that restores U(1) symmetry in these linear complex models.}
    

    
\end{abstract}

\maketitle

\section{Introduction}

    Although quantum field theory (QFT) has been formulated over one hundred years ago, many interesting fundamental aspects of the theory are not yet fully understood. Among them, a very important foundational gap in QFT is the lack of a consistent measurement framework~\cite{Sorkin,borsten}. \tc{There are several promising avenues in dealing with this issue. One such avenues is the Fewster-Verch framework, developed from an algebraic approach to quantum field theory, which formulating measurements in terms of interactions between a target and detector field using local algebras of observables~\cite{fewster1,fewster2,fewster3}.} Another possible approach to the elusive measurement problem in QFT is the use of particle detectors~\cite{PipoFTL,chicken}. These are localized nonrelativistic quantum systems that couple locally to a quantum field. First introduced by Unruh in \cite{Unruh1976} and refined by DeWitt \cite{DeWitt}, they have been extensively used in a plethora of scenarios from tools to probe the Unruh effect and Hawking radiation~\cite{HawkingGibons,Unruh-Wald,Takagi,bhDetectorsBTZ,bhDetectors,bhDetectorsAdS} to the proposal of protocols in relativistic quantum information: for example, for quantum and classical communication~\cite{Katja,Simidzija_2020} and the study of the field entanglement structure both in flat and curved spacetimes~\mbox{\cite{Valentini1991,Reznik2003,Reznik1,Pozas-Kerstjens:2015,Pozas2016,Nick,topology}}.
    
    Furthermore, particle detector models go beyond theoretical idealizations. They have been proven to be very good approximations to experimentally accessible systems. For instance, the interactions of atoms with the \tb{electric} field can be very well described by particle detector models~\cite{Pozas2016,eduardoOld,eduardo,Nicho1,richard}. In these setups, the role of the (approximately non-relativistic) localized quantum system is played by the atom that couples to the \tb{electric} quantum field. Moreover, it has been shown that simple scalar models in which two-level systems couple to real massless scalar fields are already able to reproduce most of the features of the light-matter interaction~\cite{richard} with even more accuracy than the typical Jaynes-Cummings or Rabi models used in quantum optics~\cite{ScullyBook}. 
    
    The usefulness of particle detectors to closely model real experiments is not limited to the light-matter interaction. Indeed, a new kind of particle detector that modelled the detection of neutrino field excitations has been proposed~\cite{neutrinos}. There, the detector would be associated with a fermionic degree of freedom which models nucleons and electron/positron states within the nucleus of an atom. The proposed model was able to recover the well-known results of neutrino oscillations and provided a platform to extend particle detector models to neutrino fields where fermionic fields are probed by means of a linear interaction with a localized system.
    
    The fermionic particle detector in~\cite{neutrinos}  was applied to a specific process in the emission and detection of neutrinos. Nevertheless, a more detailed study of the model and its response to general field states in different spacetimes has not yet been carried out. 
    
    On the other hand, knowing that the interaction of atoms with light can be captured by a simpler model, where a scalar field is considered, one may then wonder whether the fermionic particle detector could be well approximated by a simpler scalar theory. The natural candidate is the replacement of the fermionic field by a complex scalar field. Although a complex quantum field theory does not contain spin degrees of freedom, it has a nontrivial antiparticle content, so that there is room to explore which results of the fermionic detectors can carry to this simplified model.
    
    The goals of this paper are: 1)  study the effect of linearly coupling detectors to non-Hermitian fields extending and generalizing the results of~\cite{neutrinos} to different kinds of fermionic and bosonic fields; 2)  study in what ways simple linear detector models for  complex scalar fields can capture the features of spinor field particle detectors; and 3)  characterize the role that field statistics (fermionic vs bosonic) plays in the phenomena of particle detection. 
    
    This paper is organized as follows: In Section \ref{sec:UDW} we review a simple model for the interactions of atoms with a quantum electric field and its relation to the Unruh-DeWitt detector model, that couples a probe to a scalar field. In Section \ref{sec:Fermion}, we review the fermionic particle detector model and compute the transition probability of the detector for different field states. In Section \ref{sec:Complex} we introduce the complex scalar particle detector model and compare the main features of the model with the fermionic case. Conclusions can be found in Section \ref{sec:Conclusions}.

\section{The Light-matter interaction and the UDW Model}\label{sec:UDW}
    
    As we discussed in the introduction, it is well established that particle detectors provide good models for the light-matter interaction. The extent to which the \tb{Unruh}-DeWitt  (UDW\footnote{In some literature this model is abbreviated as UdW instead of UDW. Note that, unlike deSitter, Bryce DeWitt's last name is spelled with capital D so UDW is arguably a better moniker for the model.}) model is a good  approximation for the light-matter dynamics has been extensively discussed in past literature~\cite{Pozas2016,eduardoOld,eduardo,Nicho1,richard}. In this section we review the sense in which the interaction of atoms with light can be regarded as a particle detector model \tb{and show} the cases where it can be well approximated by an Unruh-DeWitt detector interacting with a massless scalar field. In Subsection \ref{sub:lightMatter} we review the dipole approximation for the interaction of atoms with an external electric field. In Subsection \ref{sub:UDW} we review the UDW model and in Subsection \ref{sub:vector} we present a generalized particle detector model that couples to a real quantum vector field and reduces to the light-matter interaction in a particular case.
    
    \subsection{A Simple Model for the Light-Matter Interaction}\label{sub:lightMatter}
    
    The Schr\"odinger description for a single electron hydrogen atom consists of a wavefunction description, whose time evolution is prescribed by the quantization of the following classical Coulomb Hamiltonian,
    \begin{equation}\label{eq:free}
        {H}_0 = \frac{{\bm p}^2}{2m} - \frac{e^2}{{r}},
    \end{equation}
    where the first term on the right hand side is the kinetic term, while the second term is the Coulomb potential due to the central charge in the nucleus, with  $e$ being the charge of the electron. We consider the mass of the nucleus to be much larger than the electron so that the dynamics of the centre of mass is neglected and that the reduced mass of the electron is approximately $m\approx m_e$. The \tb{negative} eigenvalues of $\hat H_0$ are discrete and, for the Hydrogen atom, are given by $E_n = E_1/n^2$, where $E_1 \approx -13,6eV$. The associated eigenfunctions will be denoted by $\psi_{n}(\bm x) = \braket{\bm x}{{n}}$, where $n$ is a multi-index that contains the quantum numbers associated with the eigenstates.
    
    The interaction of the hydrogen atom with an \tb{ electric field $\hat{\bm{E}}(t,\bm x)$} can be approached in many different ways. We refer the reader to~\cite{richard} for further detail. For our purposes, we will follow a similar approach as that of~\cite{Pozas2016}, where the dipole approximation is employed, and one obtains the following gauge invariant interaction Hamiltonian for an inertial hydrogen atom:    
    \begin{equation}\label{eqDipole}
        \hat{H}_I(t) = e\int d^3 \bm x \:\hat{\bm d}(t,\bm x) \cdot \hat{\bm E}(t,\bm x),
    \end{equation}
    where we assume our observables to be in the interaction picture, that is, we `incorporate' in them the evolution with respect to the free Hamiltonian $\hat{H}_0$. The operator $\hat{\bm d}(t,\bm x)$ is usually called the dipole \tb{density} operator\tb{, and when restricted to the discrete atomic spectrum of bound states } it can be written in terms of the eigenfunctions of $\hat{H}_0$ as
    \begin{equation}
        {\color{black}\hat{\bm d}(t,\bm x) = \sum_{nm} \:\bm \Lambda_{nm}(\bm x) e^{\ii\Omega_{nm}t}\ket{n}\!\!\bra{m},}
    \end{equation}
    where the smearing vector is defined as 
    \begin{equation}\label{eq:smearingVec}
        {\color{black}\bm \Lambda_{nm}(\bm x) = \psi^*_n(\bm x)\bm x\psi_m(\bm x),}
    \end{equation}
    and we denote the energy gap between the states $n$ and $m$ by $\Omega_{nm} = E_n-E_m$.
    
    Notice that at this stage, after quantizing the electromagnetic field, we are left with a localized nonrelativistic quantum system coupled to a quantum field. \tb{We remark that ``localized'' in this context does not necessarily refer to compact support. Instead, it refers to systems whose support is strongly localized around a region of space, as is usually considered in the literature when studying particle detector models~\cite{Pozas-Kerstjens:2015,Pozas2016,eduardoOld,eduardo,Nicho1,richard,us,us2}}. This is the defining characteristic of \tb{particle detector models studied in the field of RQI}.  Nevertheless, the model can be further simplified if one considers energy transitions between only two energy levels (see, e.g., \cite{Pozas2016,richard}), say $E_g$ and $E_e$. In this case, if we denote $\bm \Lambda(\bm x) = \bm \Lambda_{eg}(\bm x)$ and $\Omega = \Omega_{eg}$, the dipole operator can be further recast as
    \begin{equation}
         \hat{\bm d}(t,\bm x) =  \bm \Lambda(\bm x) e^{-\ii\Omega t}\hat{\sigma}^-  + \bm \Lambda^*(\bm x) e^{\ii\Omega t} \hat{\sigma}^+,
    \end{equation}
    where we have denoted $\hat{\sigma}^+ = \ket{e}\!\!\bra{g}$ and $\hat{\sigma}^- = \ket{g}\!\!\bra{e}$.

    With this simple model, one can already grasp some of the fundamental features of the interaction of a hydrogen atom with a quantum external electromagnetic field. Noting that the atom is comoving with the field quantization frame, the electric field can be expanded in terms of plane-wave modes according to
    \begin{equation}
        \hat{\bm E}(\mf x) = \ii \!\sum_{s=1}^2 \!\int\!\!\frac{ d^3 \bm k}{(2\pi)^{\frac{3}{2}}} \sqrt{\frac{{\color{black}|\bm k|}}{2}}\left(\hat{a}_{\bm k,s}^\dagger e^{-\ii \mf k\cdot \mf x}- \hat{a}_{\bm k,s}e^{\ii \mf k\cdot \mf x}\right)
        \!\bm\epsilon(\bm k,s),
    \end{equation}
    where the $s=1,2$ index labels two independent polarizations of the field associated with the vectors $\bm{\epsilon}(\bm k,s)$ for each mode. The $\hat{a}_{\bm k,s}$ operators denote the annihilation operators of the field and satisfy the canonical commutation relations with the creation operators $\hat{a}_{\bm k,s}^\dagger$,
    \begin{align}
        \big[\hat{a}^{\vphantom{\dagger}}_{\bm k,s},\hat{a}^\dagger_{\bm k',s'}\big] = \delta_{ss'}\delta^{(3)}(\bm k - \bm k').
    \end{align}
    Using the expansion for the electromagnetic field above, it is straightforward to compute the first order transition probability for an atom that starts in a given ground state to end in an excited state. These computations have been thoroughly studied in the literature, in e.g.~\cite{Pozas2016}.
    
    \subsection{The UDW Model}\label{sub:UDW}
    
         As we discussed in the introduction, the first \tb{theoretical model for a particle detector} as we know \textcolor{black}{it} today was introduced by Unruh in~\cite{Unruh1976}, where a quantum particle was used to probe a QFT. The model was later simplified by DeWitt in~\cite{DeWitt}, where he \tb{considered} \tm{a system with discrete internal energy eigenstates coupled to a scalar quantum field via a monopole coupling.} This model has become known as the Unruh-DeWitt  detector model and has been employed in a plethora of scenarios in quantum optics and quantum field theory in flat and curved spacetimes. \tc{Moreover, this detector model has been shown to capture many features of the light-matter interaction~\cite{Pozas-Kerstjens:2015,Pozas2016,eduardo,Nicho1,richard}. In this subsection, we will present the UDW model and study its response to wavepackets, when the detector is consider to interaction with a real scalar field. We will not make any assumptions regarding the fields mass so that we can study the model both in the massless case (in order to compare with the light-matter interaction that will be studied in Subsection \ref{sub:vector} and in the massive case, which will be used to compare with the linear fermionic and complex scalar models that will be studied in Sections \ref{sec:Fermion} and \ref{sec:Complex}. }
         
         Schematically, a UDW detector is a two-level quantum system whose internal degree of freedom is localized along a timelike curve $\mf z(\tau)$ in an $n+1$ dimensional spacetime, where we assume $\tau$ to be the proper time parameter of the curve. The detector's free evolution is governed by the free Hamiltonian
        \begin{equation}\label{eq:detectorH}
            \hat{H}_D = \Omega\hat{\sigma}^+\hat{\sigma}^-,
        \end{equation}
        that generates time evolution with respect to the proper time of the curve. We denote the states of the two-level system by $\ket{g}$ and $\ket{e}$ such that $\hat{H}_D\ket{g} = 0$ and \mbox{$\hat{H}_D\ket{e} = \Omega\ket{e}$}.
    
    We recall that in the simplest version of this model, the detector is assumed to interact with a free real scalar quantum field $\hat{\phi}(\mf x)$. Given an orthonormal basis of solutions to the  Klein-Gordon equation, $\{u_{\bm k}(\mf x),u_{\bm k}^*(\mf x)\}$, the free field can be written as
    \begin{equation}\label{eq:scalarField}
        \hat{\phi}(\mf x) = \int \dd^n \bm k \left(u_{\bm k}(\mf x) \hat{a}_{\bm k}+u_{\bm k}^*(\mf x) \hat{a}^\dagger_{\bm k}\right),
    \end{equation}
    where $\hat{a}_{\bm k}^\dagger$ and $\hat{a}_{\bm k}$ are the creation and annihilation operators that satisfy the canonical commutation relations
    \begin{align}
        \big[\hat{a}^{\vphantom{\dagger}}_{\bm k},\hat{a}^\dagger_{\bm k'}\big] = \delta^{(3)}(\bm k - \bm k').
    \end{align}
    This mode expansion\footnote{Although in general (in the absence of a timelike global Killing symmetry) there is no objective way to define a mode decomposition, we assume from now on that the mode decomposition chosen is physically relevant: either there is some observer that is experimentally relevant for the setup \tc{whose local notion of time is associated with the frequency decomposition} or there is some symmetry reason to choose it. The mode decomposition defines the vacuum state of the theory that we will from now on denote with $\ket{0}$.}  then defines a field Hilbert space representation associated with the choice of modes from Eq.~\eqref{eq:scalarField}. It is tradition to associate $u_{\bm k}(\mf x)$ with the positive frequency modes and $u_{\bm k}^*(\mf x)$ with the negative frequency ones due to their signature with respect to the Klein-Gordon inner product.
    
    Indeed, the interaction of a typical UDW detector with the quantum field is linear in the field  amplitude and can be put in a covariant way by means of a spacetime smearing function $\Lambda(\mf x)$, which is supported around the trajectory $\mf z(\tau)$  (See~\cite{us,us2}). The smearing function then controls both the spatial and temporal profile of the interaction. The interaction between the field and detector can be prescribed in terms of a Hamiltonian density $\hat{\mathfrak{h}}_I(\mf x) = \hat{h}_I(\mf x) \sqrt{|g|}$, where $\hat{h}_I(\mf x)$ is the scalar Hamiltonian weight and $\sqrt{|g|}$ is the determinant of the metric. The Hamiltonian weight for the interaction is given by
    \begin{equation}\label{eq:hIreal}
        \hat{h}_I(\mf x) = \lambda \Lambda(\mf x) \hat{\mu}(\tau) \hat{\phi}(\mf x),
    \end{equation}
    where $\hat{\mu}(\tau)$ is the monopole moment of the detector, given in terms of the ladder operators by \mbox{$\hat{\mu}(\tau) = e^{-\ii \Omega\tau}\sigma^- + e^{\ii \Omega\tau}\sigma^+$}. Notice that although the proper time of the curve is just defined along $\mf z(\tau)$, it is possible to define $\tau$ locally around the curve by means of Fermi normal coordinates, as was done in, e.g., \cite{us, us2}. 
    
    With this, the time evolution operator associated with the quantum system composed of the detector and quantum field can be written as
    \begin{equation}
        \hat{U} = \mathcal{T}_\tau \exp \left(-\mathrm{i}\int \dd V \hat{h}_I(\mf x)\right),
    \end{equation}
    where we assume the time ordering to happen with respect to the detector's proper time, although it has been shown in \cite{us2} that in the cases we will be interested \tb{in},
    any notion of time ordering would be equivalent. Namely, \tb{we will only be looking at the excitation probability to leading order in perturbation theory and when the detector starts in the ground state $\ket{g}$.} \tm{The time ordering with respect to $\tau$ is defined in terms of the Fermi normal coordinates $(\tau,\bm \xi)$, defined locally around the trajectory $\mf z(\tau)$. For more details, see \cite{us,us2,mine}.}
    
    Regarding the field, following~\cite{erickson}, we will assume that its initial state is a general one-particle wavepacket $\ket{\varphi}$, given by
    \begin{equation}\label{eq:psiReal}
        \ket{\varphi} = \int \dd^n \bm k f(\bm k) \hat{a}_{\bm k}^\dagger\ket{0},
    \end{equation}
    where $\ket{0}$ denotes the vacuum of the quantum field (according to the mode decomposition~\eqref{eq:scalarField}) and $f(\bm k)$ gives the momentum distribution of the one-particle excitation. The condition that the state above is normalized imposes that $f$ has unit $L^2$ norm.

    The detector excitation probability amplitude for an arbitrary final state of the field $\ket{\text{out}}$ is given, to leading order, by
    \begin{align}
        \mathcal{A}_{g\rightarrow e}(\text{out}) &=\bra{e,\text{out}}\hat{U}\ket{g,\varphi} \\
        &= \!-\ii\lambda\!\int\! \dd V \Lambda(\mf x) e^{\ii\Omega\tau} \bra{\text{out}}\!\hat{\phi}(\mf x)\!\ket{\varphi}+\mathcal{O}(\lambda^2).\nonumber
    \end{align}
    To obtain the transition probability of the detector, we must sum over all possible field final states. The \tb{leading order transition} probability then reads
    \begin{align}\label{eq:probReal}
         p_{g\rightarrow e} &= \sumint_{\text{out}} |\mathcal{A}_{g\rightarrow e}(\text{out})|^2\\
         &= {\color{black}\lambda^2}\int \dd V \dd V' \Lambda(\mf x')\Lambda(\mf x) e^{\ii\Omega(\tau-\tau')} \bra{\varphi}\hat{\phi}(\mf x')\hat{\phi}(\mf x)\ket{\varphi}.\nonumber
    \end{align}
    It is then enough to compute the two-point function of the field in the state $\ket{\varphi}$. This computation can be found in full detail in Appendix \ref{ap:real}. In the end, we find that the transition probability can be written in the following form:
    \begin{align}\label{eq:probRealState}
        p_{g\rightarrow e} &= \lambda^2 \int \dd V \dd V' \Lambda(\mf x')\Lambda(\mf x) e^{\ii\Omega(\tau-\tau')}\\
        &\quad\quad\quad\:\:\times\left(W_0(\mf x',\mf x)+F(\mf x')F^*(\mf x) + F^*(\mf x')F(\mf x)\right),\nonumber
    \end{align}
    where we have defined 
    \begin{equation}\label{eq:Freal}
    \begin{aligned}
        F(\mf x) &= \int \dd^n \bm k f(\bm k) u_{\bm k}(\mf x),\\
        W_0(\mf x',\mf x) &= \int \dd^n\bm k \:u_{\bm k}(\mf x')u_{\bm k}^*(\mf x).
    \end{aligned}
    \end{equation}
    $W_0(\mf x',\mf x)$ is the vacuum Wightman function. Indeed, as we see from Eq.~\eqref{eq:probRealState}, it is possible to identify the vacuum contribution to the excitation probability, and separate it from the one-particle content contribution, associated with the $F(\mf x')F^*(\mf x)$ and $F^*(\mf x')F(\mf x)$ terms. Note that the $F^*(\mf x')F(\mf x)$ term is usually referred to as the counter-rotating term~\cite{Funai,erickson}, this is because in the transition probability \tb{this term is} proportional to \tm{$u_{\bm k}(\mf x)$ ({\color{black} which contain a factor of $e^{- \text{i} \omega_{\bm k} t}$} in flat spacetimes)} \tc{so that when combined with the $e^{\ii\Omega \tau}$ factor, we obtain a counter-rotating contribution. On the other hand, the term $F(\mf x')F^*(\mf x)$ is proportional to \tm{$u_{\bm k}^*(\mf x)$ ({\color{black} which contain a factor of $e^{\text{i} \omega_{\bm k} t}$} in flat spacetimes)} so that when combined with the $e^{\ii\Omega \tau}$ factor, we obtain a co-rotating term}. In other words, the counter-rotating terms can be associated with the coupling of the positive phase terms of the detector with the negative frequency modes of the field. It can be shown that there is a variety of scenarios where the counter-rotating terms vanish~\cite{Funai,erickson}, and the overall contribution of the co-rotating term tends to be larger. As a matter of fact, in Appendix \ref{ap:harmlessLabel} we show that the counter-rotating contribution from the wavepacket terms is always smaller than the vacuum contribution of the field.

    So far we have considered a general field expansion. For the \tb{remainder} of the section we  will consider a typical plane-wave mode expansion \tc{in ($n+1$)-dimensional} Minkowski spacetime:
    \begin{align}\label{eq:modesReal}
        u_{\bm k}(\mf x) = \frac{1}{(2\pi)^{\frac{n}{2}}}\frac{e^{\ii \mf k\cdot \mf x}}{\sqrt{2\omega_{\bm k}}},
    \end{align}
    where we represent the event $\mf x \equiv (t,\bm x)$ in an inertial coordinate system and \mbox{$\mf k \equiv (\omega_{\bm k},\bm{k})$} with $\omega_{\bm k} = \sqrt{\bm k^2+m^2}$.  As a first exercise we will compute the response of an inertial detector in flat spacetime when the field is prepared in a one-particle wavepacket. That is, we assume the initial state of the field to be given by Eq. \eqref{eq:psiReal} where the $a_{\bm{k}}^\dagger$ is a plane wave creation operator, and we prescribe the shape of the wavepacket in momentum space by a Gaussian centered at $\bm k_0$ and localized around it with a spectral width $\sigma$:
    \begin{equation}\label{eq:fReal}
        f(\bm k) =\frac{1}{(\pi \sigma^2)^{\frac{n}{4}}}e^{-\frac{(\bm k-\bm k_0)^2}{2\sigma^2}}.
    \end{equation}
    This choice ensures that $\norm{f}_2 = 1$, and therefore the particle state $\ket{\varphi}$ from Eq. \eqref{eq:psiReal} is normalized. These calculations are analogous to those performed  in~\cite{erickson} for the linear model, although we allow the field to be massive.
    
    We assume the detector's trajectory to be $\mf z(t) = (t,\bm 0)$, so that it is comoving with the field quantization frame $(t,\bm x)$. We consider the detector to be pointlike. This corresponds to the following choice of spacetime smearing function:
    \begin{equation}
        \Lambda(\mf x) = \chi(t) \delta^{(3)}(\bm x),
    \end{equation}
   where $\chi(t)$ is a switching function and where we assumed the detector to be centered at the origin, which also corresponds to the center of the wavepacket of the state $\ket{\varphi}$ in position space. We will consider that the detector couples to the field for a very long time. The detector long-time response is computed through the so-called \emph{adiabatic limit}. Namely, consider $\chi_{{}_T}(t) \equiv \chi(t/T)$ to be a one parametric family of switching functions where $T$ is the timescale governing the duration of the interaction (e.g. $\chi_{{}_T}(t)=e^{-t^2/T^2}$). Assume $\chi_{{}_T}(t)$ to be such that its Fourier transform \tb{decays} faster than any polynomial. We then take the $T\rightarrow \infty$ limit. The adiabatic limit yields the physical behaviour of a detector switched on for long times preventing spurious UV divergences, as discussed, e.g.,  in~\cite{Satz_2007,LoukoCurvedSpacetimes,erickson}. 
   
    \begin{figure*}[t]
        \includegraphics[scale=0.44]{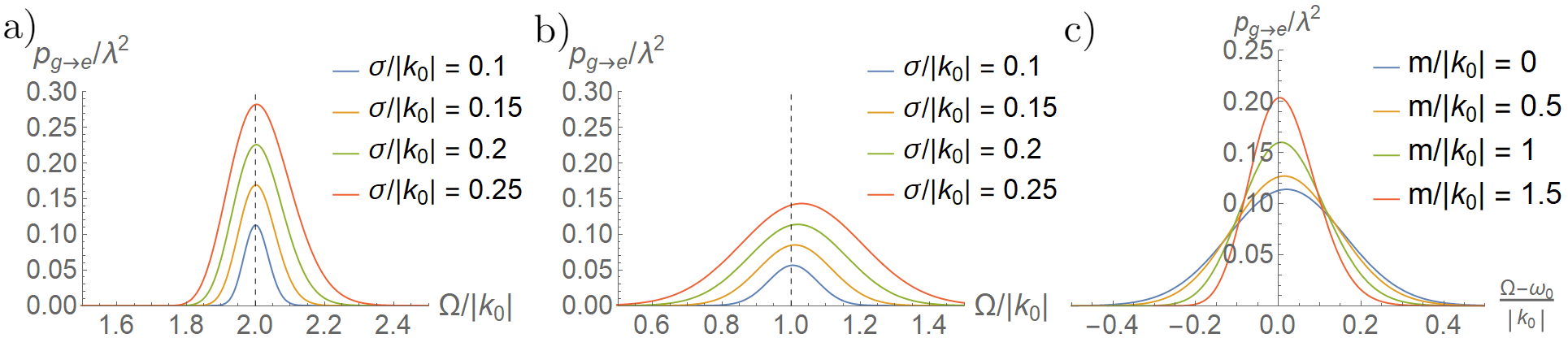}
        \caption{a) Excitation probability of the detector interacting with a field of mass $m = \sqrt{3}|k_0|$ in three space dimensions for different values of $\sigma$. The dashed line corresponds to $\Omega/|\bm k_0| = \sqrt{1+m^2/|\bm k_0|^2} = 2$, which is the resonance frequency of the detector. b) Excitation probability of the detector interacting with a massless field in three space dimensions for different values of $\sigma$. The dashed line corresponds to $\Omega = |\bm k_0|$, which is the resonance frequency of the detector. c)  Excitation probability of a detector interacting with fields of varying masses in three space dimensions for $\sigma = 0.2 |k_0|$. We plot the frequency of the detector centered at its resonance point $\Omega = \sqrt{|\bm k_0|^2+m^2}$.}\label{fig:abc}
    \end{figure*}
    
    Under the long time limit assumption, both the vacuum contribution and the counter-rotating contribution from the wavepacket vanish (see, e.g.~\cite{erickson}, and Appendix~\ref{ap:real}). That is, out of the three terms in Eq. \eqref{eq:probRealState}, only the co-rotating term $F(\mf x')F^*(\mf x)$ survives. We also see that, as expected, in the adiabatic long-time limit the excitation probability is only non-zero in the case where the detector gap is larger than the mass of the particle, i.e., $\Omega>m$. In this case, the probability can be written as
    \begin{align}\label{eq:probFinalReal}
        p_{g\rightarrow e}=&\frac{2\lambda^2 \Omega(\Omega^2 - m^2)^{\frac{n-2}{2}}}{\pi^{\frac{n}{2}-2}\sigma^{4-n}|\bm k_0|^{n-2}} e^{-\frac{|\bm k_0|^2+\Omega^2-m^2}{\sigma^2}}\\&\:\:\:\:\:\:\:\:\:\:\times I_{\frac{n-2}{2}}\left(\frac{|\bm k_0|\sqrt{\Omega^2 - m^2}}{ \sigma^{2}}\right)^2\Theta(\Omega - m),\nonumber
    \end{align}
    where $I_n$ is the modified Bessel function of the first kind~\cite{gradshteyn} and $\Theta$ denotes the Heaviside step function. This result is a slight generalization of what was obtained in \cite{erickson}, where a massless field was considered. The details regarding the resonance of the detector with the field have been thoroughly discussed in \cite{erickson}, where it was shown that localizing the momentum state sharply around $\bm k_0$ does not necessarily increase the probability of detection, and the behaviour depends explicitly on the spacetime dimension. In fact, for $n\geq 3$, we have $p_{g\rightarrow e}\rightarrow 0$ as $\sigma\rightarrow  0$ even though the energy of the wavepacket does not vanish in the monochromatic limit $\sigma\rightarrow  0$. The reasoning behind it lies in the fact that localizing the one-particle state in momentum space delocalizes it in position space. With this, the energy density of the particle at the location of the detector vanishes in the limit of $\sigma \rightarrow 0$. Nevertheless, as $\sigma$ decreases, we have a resonance effect when the detector energy gap satisfies $\Omega = \omega_0 \coloneqq  \sqrt{|\bm k_0|^2 + m^2}$.
    
     In particular, for future comparisons, it is worth pointing out that Eq. \eqref{eq:probFinalReal} for the case of $n=3$ yields
    \begin{align}\label{eq:probFinal3DReal}
        p_{g\rightarrow e} =& \frac{4\lambda^2 \sigma\Omega}{\sqrt{\pi} |\bm k_0|^2}e^{-\frac{|\bm k_0|^2 + \Omega^2 - m^2}{\sigma^2}}\\&\:\:\:\:\:\:\:\:\:\:\:\:\:\:\:\:\times\sinh^2\left(\frac{|\bm k_0|\sqrt{\Omega^2-m^2}}{\sigma^2}\right)\Theta(\Omega - m).\nonumber
    \end{align}
     We plot this function for different values of $m$ and $\sigma$ in Figs. \ref{fig:abc} a) and \ref{fig:abc} b). Notice that as $\sigma$ decreases, the more localized the particle state is in momentum space and the more delocalized the particle is in position space. This makes the excitation probability decrease as $\sigma$ decreases due to the fact that the detector only probes the field along a (tightly localized) timelike trajectory. This effect has been studied in detail in \tm{Section III B of} \cite{erickson} in the massless case.

    
    
    The effect of the field's mass on the probability can be seen in Fig.~\ref{fig:abc} c): as the mass of the field increases the resonant peak becomes sharper around the frequency $\Omega = \omega_0= \sqrt{|\bm k_0|^2 + m^2}$. Furthermore, the peak of the probability distribution can be found by setting $\Omega=\omega_0$ in Eq. \eqref{eq:probFinal3DReal} and is found to be
    \begin{equation}
        p_{g\rightarrow e} = \frac{4\lambda^2 \sigma}{\sqrt{\pi} |\bm k_0|^2}e^{-\frac{2|\bm k_0|^2}{\sigma^2}}\sinh^2\left(\frac{|\bm k_0|^2}{\sigma^2}\right)\sqrt{|\bm k_0|^2+m^2}.
    \end{equation}
    This implies that for a fixed field state, the peak of the \tb{leading order} excitation probability grows with the mass of the field. In particular, \tb{as long as we remain in the perturbative regime,} for $m\gg |\bm k_0|$, we obtain a linear growth with the mass of the field.
    
    
    \subsection{A particle detector probing a real vector quantum field}\label{sub:vector}
    
    We will now show  that the scalar light-matter interaction model presented in Subsection \ref{sub:UDW} captures the right resonant behaviour obtained through a more realistic vector coupling as described in~\ref{sub:lightMatter}. Let us consider a vector coupling of the form~\eqref{eqDipole}  modelling an atomic system dipolarly coupled to the (vector) electric field in arbitrary spacetime. \tc{In general, the electric field associated with a given timelike congruence defined by a timelike vector field $n^\mu(\mf x)$ can be written as $E^\mu(\mf x) = F^{\mu\nu}(\mf x)n_\nu(\mf x)$, where $F_{\mu\nu}$ is the (gauge invariant) electromagnetic tensor. This allows one to effectively treat the electric field as a 4-vector, which is defined in terms of a congruence of observers. \tc{The $E^\mu(\mf x)$ field defined by this congruence is then orthogonal to the field $n^\mu(\mf x)$, so that one sees that this is indeed an effective three-vector field, tangent to the rest spaces of the congruence.} Under these conditions, it is possible to write the quantum electric field as}
    \begin{equation}\label{eq:fieldEM}
        {\color{black}\hat{E}^\mu(\mf x) = \sum_{s=1,2}\int \dd^3 \bm p \left(U^\mu_{\bm p,s}(\mf x) \hat{a}_{\bm p,s}+U^{\mu*}_{\bm p,s}(\mf x) \hat{a}^\dagger_{\bm p,s}\right),}
    \end{equation}
    \tc{where $U^\mu_{\bm p,s}(\mf x)$ are vector fields obtained from the contraction of the solutions to the equations of motion of $\hat{F}^{\mu\nu}$ with the congruence defined by $n^\mu(\mf x)$. For convenience, we write $U^\mu_{\bm p,s}(\mf x) = u_{\bm p,s}(\mf x) \epsilon^\mu(\bm p,s)$, where $s=1,2$ labels the polarizations of the electric field, $u_{\bm p,s}(\mf x)$ are solutions to the scalar Klein-Gordon equation and $\epsilon^\mu(\bm p,s)$ are the normalized polarization four-vectors.}
    
    In this context, the particle detector model is still described by a two-level system moving along a trajectory $\mf z(\tau)$ whose time evolution is dictated by the Hamiltonian in Eq. \eqref{eq:detectorH}. Since it couples to a vector field, the detector must have a vector degree of freedom. To add this to the model, we promote the spacetime smearing function to a vector field $\Lambda_\mu(\mf x)$ supported locally around the trajectory $\mf{z}(\tau)$. \textcolor{black}{ In order to prescribe the electric field $\hat{E}^\mu(\mf x)$ seen locally around the detector, we must extend the four velocity of the trajectory $\mf z(\tau)$ locally around the curve. We do so according to the Fermi normal coordinates $(\tau,\bm \xi)$, in the lines of what was done in~\cite{us,us2,mine}. In essence, we choose the timelike congruence of observers $n^\mu = (\partial_\tau)^\mu$, which will yield  the electric field as seen by observers comoving with the detector.} The interaction Hamiltonian weight is then prescribed as
    \begin{align}\label{eq:hIvector}
        &\hat{h}_I(\mf x) = \lambda (\Lambda_\mu(\mf x)e^{\ii \Omega \tau}\hat{\sigma}^+ +\Lambda_\mu^*(\mf x) e^{-\ii \Omega \tau}\hat{\sigma}^-)\hat{E}^\mu(\mf x)\\
        & \:\:\:\:\:{\color{black}=\lambda (\Lambda_\mu(\mf x)n_\nu(\mf x)e^{\ii \Omega \tau}\hat{\sigma}^+ +\Lambda_\mu^*(\mf x)n_\nu(\mf x) e^{-\ii \Omega \tau}\hat{\sigma}^-)\hat{F}^{\mu\nu}(\mf x).}\nonumber
    \end{align}
    This interaction can be seen in two ways: as a generalization of Eq. \eqref{eqDipole} to curved spacetimes or as a generalization of Eq.~\eqref{eq:hIreal} to a vector interaction.
    
    To study the response of the detector to a one-particle Fock wavepacket with arbitrary polarization, let us consider that the field $\hat{E}_\mu$ starts out in the following state
    \begin{equation}\label{eq:psiVector}
        \ket{\varphi} = \sum_{s=1,2} \int d^3 \bm k f_s(\bm k) \ket{\bm k,s},
    \end{equation}
    while the detector starts in its ground state, $\ket{g}$. The normalization of the state is achieved by demanding that the two polarization momentum distributions satisfy  $\norm{f_1}_2^2 + \norm{f_2}_2^2 = 1$. Proceeding in analogy to what was done in the previous subsection, we find that the \tb{leading order} probability that the detector ends up in an excited state after the interaction is given by
    \begin{align}\label{eq:probVector}
         p_{g\rightarrow e} &= \lambda^2 \int \dd V \dd V' \Lambda_\nu^*(\mf x')^*\Lambda_\mu(\mf x) e^{\ii\Omega(\tau-\tau')}\\
         &\quad\quad\!\times\!\left(W_0^{\nu\mu}(\mf x',\mf x)+F^{\nu}(\mf x')F^{\mu*}(\mf x) + F^{\nu*}(\mf x')F^{\mu}(\mf x)\right)\!,\nonumber
    \end{align}
    where we defined
    \begin{align}
        F^\mu(\mf x) &= \sum_{s=1,2}\int \dd^3 \bm k f_s(\bm k)\epsilon^\mu(\bm k,s)u_{\bm k,s}(\mf x),\label{eq:FEM}\\
        W_0^{\nu\mu}(\mf x',\mf x) &= \sum_{s=1,2}\int \text{d}^3 \bm k \epsilon^\nu(\bm k,s)\epsilon^\mu(\bm k,s)u_{\bm k,s}(\mf x')u_{\bm k,s}^*(\mf x).\nonumber
    \end{align}
    Here, $ W_0^{\nu\mu}(\mf x',\mf x) \coloneqq \bra{0}\hat{E}^\nu(\mf x')\hat{E}^\mu(\mf x) \ket{0}$ is the two-point vacuum Wightman tensor of the vector field in the chosen quantization scheme. The details of this computation can be found in Appendix \ref{ap:vector}. Notice that we again find a vacuum contribution to the excitation probability and two terms that come form the state's `particle content'.
    
    One can apply the techniques above to treat the interaction of a two-level atom with light that was discussed in \tc{Subsection \ref{sub:lightMatter}}. {\color{black}In this case, the trajectory associated with the detector will be inertial. We then employ inertial coordinates associated to its motion which define the inertial congruence $n^\mu = (1,0,0,0)$.} We obtain the following mode expansion for the electric field considering \tb{an inertial quantization frame} $(t,\bm x)$
    \begin{equation}
        \hat{E}^\mu(\mf x) = \ii \!\sum_{s=1}^2 \!\int\!\!\frac{ d^3 \bm k}{(2\pi)^{\frac{3}{2}}} \sqrt{\frac{\bm k}{2}}\left(\hat{a}_{\bm k,s}^\dagger e^{-\ii \mf k\cdot \mf x}- \hat{a}_{\bm k,s}e^{\ii \mf k\cdot \mf x}\right)
        \!\epsilon^\mu(\bm k,s),
    \end{equation}
    where the polarization vectors satisfy $k_\mu \epsilon^\mu(\bm k,s)=0$, due to the divergenceless condition of the electric field \tc{associated with an inertial congruence in flat spacetimes}, $\nabla_\mu E^\mu = 0$. \tc{This provides a physical interpretation for $\mf k$ in spacetime: each $\bm k$ is associated with a direction of propagation of the field in the rest space associated with $n^\mu$, so that in the inertial coordinates picked, $k^\mu = (|\bm k |,\bm k)$ is the wave four-vector. It is usual to call $\bm k$ the wavevector.} The polarization four-vectors $\epsilon^\mu(\bm k,s)$ also form a basis to the (spacelike) orthogonal space to $\mf k$ and $\mf{n} = \partial_t$. Thus, they satisfy the completeness relation
    \begin{equation}
        {\color{black}\sum_{s=1}^2 \epsilon_\mu(\bm k,s) \epsilon_\nu(\bm k,s) = \eta_{\mu\nu}+n_\mu n_\nu - \frac{k_\mu k_\nu}{|\bm k|^2}.}
    \end{equation}
    
    With these tools, it is possible to compute the excitation probability for an atom coupled to the field, through a coupling of the form of Eq. \eqref{eq:hIvector}. This translates into choosing a spacetime smearing vector field $\Lambda_\mu(\mf x)$ in equation \eqref{eq:smearingVec} with $n = g$ and $m = e$. We are interested in comparing the response of the more realistic vector model (see, e.g.~\cite{Pozas2016,richard}) to those of the scalar UDW model in Subsection~\ref{sub:UDW} for a pointlike detector interacting in the long time regime. In this simplified case, we will choose a spacetime smearing vector that is a Dirac delta supported along the worldline of the detector. In other words, we assume the detector's trajectory to be $\mf z(t) = (t,\bm 0)$. That is, we have that
    \begin{equation}\label{eq:pointEM}
        \Lambda_\mu(\mf x) = \chi(t){\color{black}\delta^{(3)}(\bm x)} X_\mu,
    \end{equation}
    where $\chi(t)$ is a switching function that will be used to compute the long-time limit adiabatically\tc{, $X_\mu = \Delta\, e_\mu$ is an arbitrary spacelike vector with units of length which are encoded in the constant $\Delta$, where $e_\mu$ is a unit norm spacelike vector.  $X_\mu$ basically tells us the polarization of the electric field that the detector (which is vectorial in nature) couples to and the length associated with the dipole coupling. If one were to represent an atom, for example, it would be natural to consider $\Delta \approx a_0$, where $a_0$ is the Bohr radius.}
    
    We will also make a further choice for the field state $\ket{\varphi}$ in~\eqref{eq:psiVector} by specifying its polarization and momentum distribution:
    \begin{align}
        f_1(\bm k) &= \frac{\alpha_1}{(\pi \sigma^2)^{\frac{3}{4}}}e^{-\frac{|\bm k-\bm k_0|^2}{2\sigma^2}},\label{eq:f1EM}\\
        f_2(\bm k) &= \frac{\alpha_2}{(\pi \sigma^2)^{\frac{3}{4}}}e^{-\frac{|\bm k-\bm k_0|^2}{2\sigma^2}},\label{eq:f2EM}
    \end{align}
    where the normalization condition for the state $\ket{\varphi}$ implies that $|\alpha_1|^2 + |\alpha_2|^2 =1$.
    \begin{figure}[t]
        \centering
        \includegraphics[scale=0.64]{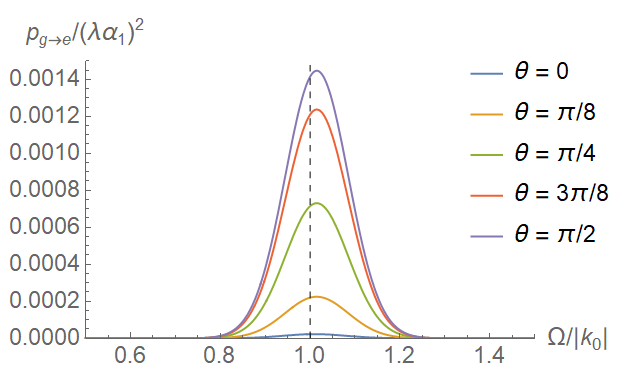}
        \caption{Transition probability for a detector probing a vector quantum field as a function of the detector gap for different values of the relative angle $\theta$ between the dipole vector $\bm X$ and the center of the momentum distribution of the wavepacket, $\bm k_0$. The plots above use $\sigma = 10^{-1} |\bm k_0|$ and $\Delta = |\bm k_0$ for the localization of the wavepacket in momentum space.}
        \label{fig:EM}
    \end{figure}
    
    \begin{figure}[t]
        \centering
        \includegraphics[scale=0.64]{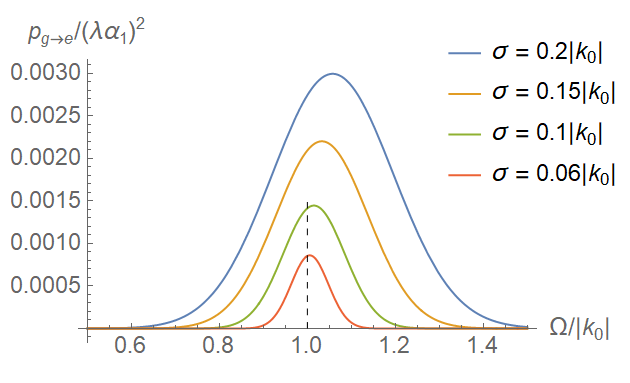}
        \caption{Transition probability for a detector probing a vector quantum field as a function of the detector gap when the relative angle $\theta$ between the dipole vector $\bm X$ and the center of the momentum distribution of the wavepacket, $\bm k_0$ is set to $\pi/2$. The plots above are for different values of $\sigma$ and $\Delta = |\bm k_0|$.}
        \label{fig:EM2}
    \end{figure}
    We perform the computations associated with this model in Appendix \ref{ap:vector}, where we find the general excitation probability for the detector as a function of the average momentum of the wavepacket $\bm k_0$. \tc{The vector $\bm k_0$ can also be associated with the average propagation three-vector of the wavepacket in the rest space associated with $n^\mu$}. In the case where $\bm k_0$ is parallel to the dipole four-vector $\mf X$ \tc{(or equivalently, to its projection in the spacelike surface orthogonal to $n^\mu$)} and only the polarization $s=1$ has components in the direction of $\mf X$, we obtain the following excitation probability,
    \begin{equation}
        p_{g\rightarrow e}^{\parallel} = \lambda^2\frac{ |\alpha_1|^2 {\color{black}\Delta^2}\sigma \Omega^3}{4 \sqrt{\pi}|\bm k_0|^2}e^{-\frac{|\bm k_0|^2+\Omega^2}{\sigma^2}}I_1\left(\frac{\Omega|\bm k_0|}{\sigma^2}\right)^2,
    \end{equation}
    where $\Delta = \norm{\bm X}$ is the length of the dipole vector and the coupling constant for a Hydrogen-like atom is $\lambda = e$. In the case where $\bm k_0$ is perpendicular to the polarization vector we have the following result for the excitation probability:
    \begin{align}
        p_{g\rightarrow e}^{\perp} &= \lambda^2 \frac{|\alpha_1|^2{\color{black}\Delta^2} \Omega^5}{16 \sqrt{\pi}\sigma^3}e^{-\frac{|\bm k_0|^2+\Omega^2}{\sigma^2}}\\
        &\:\:\:\:\:\times\left(I_1\left(\frac{\Omega|\bm{k}_0|}{2\sigma^2}\right)^2+I_0\left(\frac{\Omega|\bm k_0|}{2\sigma^2}\right)^2\right)^2.\nonumber
    \end{align}
    
    In Fig. \ref{fig:EM} we plot the excitation probability for different values of the relative angle $\theta$ between $\bm X$ and $\bm k_0$ for $\sigma = 10^{-1}|\bm k_0|$. We then see that the maximum contribution for the excitation of the detector happens when $\bm k_0$ is orthogonal to the polarization vector, as expected, because in this case the electric field is the most parallel to the detector's dipole, thus maximizing the value of the interaction term \mbox{$\bm{d}\cdot \bm{E}$}.
    
    \tc{In Fig. \ref{fig:EM2}, we plot the  excitation probability for different values of the spectral width of the momentum wavepacket, $\sigma$ with fixed angle between $\bm X$ and $\bm k_0$, $\theta = \pi/2$. It is then possible to notice that the peaks of the excitation probability are not precisely centered at $\Omega = |\bm k_0|$, similar to what was seen for the massless scalar field in~\cite{erickson}. However, as shown in~\cite{erickson}, in the monochromatic limit ($\sigma\to0$) the peak of the distribution approaches the expected resonance value, $\Omega = |\bm k_0|$. Nevertheless, in the $\sigma\rightarrow 0$ limit a result similar to what was found in~\cite{erickson} is also present: as we consider wavepackets more localized in momentum space, these packets become delocalized in position space. Given that these states have a finite total energy (again, see~\cite{erickson}), in the limit of $\sigma\rightarrow 0$, the wavepacket becomes completely delocalized and the energy density goes to zero. In particular, the fact that our particle detector model is localized implies that the effective particle content that interacts with the detector becomes negligible. For more details regarding this phenomenon, we refer the reader to the discussions in the end of Subsection III B and Appendix B of \cite{erickson}. In summary, in Fig.~\ref{fig:EM} we also see the same kind of resonance effect as seen in Fig.~\ref{fig:abc} b), where the peak of the probability distribution is centered at $\Omega \approx |\bm k_0|$, where the approximation is more precise the smaller the value of $\sigma$ is.}
    
\section{The Fermionic Particle Detector}\label{sec:Fermion}

    Fermionic UDW detectors have been successfully employed in QFT in curved spacetimes for a long time. For example, the fermionic UDW detector proposed in~\cite{Takagi,neutrinos} has been shown to capture the phenomenology of the Unruh effect for fermion fields. In all these models the coupling of the internal degree of freedom of the detector with the fermion field was, in nature, quadratic, mimicking the coupling of the electric field to charged fermions, which is also quadratic. However, having quadratic particle \tb{detectors} for fermionic fields makes it difficult to establish a direct comparison between the scalar UDW models (which are mostly linear) and the fermionic ones (which are mostly quadratic). If one is to observe different phenomenology in the response of boson versus fermion detectors, would the difference come from the quadratic nature of the coupling? Or would it be due to the statistics of the field? This has been partially studied in the past comparing the quadratic UDW boson model to the fermionic detector, showing that the answers to \tb{these} questions are not trivial~\cite{Takagi,Sachs1}. Moreover, the quadratic fermion UDW detector model was proposed because in order to respect U(1) symmetry, if the detector is U(1) invariant \tc{it has to couple to a bosonic operator (in this case $\bar{\psi}\psi$)~\cite{Takagi,Sachs1,renBosonsFermions}}. This already makes it very difficult to see distinctions coming from particle vs anti-particle phenomenology as we will discuss later on.   
    
    On the other hand, in the recent paper~\cite{neutrinos} another kind of fermionic particle detector model was introduced with the aim to probe neutrino fields. In that model, the detector would itself be regarded as a fermionic system, and the ground and excited states were thought of as associated with different states of nucleons in the nucleus of an atom. The goal of this section is  to review and generalize the model introduced in~\cite{neutrinos}, and explicitly compute the transition probability for a fermionic detector interacting with a spin $1/2$ field. In particular, we will study in depth its response to a one-particle Fock wavepacket, \tc{so that we can compare our results with those obtained in Subsection \ref{sub:UDW} for the massive real scalar field. This comparison will allow us to analyze the differences in particle detection due to the statistics of the field and to its anti-particle content. First by studying in this section the combined effect of the statistics and anti-particle content of the field in the response of the detector, and later on in Section \ref{sec:Complex} discriminating the two effects by comparing with a scalar model with anti-particle content but bosonic statistics.}
    
    \subsection{The Fermi Interaction}\label{sub:4fermi}
    
    Before generalizing to particle detectors for arbitrary fermion fields, in this subsection we present the technical details of the detector model introduced in~\cite{neutrinos}. \tc{In particular, we will introduce the physical scenario that this particle detector model is inspired from. We will also discuss the mechanisms which preserve the U(1) symmetry in these models, despite the fact that they are linearly coupled.}

    \tc{The motivation for the model of \cite{neutrinos} comes from the decay of nucleons via the weak interaction.} This process is very well described by the four-Fermi theory. In this formalism, the proton, neutron, electron and neutrino are treated as fermionic fields that interact according to the following Lagrangian density,
    \begin{equation}\label{eq:4FLag}
        \mathcal{L}_{\text{4F}} = -\dfrac{G_F}{\sqrt{2}}\Big(\left(\bar{\nu}_e\gamma^\mu (1 - \gamma^5)e\right)\left(\bar{n}\gamma_\mu (1 - \gamma^5)p\right) + \text{H.c.}\Big),
    \end{equation}
    where $p$, $n$, $e$ and $\nu_e$ are the fermionic quantum fields associated with the proton, neutron, electron and electron neutrino, respectively, $\gamma^\mu$ denotes the gamma matrices, with $\gamma^5 = \ii \gamma^0\gamma^1\gamma^2\gamma^3$ and $G_F \approx 1.16 \times 10^{-5}\, \text{GeV}^{-2}$ is the Fermi constant. It is important to remark that the weak interaction only couples to the left-handed fields, which justifies the projector on this subspace, $P_L = (1-\gamma^5)$, in the Lagrangian above.

    In \cite{neutrinos,VanzellaHadrons} it has been argued that a limiting case of the theory in Eq. \eqref{eq:4FLag} can be obtained by treating the neutron-proton part of the system as a two-level system associated with the states $\ket{n}$ for the neutron and  $\ket{p}$ for the proton. This simplified two-level system is then defined around the nucleus' trajectory and is smeared by a four-current $j_\mu(\mf x)$. The procedure of applying this simplification to Eq. \eqref{eq:4FLag} amounts to the following replacement
    \begin{equation}\label{eq:pn}
        \bar{n}\gamma_\mu (1 - \gamma^5)p \longrightarrow j_{\mu}(\mf x)e^{\ii\Delta M \tau}\ket{n}\!\!\bra{p},
    \end{equation}
    where $\Delta M$ denotes the mass difference between the neutron and proton and $\tau$ is the proper time of the  trajectory of the nucleus. 
    
    The next step towards obtaining \tb{a} fermionic detector model is to reduce the electron field to a two-level system smeared by a spinor field. Namely, we want to approximate the electron field by only considering one electron state to be part of the detector description. We will denote such state by $\ket{1}_e$ and assume that the excitation is such that---to a good approximation---it behaves as an energy eigenstate, i.e., its time evolution is given by $e^{-\ii \tm{\omega_e} \tau}\ket{1}_e$, where $\omega_e = \sqrt{|\bm k_e|^2+m^2}$ is the average energy of the electron involved in the process. Effectively, this means that we can write the electron field in Eq. \eqref{eq:4FLag} as
    \begin{equation}\label{eq:el}
        (1-\gamma^5) \hat{e} \longrightarrow u(\mf x)e^{-\ii\tm{\omega_e} \tau} \hat{a}_e,
    \end{equation}
    where $\hat{a}_e$ is the annihilation operator associated with the state $\ket{1}_e$, that is, $\hat{a}_e^\dagger \ket{0}_e = \ket{1}_e$ and $u(\mf x)$ is the left-handed projection of the spinor field associated with the spatial profile of the state. In this approximation, the quantum degree of freedom of the electron field can be associated with the states $\ket{0}_e$ and $\ket{1}_e$ related to the absence of the electron (vacuum of the field) and its presence, respectively. Notice that the reduction in Eq.~\eqref{eq:el} should not be thought of as an approximation. Rather, it represents the particular choice of process that gives rise to the physical system that we will be able to characterize as a simple two-level fermionic detector model.
    
    To probe the neutrino field, we can focus on the $\beta$-decay, where a neutron can decay into a proton, electron and anti-neutrino,
    \begin{equation}
        n \longrightarrow p+e^- + \bar{\nu}_e.
    \end{equation}
    Particularizing  to this process allows one to obtain the appropriate smearing field $u(\mf x)$ and also the state $\ket{1}_e$ associated with the electron field. In flat spacetimes and according to the usual mode decomposition for spinor fields, one can write the $\hat{e}(\mf x)$ as
    \begin{equation}
        \hat{e}(\mf x) = \sum_{s=1}^2\int \dd^3 \bm k \left(u_{s}(\bm k)e^{\ii \mf k \cdot \mf x}\hat{a}_{\bm k,s}+v_{s}(\bm k)e^{-\ii \mf k \cdot \mf x}\hat{b}^\dagger_{\bm k,s}\right),
    \end{equation}
    where $u_{s}(\bm k)e^{\ii \mf k \cdot \mf x}$ and $v_{s}(\bm k)e^{-\ii \mf k \cdot \mf x}$ are a basis of solutions to Dirac's equation and $\hat{a}^\dagger_{\bm k,s}$ and $ \hat{b}^\dagger_{\bm k,s}$ are the creation operators associated with particles and anti-particles, respectively. We simplify the the electron's description by assuming that the state for an electron involved in the $\beta$-decay is associated with momentum profile functions $\varphi_s(\bm k)$ centred around a momentum $\bm k_e$ for $s=1,2$. That is, the electron one-particle state involved in the process is of the form
    \begin{equation}
        \ket{1}_e = \sum_{s=1}^2\int \dd^3 \bm k \varphi_s(\bm k)\hat{a}^\dagger_{\bm k,s}\ket{0}_e,
    \end{equation}
    where $\ket{0}_e$ denotes the vacuum of the electron field. With this in mind, it is possible to consider a description for the electron field that only takes into account the energy modes associated with the state $\ket{1}_e$, as we have discussed above. Notice that this description gets rid of the antiparticle content, which again should not be thought of as an approximation, but rather as a reflection of the fact that it does not play any role in the $\beta$-decay.

    Thus, the particle detector model is capturing the physics  of the proton/neutron+electron system in what concerns the interaction with the neutrino field. Namely, it interacts with the neutrino field according to Eq.~\eqref{eq:4FLag} after the replacements in Eqs.~\eqref{eq:pn} and \eqref{eq:el}. After all the replacements, we obtain the following Lagrangian density for the description of the system
    \begin{align}
        \mathcal{L}_{\text{4F}} &= -\dfrac{G_F}{\sqrt{2}}\Big(e^{\ii(\Delta M - \omega_e)\tau}\bar{\nu}_e\slashed{j}(\mf x)u(\mf x)\hat{a}_e{\color{black}\ket{n}\!\!\bra{p}}
        \\&\:\:\:\:\:\:\:\:\:\:\:\:\:\:\:\:\:\:\:\:\:\:\:\:\:+e^{-\ii(\Delta M - \omega_e)\tau}\bar{u}(\mf x)\slashed{j}(\mf x){\nu}_e\hat{a}_e^\dagger{\color{black}\ket{p}\!\!\bra{n}}\Big).\nonumber
    \end{align}
    Notice that in the model above there are only nonzero transition probabilities between the detector states \mbox{$\ket{0}_e\otimes \ket{n}$} and $\ket{1}_e\otimes \ket{p}$. Indeed, this must be the case in order for this description to conserve charge. The last step in building the neutrino particle detector model is to project the theory above into the subspace spanned by $\ket{g} = \ket{1}_e\otimes \ket{p}$ and $\ket{e} = \ket{0}_e\otimes \ket{n}$, whose energies differ by $\Omega = \Delta M - \omega_e$. We then obtain a two-level system coupled to a fermionic field by means of a smearing four-current density $j_\mu(\mf x)$ and a smearing spinor field $u(\mf x)$. We can then write the final interaction Hamiltonian weight in terms of the SU(2) ladder operators $\hat{\sigma}^- = \ket{g}\!\!\bra{e}$ and $\hat{\sigma}^+ = \ket{e}\!\!\bra{g}$:
    \begin{align}\label{eq:neutrinoH}
        \hat{h}_I(\mf x) &= \dfrac{G_F}{\sqrt{2}}\left(e^{\ii\Omega\tau}\bar{\nu}_e\slashed{j}(\mf x)u(\mf x)\sigma^++e^{-\ii\Omega\tau}\bar{u}(\mf x)\slashed{j}(\mf x){\nu}_e\sigma^-\right).
    \end{align}
    As one can easily see, after these reductions we can write the \tb{neutrino particle detector} model in a way that resembles the standard UDW-like models presented before, but in this case one must provide both a spacetime-smearing four-current associated with the nucleons and a spacetime-smearing spinor field associated with the emitted/absorbed electron. It is also important to comment that the nucleus is much more localized than any other relevant lengthscale of the problem, and it is then natural to consider the spatial profile of the hadronic current to be pointlike.
    
        Notice that, albeit simple, this model captures the fundamental features of the physical system modelled is that it has been shown to reproduce the results of neutrino oscillations~\cite{neutrinos}. Namely, the $\beta$-decay of the neutron, i.e., 
    \begin{equation}\label{betaDecay}
        n \rightarrow p + e^- + \bar{\nu}_e,
    \end{equation}
    is described by the deexcitation of the detector system. The energy spectrum of the emitted electron defines the spatial profile of the $u(\mf x)$ field. Similarly, the localization of the neutron/proton system defines the shape of the hadronic current $j^\mu(\mf x)$. In this model, the emission of an anti-neutrino would then be associated with the deexcitation of the detector, while the absorption of an anti-neutrino will be associated with its excitation. The deexcitation of this detector model can also happen due to the stimulated neutron decay via the absorption of a neutrino, 
    \begin{equation}\label{stimulatedBetaDecay}
        n+\nu_e \rightarrow p + e^-,
    \end{equation}
    but this process would require the state of the neutrino field to contain a one-particle excitation to begin with. We will study these situations in a more general setup in the next subsection, where we will generalize this model to curved spacetimes and compute the excitation probability for one-particle field states.
    {\color{black}
\subsection{Discussion of the symmetries of the neutrino detector model}}\label{sub:referee}
    
    \tc{We are now in a position to discuss the U(1) symmetry issues associated with this model. Notice that the neutrino field is a neutral field, implying that it does not transform under local or global U(1) transformations associated with electromagnetism. Nevertheless, being a fermionic field, the neutrino field does posses an effective global U(1) symmetry in this model, associated with the conservation of weak isospin\footnote{\tb{Usually weak isospin is associated with an SU(2) symmetry and involves exchanges of $W$ bosons, however, in the Fermi theory presented above, the $W$ boson is effectively traced out, so that isospin is always conserved and coincides with a global U(1) symmetry of the theory for the four fermionic fields, with different charges associated to each field.}}, which will be discussed later. However, the individual constituents of our particle detector: neutron, proton and electron, have different electric charges and transform differently under the U(1) gauge symmetry. That is, the creation operator $\hat{a}^\dagger_e = \ket{1}_e\!\!\bra{0}_e$ and the raising operator $\ket{n}\!\!\bra{p}$ transform non-trivially under the electromagnetic U(1) symmetry. However, due to the fact that the neutron is neutral, and the electron and proton have opposite charges, the effective raising operator for the neutrino particle detector model, $\hat{\sigma}^+ = \ket{0}_e\!\!\bra{1}_e\otimes \ket{n}\!\!\bra{p}$ transforms trivially under the action of the electromagnetic U(1). Indeed, both the ground and exited states of the detector model are electrically neutral. Therefore, this model is compatible with electromagnetism and does not introduce any symmetry breaking.}
    
    \tc{Thus, in order to  discuss the U(1) symmetry of the model, it is enough to consider the weak isospin. The weak isospin of the four particles considered in the interaction are $+1/2$ for the neutrino and proton and $-1/2$ for the neutron and electron. It is important to remark that although the weak isospin is not always conserved in the interactions of the Standard Model, it is conserved in the effective description provided by the four-Fermi theory. The weak isospin symmetry can be associated with a global U(1) symmetry of the full theory that considers the proton, neutron and electron field as dynamical. In fact, in the next section we will generalize this concept and consider a general U(1) symmetry that can be associated to a fermionic field. With the weak isospin of the fields considered above, we see that under a U(1) transformation, the proton and electron contribution cancel each other, while the neutron and neutrino have opposite weak isospin, implying that the neutron and anti-neutrino both have isospin $-1/2$. In fact, in this model one can think of the anti-neutrino as a (weak isospin) charged field interacting with a detector of same charge. The fact that the charges of the detector and field are the same is important to ensure conservation of the U(1) symmetry of the theory used to derive the model.}
    
    \tc{The discussion above should make evident that this particle detector model does not break any intrinsic symmetry of the four-Fermi theory from which it is derived. However, in order to simplify the discussion regarding the detector, it is usual to consider that the U(1) transformations that would act on the operators $\hat{\sigma}^+$ and $\hat{\sigma}^-$ act on the fermionic field and current associated to the detector. That is, instead of considering a weak isospin transformation that maps \mbox{$\ket{n} \rightarrow e^{\frac{\text{i}}{2}\theta}\ket{n}$}, \mbox{$\ket{p} \rightarrow e^{-\frac{\text{i}}{2}\theta}\ket{p}$} and \mbox{$\ket{1}_e \rightarrow e^{\frac{\text{i}}{2}\theta}\ket{1}_e$}, we consider the transformations to equivalently act (at a classical level) on the smearing spinor and hadronic currents that are associated with the corresponding particles. For global transformations, it is then equivalent to consider the isospin transformations to act as $\bar{u}(\mf x) \rightarrow e^{\frac{\text{i}}{2}\theta}\bar{u}(\mf x)$ and $j^\mu(\mf x) \rightarrow e^{-\text{i}\theta} j^\mu(\mf x)$, leaving the operators unchanged. With this, we have $\bar{u}(\mf x)j^\mu(\mf x) \rightarrow e^{-\frac{\text{i}}{2}\theta}\bar{u}(\mf x) j^\mu(\mf x)$, while the anti-neutrino field transforms according to $\nu_e(\mf x)\rightarrow e^{\frac{\text{i}}{2}\theta}\nu_e(\mf x)$, leaving the interaction Hamiltonian density unchanged. This equivalent way of rephrasing the global symmetry of the theory makes the two-level system of the model invariant under gauge transformations.}
    
    \textcolor{black}{It is also interesting to study the $\mathbb{Z}_2$ charge conjugation symmetry of the theory, and how it affects the model. \tm{Charge conjugation is implemented in a fermionic field as $\psi(\mf x)\mapsto \psi^c$ is the charge conjugated field operator. Notice that the interaction Hamiltonian weight from the linear fermionic detector is not invariant under this transformation. This is no surprise. In fact, the Fermi interaction Lagrangian in Eq. \eqref{eq:4FLag} will not be invariant under charge conjugation if one only transforms the neutrino field. In order to discuss the charge conjugation invariance, all fields must be transformed under charge conjugation. Given that our fermionic particle detector is built as an effective limit of these fields, it is only natural to expect that the detector must also transform under charge conjugation. In fact, the complete action of the $\mathbb{Z}_2$ charge conjugation symmetry on the detector field system should act in the field according to $\psi(\mf x)\mapsto \psi^c$ and in the detector components according to $\Lambda(\mf x)\mapsto \Lambda^c(\mf x) = -\ii \gamma^2\Lambda^*(\mf x)$, $\sigma^{\pm}(\tau)\mapsto (\sigma^{\pm}(\tau))^\dagger = \sigma^{\mp}(\tau)$. Indeed, understood this way, the charge conjugation applied to the detector-field system would leave the interaction Hamiltonian invariant, and hence the theory is symmetric under charge conjugation. However, as }\tc{we will see in detail in Subsection \ref{sub:fermion}, the linear fermionic detector model presented in Subsection \ref{sub:4fermi} is able to distinguish between particle and anti-particle content. This might look conflicting with the \tm{previous} discussion, where we show that the interaction Hamiltonian does not break any symmetry of the theory. Indeed, it is the detector states are responsible for breaking such symmetry: the ground state is associated with a proton and electron state, and thus has no total weak isospin, while the ground state is associated with the neutron, which has weak isospin $-1/2$.}}

    \tc{The last step to complete the discussion regarding symmetries of the theory is to discuss Lorentz symmetry. Notice that the fact that $u(\mf x)$ is a spinor field, $j^\mu(\mf x)$ is a vector field (and thus $\slashed{j}(\mf x)$ is an operator in the spinor bundle) and $\nu_e(\mf x)$ is also a spinor field, allows us to conclude that the contractions $\bar{\nu}_e\slashed{j}(\mf x)u(\mf x)$ and $\bar{u}(\mf x)\slashed{j}(\mf x){\nu}_e$ are invariant under local Lorentz transformations, producing a Lorentz scalar Hamiltonian weight, as expected.}

    \tm{Finally, {we discuss a subtlety regarding} the gauge behaviour of the model. In order to understand what happens when a non-relativistic quantum system couples with a quantum field with a gauge degree of freedom, let us first make an argument for the well-understood case of the light-matter interaction. When one considers the dipole coupling of an atom with the electric field from Eq. \eqref{eqDipole}, the dipole operator is given in terms of the wavefunctions of the electron (see, e.g.,~\cite{Edu2013,Pozas2016,richard}). These wavefunctions are gauge dependent: under a gauge transformation $A_\mu \mapsto A_\mu + \partial_\mu f$, the electron wavefunction transforms according to $\psi(\mf x) \mapsto e^{-\text{i} e f(\mf x)}\psi(\mf x)$. Although this does leave the interaction Hamiltonian $\hat{\bm d} \cdot \hat{\bm E}$ invariant, the electric potential term and kinetic term in the free Hamiltonian in Eq. \eqref{eq:free} will change in different ways. The potential term will change by $\partial_t f$ and the kinetic term involves space derivative operators, so that it yields different results when applied to $\psi(\mf x)$ or $e^{-\text{i} e f(\mf x)}\psi(\mf x)$. Also notice that this issue is present even if the external electric field is classical. In fact this model, as stated, is gauge dependent. The way to properly address gauge invariance in the light-matter interaction is to show in what way the dipole coupling emerges by an approximate gauge transformation from the minimal coupling Hamiltonian, and that requires to take into account the center of mass delocalization of the atom and the fact that there will be a coupling of the trajectory of the detector to the field via a R\"ontgen term~\cite{richard}. Nevertheless, in~\cite{richard} it was shown that under the typical conditions that one deals with for atoms moving along classical trajectories for times much longer than the light-crossing time of the atom, the effective dipole model accurately predicts the outcome of experiments.}

    
    \tm{A gauge problem also arises in the fermionic particle detector model presented in Subsection \ref{sub:4fermi}. In this model the prescription of the `electron wavefunction' $u(\mf x)$, together with $\slashed{j}(\mf x)$ depend on a gauge choice. However, the issue is not as grave as the one that shows up in the light-matter interaction for two reasons: first, our interaction Hamiltonian is gauge independent. Second, we are not coupling to a gauge field, only to a field that has (weak isospin) charge. In fact, in our specific example, the prescription of the `electron wavefunction' $u(\mf x)$ and hadronic current $j^\mu(\mf x)$ is given in terms of the energy spectrum of the particles involved in the $\beta^-$-decay process, as done in Eqs.~\eqref{eq:el} and \eqref{eq:pn}.}
    
    \tm{However, one should not overlook the gauge dependence of the predictions of the model. Indeed, a possible issue arises in the prescription of these smearing functions in a given fixed gauge. While considering this issue is of paramount importance to match the predictions to a specific experiment, one can argue that---same as in the light-matter interaction---the gauge dependence of the predictions will not affect the qualitative behaviour of the results. The fact that the interaction Hamiltonian is invariant under Lorentz, U(1) and charge-conjugation transformations provides a good argument for the result independence of this residual uncontrolled degree of freedom. Nevertheless, we emphasize again that in order to match specific experiments, these issues should not be overlooked. We leave a detailed study of the gauge freedom in these models for future works.}
    
    \vspace{5mm}
    
    \subsection{A particle detector probing a spin 1/2 field}\label{sub:fermion}
    
    A general free massive spin $1/2$ quantum field in a $(3+1)$ dimensional spacetime can be expanded as
    \begin{equation}\label{eq:fermionField}
        \hat{\psi}(\mf x) = \sum_{s=1}^2 \int \dd^3 \bm p\left( u_{\bm p,s}(\mf x)\hat{a}_{\bm p,s} + v_{\bm p,s}(\mf x) \hat{b}^\dagger_{\bm p,s}\right),
    \end{equation}
    where the spinors $u_{\bm p,s}(x)$ and $v_{\bm p,s}(x)$ constitute a basis of solutions to Dirac's equation in the given spacetime. Here $s$ labels the spin of the solutions, while $\hat{a}_{\bm p,s}$ are the annihilation operators associated with particles and $\hat{b}^\dagger_{\bm p,s}$ are the creation operators associated with antiparticles. We choose a normalization for $u_{\bm p,s}(\mf x)$ and $v_{\bm p,s}(\mf x)$ such that the creation and annihilation operators satisfy the anti-commutation relations
    \begin{align}
        \big\{\hat{a}^{\vphantom{\dagger}}_{\bm p},\hat{a}^\dagger_{\bm p'} \big\} &= \delta^{(3)}(\bm p - \bm p'),&&& \big\{\hat{b}^{\vphantom{\dagger}}_{\bm p},\hat{b}^\dagger_{\bm p'} \big\} &= \delta^{(3)}(\bm p - \bm p'),
    \end{align}
    \tm{with all other anti-commutators vanishing.}
    
    \color{black} According to the discussion of the previous subsection, a detector model that couples to a fermionic field linearly must itself have a fermionic degree of freedom in order to  preserve the \tb{Lorentz symmetry} of the theory. We then include this fermionic degree of freedom in the spacetime smearing function of the detector, regarding $\Lambda(\mf x)$ as a classical fermionic field with support centered around the detector's worldline. Once again, the detector will be described by a two-level quantum system governed by the (internal) free Hamiltonian from Eq.~\eqref{eq:detectorH}. 
    
    The interaction Hamiltonian weight for the field-detector system within this model can be written as a straightforward generalization of Eq.~\eqref{eq:neutrinoH}:
    \begin{equation}\label{hIfermion}
        \hat{h}_I(\mf x) \!= \!\lambda\!\left(\!e^{\ii\Omega \tau}\hat{\sigma}^+\hat{\bar{\psi}}(\mf x)\slashed{j}(\mf x)\Lambda(\mf x) + e^{-\ii\Omega \tau}\hat{\sigma}^-\bar{\Lambda}(\mf x)\slashed{j}(\mf x)\hat{\psi}(\mf x)\!\right)\!.
    \end{equation}
    \tc{Following the discussion in Subsection~\ref{sub:referee}, we allow $\Lambda(\mf x)$ to transform under $U(1)$ transformations in the opposite way as that of the $\bar{\psi}(\mf x)$ field\tm{. This  makes the theory U(1) invariant.} Furthermore, we remind the reader that being a spinor, $\bar{\Lambda}(\mf x)$ transforms under Lorentz transformations in the inverse way that $\hat{\nu}(\mf x)$ does (and the analogous statement for the conjugate fields), so that the theory is also invariant under} any local Lorentz transformation in the spinor bundle. By comparison with the previous section, we assume $\lambda$ to have units of $[E]^{-2}$\tc{, $\Lambda(\mf x)$ to have units of a spin $1/2$ field, that is, $[E]^{\frac{3}{2}}$, and $j^\mu(\mf x)$ to have units of the square of a spinor, or, equivalently, of a density in space, that is, $[E]^{3}$.}
    
    \tm{Before studying the response of this particle detector model to particle and anti-particle states, it is important to make a remark about causality: It is well known that due to their intrinsic non-relativistic nature, smeared particle detectors probe the quantum field simultaneously at multiple spacelike separated points within the smearing region. This issue has been thoroughly studied in \cite{us2,PipoFTL}. However, the current studies regarding this feature of particle detectors consider real scalar fields that satisfy canonical \emph{commutation} relations. Up to this point there has been no study of causality violations and communication using detectors linearly coupled to fermionic particle detectors. For this reason, one should be careful when using these models for relativistic quantum informational protocols until a proper analysis of causality within this model is carried out. Nevertheless, for the purpose of this manuscript, we will only consider the long-time response of one particle detector to particle and anti-particle excitations where these issues are not relevant.}
    
    An arbitrary one-particle state for the fermionic field in this theory can be written in terms of four functions, $f_1,f_2,g_1,g_2$ defined in momentum space and associated with the two different spin polarizations of the field and \tb{its} particle and anti-particle content respectively. The one-particle Fock state is explicitly given by
    \begin{equation}\label{eq:psiFermion}
        \ket{\varphi} = \sum_{s=1}^2\int \dd^3 \bm k (f_s(\bm k) \ket{\bm k,s}+g_s^*(\bm k)\ket{\bar{\bm k},s}),
    \end{equation}
    where we denote $\ket{{\bm k},s} = \hat{a}_{\bm k,s}^\dagger \ket{0}$ and $\ket{\bar{{\bm k}},s} = \hat{b}_{\bm k,s}^\dagger \ket{0}$. In order for this state to be normalized, we then get the following condition on the $L^2$ norm of the functions $f_s$ and $g_s$,
    \begin{equation}
        \sum_{s=1}^{2} \left(\norm{f_s}^2_2+\norm{g_s}^2_2\right) = 1.
    \end{equation}
    
    The \tb{leading order} transition probability for a linear fermion detector when interacting with \tb{the field in }a general one-particle state of the form~\eqref{eq:psiFermion} is computed in detail in Appendix~\ref{ap:spinor}. It reads
    \begin{widetext}
    \begin{equation}\label{eq:pgeFermion}
        p_{g\rightarrow e} = \lambda^2\int \dd V \dd V'  e^{-\ii\Omega(\tau-\tau')} \bar{\Lambda}(\mf x)\slashed{j}(\mf x)\left(W_0(\mf x,\mf x') +G(\mf x)\bar{G}(\mf x') - F(\mf x')\bar{F}(\mf x)\right)\slashed{j}(\mf x')\Lambda(\mf x'),
    \end{equation}
    \end{widetext}
    where we have defined the spinors $F(\mathsf{x})$ and $G(\mf x)$ and the (particle sector) two-point vacuum  Wightman tensor $W_0$ as
    \begin{align}
        F(\mf x) &= \sum_{s=1}^2\int \dd^3 \bm k f_s(\bm k)u_{\bm k,s}(\mf  x),\nonumber\\
        G(\mf x) &= \sum_{s=1}^2\int \dd^3 \bm k g_s(\bm k)v_{\bm k,s}(\mf  x),\label{eq:Fspinor}\\
        W_0(\mf x,\mf x') &= \bra{0}{\hat{\psi}}(\mf x)\hat{\bar{\psi}}(\mf x')\ket{0} =\sum_{s=1}^2\int \dd^3 \bm p\: u_{\bm p,s}(\mf x)\bar{u}_{\bm p,s}(\mf x').\nonumber
    \end{align}
    
    As we have seen in the previous sections, the vacuum Wightman term corresponds to the vacuum excitation probability of the detector, while the $F(\mf x)$ and $G(\mf x)$ functions correspond to the particle and anti-particle content contributions, respectively. At this stage, it is possible to see a major difference between the previous (bosonic) cases and the fermionic spin 1/2 field. Namely, {from the minus sign in Eq. \tm{\eqref{eq:pgeFermion}}, }we see that the particles contribute negatively to the excitation probability. This effect can be understood by looking at the physical system that the detector models. Indeed, the excitation of the detector is associated with the term $\hat{\sigma}^+(\tau)\hat{\bar{\psi}}(\mf x)\slashed{j}(\mf x)\Lambda(\mf x)$ in the \tc{interaction}. This term \tc{represents the} annihilation of a $\psi$ anti-particle or the creation of a $\psi$ particle \tc{promoting the excitation of the detector. In the case of the neutrino particle detector presented in Subsection \ref{sub:4fermi} these would be associated with the following inverse $\beta$ decay processes:}
    {\color{black}
    \begin{align}
        \bar{\nu}_e+p + e^- &\rightarrow n, \\
        p + e^- &\rightarrow n + \nu_e.\label{hard}
    \end{align}}
    \tc{Due to the Dirac statistics satisfied by the fermionic field, if there is already particle content in the field, the process from Eq. \eqref{hard} becomes less likely, due to the Pauli exclusion principle on states that are already occupied. This reasoning provides intuition for the minus sign in Eq. \eqref{eq:pgeFermion}. In fact, in Section \ref{sec:Complex}, we will study a field with arbitrary statistics (that can be switched from bosonic to fermionic), and will be able to connect the negative sign with Fermi-Dirac statistics.}
     
    On the other hand, the deexcitation of the detector has the opposite behaviour: the roles of the functions $G(\mf x)$ and $F(\mf x)$ are swapped in Eq. \eqref{eq:pgeFermion} and the particle sector vacuum Wightman is swapped with the anti-particle sector Wightman tensor, $\overline{W}_0$,
    \begin{equation}\label{eq:Woconjugate}
        \overline{W}_0(\mf x,\mf x') = \bra{0}\hat{\bar{\psi}}(\mf x)\hat{\psi}(\mf x')\ket{0} = \sum_{s=1}^2\int \dd^3 \bm p\: v_{\bm p,s}(\mf x')\bar{v}_{\bm p,s}(\mf x) .
    \end{equation}
    These changes are associated with the fact that the part of the interaction Hamiltonian corresponding to the deexcitation is the term $\hat{\sigma}^-(\tau)\bar{\Lambda}(\mf x)\slashed{j}(\mf x)\hat{\psi}(\mf x)$, related to the emission of a $\psi$ anti-particle, or the absorption of a $\psi$ particle.
    
    Another interesting aspect of this detector model can be seen when one considers the vacuum excitation and deexcitation probabilities. In these cases we have that the \tb{leading order} excitation and deexcitation probabilities will be given by
    \begin{widetext}
    \begin{align}
        p_{g\rightarrow e} = \lambda^2\int \dd V \dd V'  e^{-\ii\Omega(\tau-\tau')}\bar{\Lambda}(\mf x)\slashed{j}(\mf x) \bra{0}{\psi}(\mf x)\bar{\psi}(\mf x')\ket{0}\slashed{j}(\mf x')\Lambda(\mf x'),\\
        p_{e\rightarrow g} = \lambda^2\int \dd V \dd V'  e^{\ii\Omega(\tau-\tau')} \bra{0}\bar{\psi}(\mf x)\slashed{j}(\mf x)\Lambda(x)\bar{\Lambda}(\mf x')\slashed{j}(\mf x'){\psi}(\mf x')\ket{0}.
    \end{align}
    \end{widetext}
    We then see that the probabilities above depend on different Wightman functions of the field. This is in contrast with the case of real field detectors, where the Wightman function associated with the negative frequency solutions would be the conjugate of the one associated with positive frequencies. Let us compare the excitation versus de-excitation probability of a fermionic detector interacting with vacuum in a flat spacetime. We can rewrite the transition probabilities as
    \begin{widetext}
    \begin{align}
        p_{g\rightarrow e} &= \lambda^2\int \dd V \dd V'  e^{-\ii\Omega(\tau-\tau')} \bar{\Lambda}(\mf x)\slashed{j}(\mf x)\int \dd^3 \bm p\frac{\slashed{\mf{p}} + m}{2\omega_{\bm p}} e^{\ii \mf p\cdot (\mf x-\mf x')}\: \slashed{j}(\mf x')\Lambda(\mf x'),\label{eq:pomega}\\
        p_{e\rightarrow g} &= \lambda^2\int \dd V \dd V'  e^{-\ii\Omega(\tau-\tau')} \bar{\Lambda}(\mf x)\slashed{j}(\mf x)\int \dd^3 \bm p\: \frac{-\slashed{\mf{p}} + m}{2\omega_{\bm p}} e^{-\ii \mf p\cdot (\mf x-\mf x')}\slashed{j}(\mf x')\Lambda(\mf x')\\
        &= \lambda^2\int \dd V \dd V'  e^{\ii\Omega(\tau-\tau')} \bar{\Lambda}(\mf x')\slashed{j}(\mf x')\int \dd^3 \bm p\: \frac{-\slashed{\mf{p}} + m}{2\omega_{\bm p}} e^{\ii \mf p\cdot (\mf x-\mf x')}\slashed{j}(\mf x)\Lambda(\mf x).\label{eq:pomegade}
    \end{align}
    \end{widetext}
    From the expressions above, it is clear that unlike what happens in the scalar case, the excitation probability of the detector, is not related to the deexcitation probability simply by a level inversion $\Omega \rightarrow -\Omega$. This happens because the effect of changing the sign of the energy gap in the Hamiltonian from Eq. \eqref{hIfermion} is not only to swap the roles played by $\hat{\sigma}^+$ and $\hat{\sigma}^-$. The swap $\Omega \rightarrow - \Omega$ would also change whether the excitation would be associated with the absorption of a fermionic particle or antiparticle. Mathematically, this can be be traced back to the fact that both $\hat{\sigma}^+$ and $\hat{\sigma}^-$ can couple to either the field $\hat{\psi}$ or to its conjugate $\hat{\bar{\psi}}$. \textcolor{black}{It is important to remark that while the full model does not break any symmetries of the theory, the excitation and deexcitation probabilities of the detector consider specific states for the probe. As mentioned in Subsection \ref{sub:referee}, the choice of states is the responsible for breaking the $\mathbb{Z}_2$ charge conjugation of the theory, as can be seen in Eqs. \eqref{eq:pomega} and \eqref{eq:pomegade}.}
    
    To finish this Section, we consider an example of fermionic particle detection in flat spacetime, with a pointlike inertial detector.
    We assume the initial state of the field to be the one-particle Fock wavepacket $\ket{\varphi}$, given in Eq. \eqref{eq:psiFermion} with the following choices of momentum distribution for each spin polarization,
    \begin{align}
        f_1(\bm k) &= \frac{\alpha_1}{(\pi \sigma^2)^{\frac{3}{4}}}e^{-\frac{|\bm{k}-\bm{k}_0|^2}{2\sigma^2}},&&&
        f_2(\bm k) &= \frac{\alpha_2}{(\pi \sigma^2)^{\frac{3}{4}}}e^{-\frac{|\bm k-\bm k_0|^2}{2\sigma^2}},\nonumber \\
        g_1(\bm k) &= \frac{\beta_1}{(\pi \sigma^2)^{\frac{3}{4}}}e^{-\frac{|\bm k-\bm k_0|^2}{2\sigma^2}},&&&
        g_2(\bm k) &= \frac{\beta_2}{(\pi \sigma^2)^{\frac{3}{4}}}e^{-\frac{|\bm k-\bm k_0|^2}{2\sigma^2}}.\label{eq:fSpinor}
    \end{align}
    Here $|\alpha_1|^2 + |\alpha_2|^2 + |\beta_1|^2 + |\beta_2|^2 = 1$. These coefficients control the particle/anti-particle content in each spin polarization.
    
    We then pick the following hadronic current $j^\mu(\mf x)$ and spinor smearing field $\Lambda(\mf x)$ for the detector:
    \begin{align}\label{eq:spinorChoices}
        j^\mu(\mf x) &= \delta^{(3)}(\bm x) u^\mu, &&& \Lambda(\mf x) &=  \Delta^{\frac{3}{2}}\chi(t)\left(\begin{array}{c}
        A_1\\A_2\\B_1\\B_2
        \end{array}\right),
    \end{align}
    where $u^\mu$ is the detector's four-velocity, $\Delta$ is a constant with units of energy, $\chi(t)$ is a switching function and the spinor field $\Lambda$ is written in the Dirac basis and, without loss of generality, it is chosen to be spatially independent, since any position dependence would be washed out by the Dirac delta factor in $j^\mu(\mf x)$. We will also assume that the basis chosen for the spinors is such that the $z$ component of spin is aligned with the mean momentum in the distribution (the peak of the Gaussian spectrum, $\bm k_0$).  The constant $\Delta$ takes into account the fact that the components of the spinor $\Lambda$ chosen above must have units of $[E]^{\frac{3}{2}}$ due to the spin $1/2$ nature of the field. $\Delta$ is then associated with a scale of energy for the interaction. For example, in the case of a detector probing the neutrino field, we would have $\Delta \approx m_e$, as was discussed in Subsection \ref{sub:4fermi}. The normalization condition for the spinor $(A_1,A_2,B_1,B_2)$ imposes that $|A_1|^2+|A_2|^2+|B_1|^2+|B_2|^2 =1$.
    
    In Appendix~\ref{ap:spinor} we compute in full detail the long time excitation probability for the detector in the adiabatic limit when it responds to the wavepacket in Eq.~\eqref{eq:psiFermion}. In the long time regime, we obtain the following result:
    \begin{align}\label{probFermion}
        p_{g\rightarrow e} = &\frac{4 \lambda^2 \sigma\Omega\Delta^3(\Omega+m)}{\sqrt{\pi}|\bm k_0|^2 }|\mathlarger{\gamma}^+(\Omega,m,,\bm k_0,\sigma)|^2\nonumber\\
        &\times e^{-\frac{|\bm k_0|^2+\Omega^2-m^2}{\sigma^2}}\sinh^2\left(\frac{|\bm k_0|\sqrt{\Omega^2-m^2}}{\sigma^2}\right)\nonumber\\
        &\:\:\:\:\:\:\:\:\:\:\:\times \Theta(\Omega - m),
    \end{align}
    where $\gamma^+$ is an adimensional function given by
    \begin{align}\label{eq:f}
        \mathlarger{\gamma}^+(\Omega,m &,\bm k_0,\sigma) =  \beta_1 B_1^* + \beta_2 B_2^*\\
        &\!\!\!+(\beta_1A_1^*-\beta_2A_2^*)\sqrt{\frac{\Omega-m}{\Omega + m}}\,\mathlarger{f}\!\left(\frac{\bm |\bm k_0|\sqrt{\Omega^2-m^2}}{\sigma^2}\right),\nonumber
    \end{align}
    and $f(u) = \text{coth}(u) - 1/u$ is the pole-removed hyperbolic cotangent function and satisfies $0\leq f(u)\leq 1$ for \mbox{$u\geq 0$}. Given the normalization for the spinor $\Lambda$ and for the state $\ket{\varphi}$, we see that  $|\gamma^+|^2$ is bounded by $1$. Also notice that in the adiabatic long time limit, only the anti-particle portion of the state $\ket{\varphi}$ contributes to the probability. That is, the particle content of $\ket{\varphi}$ does not influence the excitation of this detector in this regime for the choice of process that motivates our detector, as discussed in Subsection~\ref{sub:4fermi}.
    
    Notice that the dependence of $\gamma^+$ on $\Omega,m,\bm k_0$ and $\sigma$ only appears in the term in the second line of Eq.~\eqref{eq:f}. This term contributes the most near the monochromatic limit $\sigma\ll|\bm k_0|$, where $f\rightarrow 1$. Conversely, when $\sigma\gg |\bm k_0|$, we have $f\rightarrow 0$ and this term can be neglected, simplifying the expression for $\gamma^+$. This term could also be neglected if we were to consider the mass of the fermionic field to be large with respect to $|\bm k_0|$, due to the prefactor $[(\Omega - m)/(\Omega+m)]^{\frac{1}{2}}$. However, assuming $m$ to be large would go against our physical motivation that uses neutrino fields, known to have extremely low mass and to be extremely relativistic. That is, in general all terms in $\gamma^+$ contribute, but we always have $|\gamma^+|^2\leq 1$. 
    

    Notice the similarity between Eq. \eqref{probFermion} for the fermionic field and Eq. \eqref{eq:probFinal3DReal} for the massive real scalar case in three spatial dimensions. However, if we compare the two expressions we see four major differences between the two models: 
    \begin{enumerate}
        \item In the fermionic case we have a dimensionful coupling constant (which, in the detector model inspired by the four-fermion interaction discussed in Subsection~\ref{sub:4fermi}, would be given by $G_F/\sqrt{2}$).
        \item The probability from Eq. \eqref{probFermion} crucially depends on the relative orientation between the spin polarizations of the detector spinor smearing and the wavepacket. This is the spinor analogue to the dependence on orientation between dipole and polarization of the wavepacket for the case of the vector field detector.  
        \item The contribution to the excitation probability for the fermionic detector comes from the antiparticle content of the field, being completely blind to the particle content of the state. This has no analogue in the detector models for real fields we have seen so far. This is not as surprising when one remembers that famous particle physics maxim that says \textit{particles of a real field are their own anti-particles}.
        \item  Unlike the models for probing real fields, the vacuum excitation and deexcitation probabilities are not connected by the $\Omega \rightarrow -\Omega$ exchange. This is due to the fact that the detector operator $\sigma^-$ couples with the field in~\eqref{hIfermion}, while $\sigma^+$ couples to its conjugate. As explained above this is because of the particular process we are using to build the fermionic detector.
        \item Finally, unlike the scalar case, the numerator of Eq.~\eqref{probFermion} is mass dependent. 
    \end{enumerate}
    However, it is arguably unfair to compare the fermionic detector model to other models where the detector couples to real fields. In order to establish a fair comparison where differences are truly rooted in the fermionic nature of the field as opposed to bosonic, we would need to compare the fermionic model with that of a complex bosonic field detector.
    
    

\section{A Detector Model for a Complex Scalar Field}\label{sec:Complex}

    In order to truly understand the differences between fermionic and bosonic detector models, and compare the response of detectors coupled linearly to these two kinds of fields, we first need to build a linear bosonic detector model where we have distinct particle and anti-particle sectors. 
    In this section we propose and study a particle detector model that couples linearly to a complex scalar field. \tc{By contrasting this model to the real scalar particle detector we will better understand the effect that the anti-particle sector of the field has in the detection of a one-particle wavepacket. It is important to remark that this particle detector model is not directly inspired by any physical process. Instead, our goal is to propose and study a toy model that is the simplest that can mimic fundamental features of the fermionic particle detector mode studied in Section \ref{sec:Fermion}. The theoretical model we will propose in this section will then be used to study two different relevant comparisons:} 1) We will build a particle detector model that couples linearly to a complex scalar field and compare it with the fermionic detector developed in Sec.~\ref{sec:Fermion} \tc{and with the UDW model studied in Subsection \ref{sub:UDW}. This will allow us to further our study of the general behaviour of linear particle detectors that couple differently to the particle and anti-particle sectors of a field.} 2) We will compare the cases of a (true bosonic) complex scalar field, with a Grassmann scalar field\footnote{A Grassmann scalar field is a spin-0 field where the complex algebra is anti-commutative instead of commutative (effectively behaving as spin-0 fermions that satisfy Fermi-Dirac statistics).}~\cite{popBound,altlandCMbook} \tm{defined with anti-commutation relations for the creation and annihilation operators rather than the usual commutation relations}. \tc{\tm{At this stage it is important to remark that this Grassmann scalar field does not represent any physical relativistic quantum field. Moreover, it} violates the spin-statistics theorem \tm{in the cases where it applies}  
    \tm{and does not satisfy the fundamental microcausality condition\footnote{\tm{In more detail, if the particle and anti-particle creation and annihilation operators both satisfy anti-commutation relations, it is easy to show that $\psi(\mf x)$ will not anti-commute with itself when evaluated at spacelike separated points, thus violating the microcausality condition for a fermionic field. A way to avoid microcausality violations in the Grassmann scalar case modifying the signs of the anti-commutators of the anti-particle sector of the field was proposed in past literature~\cite{fermionicScalars}. However this introduces unphysical states with negative norm, which would give rise to unphyiscal negative probabilities when computing detector responses. More important for us, if we did this modification of the anti-commutators we would not be able to reproduce the response of the fermionic spin $1/2$ field we have seen in Section \ref{sec:Fermion}, defeating the purpose of proposing an effective model that captures the features of the fermionic detector response.}}. In this Section, however, we allow for the complex scalar field to be able to satisfy both the physical commutation relations and the unphysical anti-commutation relations, as a toy model to compare and identify which features of the fermionic particle detector model} \tm{interacting with a localized particle state are related to the Fermi-Dirac statistics and which ones are related to the presence of particles and anti-particles. In essence, as  mentioned above, we want to study the properties of a simplified version of the fermionic particle detector of~\cite{neutrinos} and to decide whether these simplifications can be used in order to mimic the behaviour of the detector studied in Section \ref{sec:Fermion}. A note of warning however should be given against the use of our brute-forced Grassmann scalar field with anti-commutation relations for the study of any feature of the detectors where relativity, and in particular the causal structure of spacetime, is important (e.g., communication~\cite{Katja,Simidzija_2020}, entanglement harvesting~\cite{Valentini1991,Reznik2003,Reznik1,Pozas-Kerstjens:2015,Pozas2016,Nick,topology}, etc). We only consider this unphysical Grassmann scalar field for the purposes of assessing the impact of anti-commutation of the creation and annihilation operators on the response of a single detector to excitations of the field.}} 
    
    In general, a free complex scalar quantum field in an \mbox{$n+1$} dimensional spacetime can be expanded in terms of a basis of solutions to the classical equations of motion according to 
    \begin{align}
        \hat{\psi}(\mf x) &=\int \dd^n \bm p \left(u_{\bm p}(\mf x) \hat{a}_{\bm p}+u_{\bm p}^*(\mf x) \hat{b}^\dagger_{\bm p}\right),\label{eq:fieldComplex}\\
        \hat{\psi}^\dagger(\mf x) &= \int \dd^n \bm p \left(u_{\bm p}^*(\mf x) \hat{a}^\dagger_{\bm p}+u_{\bm p}(\mf x) \hat{b}_{\bm p}\right),
    \end{align}
    where $\{u_{\bm p}(\mf x),u^*_{\bm p}(\mf x)\}$ is a basis of solutions to Klein-Gordon's equations, $\hat{a}_{\bm p}$ are the annihilation operators associated with particles and $\hat{b}_{\bm p}$ are the ones associated with the annihilation of anti-particles. At this stage we could impose commutation relations for the creation and annihilation operators and have a regular complex scalar field. However, in order to generalize the model to consider Grassmann fields as well \tc{in order to better compare with the fermionic particle detector}, we will impose arbitrary sign commutation relations
    \begin{align}\label{eq:generalComm}
        \big[\hat{a}^{\vphantom{\dagger}}_{\bm p},\hat{a}^\dagger_{\bm p'} \big]_{\pm} &= \delta^{(3)}(\bm p - \bm p'),&&& \big[{\hat{b}^{\vphantom{\dagger}}_{\bm p}},{\hat{b}^\dagger_{\bm p'} }\big]_{\pm} &= \delta^{(3)}(\bm p - \bm p'),
    \end{align}
    so that `$+$' denotes anti-commutation and `$-$' commutation \tm{and all other commutators/anti-commutators vanish}. This general approach allows us to distinguish results that depend on the fact that the field has anti-particle content from the features that depend on whether it satisfies Bose-Einstein or Fermi-Dirac statistics.
    
    To build the interaction of the detector with the field, we proceed along the lines of what was done in Section \ref{sec:Fermion}. We propose a particle detector model that has a complex degree of freedom that will be encompassed in the detector's smearing function. This means that the detector will also change under $U(1)$ transformations, and the interaction of the detector and the field will remain invariant under $U(1)$. Similarly to the other detector models studied in this manuscript, we will assume the detector to be described by a two-level system whose internal quantum degree of freedom is supported along a trajectory $\mf z(\tau)$. The detector's free Hamiltonian will then be given by Eq. \eqref{eq:detectorH}. The interaction with the field will then be described by the  Hamiltonian weight\vspace{5mm}
    \begin{equation}\label{hIcomplex}
        \hat{h}_I(\mf x) =  \lambda\left(\Lambda(\mf x)e^{\ii\Omega \tau}\hat{\sigma}^+\hat{\psi}^\dagger(\mf x) + \Lambda^*(\mf x)e^{-\ii\Omega \tau}\hat{\sigma}^-\hat{\psi}(\mf x)\right),\vspace{5mm}
    \end{equation}
    where $\Lambda(\mf x)$ is a complex smearing function that regulates the coupling of the detector with the field and transforms under $U(1)$ \tc{in the same way to that of the complex field} so that the interaction Hamiltonian is \mbox{$U(1)$-invariant}, as expected in order to preserve the symmetries of the field theory. \tc{This is very similar to what was discussed in the fermionic particle detector model in Subsection \ref{sub:referee}, however, the weak isospin discussion does not fit here due to the lack of a real physical process associated to this model. Nevertheless, for all purposes, one can think that the complex field is an electrically charged field, and so is the particle detector\tm{\footnote{\tm{Although in most cases charged fields are only observable in pairs, there are important physical processes that can measure these fields via interaction with \emph{another} charged field. This is the case of anti-neutrino detection via the $\beta^-$-decay process and its inverse, where the charge considered is the weak isospin.}}}. Then the intuitive picture for the probe deexcitation would be associated with the decay of a localized particle associated with the detector, producing a particle in the field. Conversely, the excitation of the detector would correspond to the absorption of a field quanta and the creation of the localized particle associated with the detector's exited state.}
    
    We will consider a general one-particle state with arbitrary particle and anti-particle content:
    \begin{equation}\label{eq:psiComplex}
        \ket{\varphi} = \int \dd^n \bm k\, (f(\bm k)\hat{a}^\dagger_{\bm k} +g^*(\bm k)\hat{b}^\dagger_{\bm k})\ket{0},
    \end{equation}
    where $f(\bm k)$ and $g(\bm k)$ are the momentum distributions of the particle and anti-particle components, respectively. Normalization of the state $\ket{\varphi}$  implies that the $L^2$ norm of $f$ and $g$ satisfy $\norm{f}_2^2+\norm{g}_2^2 = 1$.
    
    We can then compute the transition probability for the detector to go from the ground to the excited state using similar techniques to what was performed in the previous sections
    \begin{equation}\label{eq:probComplex}
        p_{g\rightarrow e} = \lambda^2\!\!\int \dd V \dd V' \Lambda^*(\mf x')\Lambda(\mf x) e^{\ii\Omega(\tau-\tau')}\! \bra{\varphi}\!\hat{\psi}(\mf x')\hat{\psi}^\dagger(\mf x)\!\ket{\varphi}\!.
    \end{equation}
    Notice that in this case, the excitation probability and the \tb{leading order}  deexcitation probability have a similar interpretation to that of the fermionic model in Section \ref{sec:Fermion}: the excitation of the detector is associated with either the absorption of a $\psi$ anti-particle or emission of a $\psi$ particle. Conversely, the probability of deexcitation $p_{e\rightarrow g}$ will be associated with the absorption of particles or emission of anti-particles. It is explicitly given by
    \begin{equation}
        p_{e\rightarrow g} = \lambda^2\!\!\int \dd V \dd V' \Lambda(\mf x')\Lambda^*(\mf x) e^{\ii\Omega(\tau'-\tau)}\! \bra{\varphi}\!\hat{\psi}^\dagger(\mf x')\hat{\psi}(\mf x)\!\ket{\varphi}\!.
    \end{equation}

    The general excitation probability from Eq. \eqref{eq:probComplex} can be computed by plugging in the value for the Wightman function of the state $\ket{\varphi}$. In Appendix \ref{ap:complex} we perform the explicit computations and find 
    \begin{align}\label{eq:probComplexGeneral}
         p_{g\rightarrow e} &= \lambda^2\int \dd V \dd V' \Lambda^*(\mf x')\Lambda(\mf x) e^{\ii\Omega(\tau-\tau')}\\&\quad\quad\times\Big(W_0(\mf x,\mf x')+G^*(\mf x')G(\mf x) \mp F(\mf x')F^*(\mf x)\Big),\nonumber
    \end{align}    
    where the $\mp$ is associated with anti-commutation (Grassmann scalar case) and commutation (complex bosonic scalar) relations respectively. We also defined
    \begin{align}
        F(\mf x) &= \int \dd^n \bm k f(\bm k) u_{\bm k}(\mf x),\nonumber\\
        G(\mf x) &= \int \dd^n \bm k g(\bm k) u_{\bm k}(\mf x),\label{eq:Fcomplex}\\
      \nonumber  W_0(\mf x',\mf x) &= \bra{0}\hat{\psi}^\dagger(\mf x')\hat{\psi}(\mf x)\ket{0} =  \int \dd^n \bm p \:u_{\bm p}(\mf x')u_{\bm p}^*(\mf x).\nonumber
    \end{align}
    Here we see that whether our field has Fermi-Dirac or Bose-Einstein statistics will make the excitation probability depend on the particle or antiparticle content in a different way. If the bosonic commutation relations are chosen, then both the particle and antiparticle content increase the excitation probability. On the other hand, for our choice of model, if we choose anti-commutation relations for $\hat{\psi}$, the particle content decreases the overall excitation probability. This is analogous to the feature seen in the neutrino particle detector model, where the particle content of the neutrino field would inhibit the process of excitation of the proton. It is then possible to use this physically motivated\footnote{Notice that this is a consequence of the fact that linear fermion detectors have to be fermionic themselves.} asymmetric response to distinguish particle and anti-particle excitations.  
    
    Before going into the explicit example for a pointlike detector, let us compare the vacuum excitation and deexcitation probabilities in this general case. Notice that, for a complex scalar field, the vacuum Wightman function satisfies
    \begin{equation}\label{eq:vac}
        \bra{0}\!\hat{\psi}(\mf x')\hat{\psi}^\dagger(\mf x)\!\ket{0}\! = \!\bra{0}\!\hat{\psi}^\dagger(\mf x')\hat{\psi}(\mf x)\!\ket{0} \!=\! \int \dd^n\bm k u_{\bm k}(\mf x) u_{\bm k}^*(\mf x').
    \end{equation}
    This implies that both the vacuum excitation and spontaneous emission probabilities for this detector model coincide with the scalar ones. In turn, this means that this complex scalar model (regardless of statistics) preserves the $\Omega\rightarrow -\Omega$ symmetry between the vacuum excitation probability and the spontaneous decay probability. Keep in mind that this symmetry is broken in the fermionic particle detector from Section \ref{sec:Fermion} even in the vacuum. However, Eq. \eqref{eq:vac} is not satisfied by a general field state $\ket{\varphi}$, that is, $\bra{\varphi}\!\hat{\psi}(\mf x')\hat{\psi}^\dagger(\mf x)\!\ket{\varphi}\! \neq\! \bra{\varphi}\!\hat{\psi}^\dagger(\mf x')\hat{\psi}(\mf x)\!\ket{\varphi}$. This  in turn means  that besides the \tc{special case of charge\tm{-conjugation} invariant states (e.g., the vacuum)}, the complex scalar detector (regardless of statistics) does not have the symmetry  $\Omega\rightarrow -\Omega$ connecting excitation and de-excitation processes, analogous to what we had for the spinor detector.
    
    We can now reduce this general setup to a specific example that can be solved analytically. We consider a pointlike detector undergoing an inertial trajectory in Minkowski spacetime. We also assume the interaction between the detector and field to be switched on for arbitrary long times in order to clean up any finite-time switching effects. We assume the field to start in a general one-particle/antiparticle state given by Eq. \eqref{eq:psiComplex}, where we center the distribution around momentum $\bm{k}_0$ by choosing
    \begin{align}
        f(\bm k) &= \frac{\alpha}{(\pi\sigma^2)^{\frac{n}{4}}}e^{\frac{(\bm k-\bm k_0)^2}{2\sigma^2}},\\
        g(\bm k) &= \frac{\beta}{(\pi\sigma^2)^{\frac{n}{4}}}e^{\frac{(\bm k-{\bm k}_0)^2}{2\sigma^2}},
    \end{align}
    where the normalization condition on $\ket{\varphi}$ implies that $|\alpha|^2 + |\beta|^2 = 1$ and the localization in momentum space is given by the parameter $\sigma$. In this case, it is easy to see that the excitation probability of the detector (Eq. \eqref{eq:probComplexGeneral}) will be the same that we had for a scalar field, but with a factor of $|\beta|^2$, because for the choice of interaction in~\eqref{hIcomplex}, only the anti-particle content is associated with the co-rotating term. The probability reads
    \begin{align}\label{p1}
        p_{g\rightarrow e}=&\frac{2\lambda^2 |\beta|^2\Omega(\Omega^2 - m^2)^{\frac{n-2}{2}}}{\pi^{\frac{n}{2}-2}\sigma^{4-n}|\bm k_0|^{n-2}} e^{-\frac{|\bm k_0|^2+\Omega^2-m^2}{\sigma^2}}\\&\:\:\:\:\:\:\:\:\:\:\times I_{\frac{n-2}{2}}\left(\frac{|\bm k_0|\sqrt{\Omega^2 - m^2}}{ \sigma^{2}}\right)^2\Theta(\Omega - m).\nonumber 
    \end{align}
    In particular, in the case of three spatial dimensions this probability is
    \begin{align}\label{p2}
        p_{g\rightarrow e} =& \frac{4\lambda^2 |\beta|^2\sigma\Omega}{\sqrt{\pi} |\bm k_0|^2}e^{-\frac{|\bm k_0|^2 + \Omega^2 - m^2}{\sigma^2}}\\&\:\:\:\:\:\:\:\:\:\:\:\:\:\:\:\:\times\sinh^2\left(\frac{|\bm k_0|\sqrt{\Omega^2-m^2}}{\sigma^2}\right)\Theta(\Omega - m).\nonumber
    \end{align}
    Notice that the results of Eqs. \eqref{p1} and \eqref{p2} are very similar to the scalar case. This is particular to the adiabatic (long time) limit taken, where the overall behaviour of the response of all the models studied here is very similar. Nevertheless, the excitation probability of the detector depends only on the anti-particle content. In fact, if the field state contains only particles, the excitation probability vanishes in this case, same as for the fermionic detector in Section~\ref{sec:Fermion}. 
    
    Furthermore, looking at Eq.~\eqref{eq:probComplexGeneral} and comparing it with Eq.~\eqref{eq:pgeFermion}, we see that the use of a Grassmann scalar field (anti-commutation of the creation and annihilation operators) allows us to treat this model as a scalar simplification of the spinor field detector, much in the same way that the Unruh-DeWitt model reproduces the fundamental features of the light-matter interaction when the exchange of angular momentum is not relevant~\cite{Pozas2016}. The only two fundamental features that are not captured by this scalar model when compared to the fermionic one are: 1) The fact that the vacuum excitation and deexcitation are symmetric over the exchange $\Omega\rightarrow -\Omega$, while the fermionic one is not. And 2) The spin polarization of the states are not grasped by the scalar model, as one would expect. 
   

\section{Conclusions}\label{sec:Conclusions}

    We have systematically compared four different particle detector models that couple linearly to fields of different spin (scalar, vector, spinor), paying special attention to the role that particle and anti-particle sectors play in these models. In all cases we have studied the response  of the detectors when the field is in a general one-particle Fock wavepacket with respect to an arbitrary mode expansion. From this general setup we have seen that particle detector models that couple linearly to complex fields (in a similar way to the neutrino detector proposed in~\cite{neutrinos}) can distinguish the particle and anti-particle content of the field state. In fact, the complex field detector models studied here can be tuned to be sensitive to the absorption of particles or anti-particles.
    
    Moreover, we have studied the behaviour of the complex scalar model detector we proposed when the field satisfies either Bose-Einstein or Fermi-Dirac statistics (Grassmann scalar field). We showed that the linear scalar complex model with fermionic commutation relations yields a good approximation to the fermionic particle detector model.  This is completely analogous to the way in which the real scalar UDW model can be used to approximately model the light-matter interaction (see, e.g.~\cite{eduardo,richard}).
    
    In summary, this work introduces and studies a new simple particle detector model for probing complex scalar fields that can, in principle, approximate experiments where the symmetry between particles and anti-particles is broken (like neutrino detection). This model is able to grasp the fundamental features of detectors that only detect either particles or antiparticles which are common in experimental high-energy physics. In doing so it also paves the way for the study of relativistic quantum information setups (e.g., entanglement harvesting, relativistic communication, energy teleportation, etc) outside of the realms of the real scalar field or quadratically coupled complex or fermion detectors used in past literature.

\section{Acknowledgements}
    The authors thank Jonas Neuser, Alberto V. Saa and Bruno de S. L. Torres for insightful discussions. Research at Perimeter Institute is supported in part by the Government of Canada through the Department of Innovation, Science and Industry Canada and by the Province of Ontario through the Ministry of Colleges and Universities. E. M-M. is funded by the NSERC Discovery program as well as his Ontario Early Researcher Award.
\onecolumngrid

\appendix

\section{The Real Scalar Field Detecting a Particle}\label{ap:real}

In this appendix we provide the details for the \tb{leading order} excitation probability for a UDW detector probing a one-particle state in a real scalar field theory, first for a detector undergoing general trajectories in a general globally hyperbolic curved spacetime and then we specialize to the case of an inertial detector in flat spacetimes under the long time limit as in Eq.~\eqref{eq:probRealState}. Upon choosing a quantization scheme, the quantized version of a real scalar field in curved spacetimes can be expanded in terms of creation and annihilation operators according to Eq.~\eqref{eq:scalarField}. The interaction between a two-level particle detector and the quantum field is prescribed in terms of the interaction Hamiltonian density from Eq.~\eqref{eq:hIreal} as in~\cite{us,us2}. We consider the initial state of the field to be the pure state $\ket{\varphi}$ from Eq.~\eqref{eq:psiReal} given in terms of the momentum spectral profile function $f(\bm k)$. Normalization of the state implies $\norm{f}_2^2=1$. The probability for the detector to transition from ground to excited state is given in Eq. \eqref{eq:probReal}. It is then enough to calculate the two-point function for the field in the one-particle state $\ket{\varphi}$. In order to proceed, let us first state the following preliminary results,
\begin{align}
    \bra{0}\hat{a}_{\bm k'}\hat{a}^\dagger_{\bm p'}\hat{a}_{\bm p} \hat{a}^\dagger_{\bm k}\ket{0} &= \bra{0}\hat{a}_{\bm k'}\hat{a}^\dagger_{\bm p'}\ket{0}\delta(\bm p-\bm k) = \delta(\bm p'-\bm k')\delta(\bm p-\bm k).\\
    \bra{0}\hat{a}_{\bm k'}\hat{a}_{\bm p'}\hat{a}_{\bm p}^\dagger \hat{a}^\dagger_{\bm k}\ket{0} &= \bra{0}\hat{a}_{\bm k'}\hat{a}_{\bm p'}\ket{\bm p,\bm k} =  \delta(\bm p-\bm p')\delta(\bm k-\bm k')+\delta(\bm k-\bm p')\delta(\bm k'-\bm p)
\end{align}
With these, we obtain
\begin{align}
    \bra{\varphi}\hat{\phi}(\mf x')\hat{\phi}(\mf x)\ket{\varphi} &= \int \dd^n\bm k \dd^n \bm k'\dd^n \bm p \dd^n \bm p' f^*(\bm k')f(\bm k) \bra{0}\hat{a}_{\bm k'}\left(u_{\bm p'}(\mf x') \hat{a}_{\bm p'}+u_{\bm p'}^*(\mf x') \hat{a}^\dagger_{\bm p'}\right)\left(u_{\bm p}(\mf x) \hat{a}_{\bm p}+u_{\bm p}^*(\mf x) \hat{a}^\dagger_{\bm p}\right)\hat{a}^\dagger_{\bm k} \ket{0}\nonumber\\
    &= \int \dd^n \bm k \dd^n \bm k'\dd^n \bm p \dd^n \bm p' f^*(\bm k')f(\bm k) \left(u_{\bm p'}(\mf x')u_{\bm p}^*(\mf x) \bra{0}\hat{a}_{\bm k'}\hat{a}_{\bm p'} \hat{a}^\dagger_{\bm p}\hat{a}^\dagger_{\bm k}\ket{0} +u_{\bm p'}^*(\mf x')u_{\bm p}(\mf x)\bra{0}\hat{a}_{\bm k'} \hat{a}^\dagger_{\bm p'} \hat{a}_{\bm p}\hat{a}^\dagger_{\bm k}\ket{0} \right)\nonumber\\
    &= \int \dd^n\bm k \dd^n \bm k'\dd^n \bm p \dd^n \bm p' f^*(\bm k')f(\bm k)\times\\&\:\:\:\:\:\:\:\:\:\:\:\:\:\:\:\:\times \left(u_{\bm p'}(\mf x')u_{ \bm p}^*(\mf x) (\delta(\bm p-\bm p')\delta(\bm k-\bm k')+\delta(\bm k-\bm p')\delta(\bm k'-\bm p)) +u_{\bm p'}^*(\mf x')u_{\bm p}(\mf x) \delta(\bm p'-\bm k')\delta(\bm p-\bm k)\right)\nonumber\\
    &=\int \dd^n\bm k \dd^n \bm k' f^*(\bm k')f(\bm k) \left(\int \dd^n \bm p \:u_{\bm p}(\mf x')u_{\bm p}^*(\mf x)\delta(\bm k-\bm k') +u_{\bm k}(\mf x')u_{\bm k'}^*(\mf x)+u_{\bm k'}^*(\mf x')u_{\bm k}(\mf x) \right).\nonumber
\end{align}
Thus, the transition probability can be cast as
\begin{align}
    \frac{p_{g\rightarrow e}}{\lambda^2} =& \int \dd V \dd V' \Lambda(\mf x)\Lambda(\mf x')e^{\ii\Omega(\tau-\tau')}\!\!\!\int \dd^n\bm k \dd^n \bm k'\! f^*(\bm k')f(\bm k)\!\! \left(\int \dd^n \bm p \:u_{\bm p}(\mf x')u_{\bm p}^*(\mf x)\delta(\bm k-\bm k')\! +\!u_{\bm k}(\mf x')u_{\bm k'}^*(\mf x)\!+\!u_{\bm k'}^*(\mf x')u_{\bm k}(\mf x)\! \right)\nonumber\\
    =& \int \dd V \dd V'\int \dd^n \bm p \Lambda(\mf x')\Lambda(\mf x)e^{\ii\Omega(\tau-\tau')} u_{\bm p}(\mf x')u_{\bm p}^*(\mf x)\\
    &\:\:\:\:\:\:\:\:\:\:\:\:\:\:\:\:\:\:\:\:\:\:\:\:\:\:\:\:\:\:\:\:+\int \dd V \dd V'\int \dd^n\bm k \dd^n \bm k' \Lambda(\mf x)\Lambda(\mf x')e^{\ii\Omega(\tau-\tau')} f^*(\bm k')f(\bm k) u_{\bm k}(\mf x')u_{\bm k'}^*(\mf x)\nonumber\\
    &\:\:\:\:\:\:\:\:\:\:\:\:\:\:\:\:\:\:\:\:\:\:\:\:\:\:\:\:\:\:\:\:\:\:\:\:\:\:\:\:\:\:\:\:\:\:\:\:\:\:\:\:\:\:\:\:\:\:\:\:\:\:\:\:+\int \dd V \dd V'\int \dd^n\bm k \dd^n \bm k' \Lambda(\mf x)\Lambda(\mf x')e^{\ii\Omega(\tau-\tau')} f^*(\bm k')f(\bm k) u_{\bm k'}^*(\mf x')u_{\bm k}(\mf x).\nonumber
\end{align}
With the definitions in \eqref{eq:Freal}, we can then recast the transition probability as the sum of a vacuum contribution plus the `particle' contribution
\begin{equation}
\begin{aligned}
     p_{g\rightarrow e} &= \lambda^2\int \dd V \dd V' \Lambda(\mf x')\Lambda(\mf x) e^{\ii\Omega(\tau-\tau')}\left(W(\mf x',\mf x)+F(\mf x')F^*(\mf x) + F^*(\mf x')F(\mf x)\right)\\
     &= p_{g\rightarrow e}^{vac} + p_{g\rightarrow e}^{particle}
\end{aligned}
\end{equation}

The case of pointlike detectors with a switching function $\chi(\tau)$ can then be obtained by using the following spacetime smearing function
\begin{equation}
    \Lambda(\mf x) = \chi(\tau)\frac{\delta^{(n)}(\bm x - \bm z(\tau))}{u^0(\tau)\sqrt{-g}},
\end{equation}
where $\mf{u}(\tau)$ denotes the four-velocity of the detector's trajectory and the above expression can be used in any coordinate system $\mf x = (t,\bm x)$. The \tb{leading order}  excitation probability for a point-like detector then reads
\begin{equation}
    p_{g\rightarrow e} = \lambda^2\int \dd \tau \dd \tau' \chi(\tau')\chi(\tau) e^{\ii\Omega(\tau-\tau')}\left(W(\tau',\tau)+F(\tau')F^*(\tau) + F^*(\tau')F(\tau)\right).
\end{equation}

\subsection{A Simple Example}

Consider now an inertial pointlike detector in Minkowski spacetime such that its interaction is switched on indefinitely. In this case we use the mode expansion from Eq. \eqref{eq:modesReal} for $\hat{\phi}(\mf x)$ in an inertial coordinate system. We will work with the one-particle state from Eq. \eqref{eq:psiReal}, with the choice of momentum profile function $f(\bm k)$ from Eq. \eqref{eq:fReal}.  In this setup, the $F(\mf x)$ function from Eq. \eqref{eq:Freal} takes the shape
\begin{equation}
    F(\mf x) = \frac{1}{(2\pi)^{n/2}(\pi \sigma^2)^{\frac{n}{4}}}\int \dd^n\bm k e^{-\frac{|\bm k-\bm k_0|^2}{2\sigma^2}}\frac{e^{\ii \mf k \cdot \mf x}}{\sqrt{2\omega_{\bm k}}}.
\end{equation}
We will work with a massive scalar field interacting with the quantum field in the infinitely long time, so that $\omega_{\bm k} = \sqrt{\bm k^2 + m^2}$ and $\chi(\tau) = \chi(t) =  1$ (understood as taking the adiabatic limit). Under these assumptions the \tb{leading order}  excitation probability for an inertial detector located at the origin can be written as
    \begin{align}
    p_{g\rightarrow e} = &\frac{\lambda^2}{(2\pi)^{n}}\int \dd t \dd t' e^{\ii\Omega(t-t')} \Bigg(\int \dd^n\bm k \frac{e^{-\ii\Omega_{\bm k}(t'-t)}}{2\omega_{\bm k}}\nonumber\\
    &\:\:\:\:\:\:\:\:\:\:\:\:\:\:\:\:\:\:\:\:\:\:\:\:\:\:\:\:\:\:\:\:\:\:\:\:\:\:\:\:\:\:\:\:\:\:\:\:\:\:\:\:\:\:+\frac{1}{(\pi \sigma^2)^{\frac{n}{2}}}\left(\int \dd^n\bm k e^{-\frac{|\bm k-\bm k_0|^2}{2\sigma^2}}\frac{e^{-\ii \Omega_{\bm k}  t' }}{\sqrt{2\omega_{\bm k}}}\right)\left(\int \dd^n \bm k' e^{-\frac{|\bm k'-\bm k_0|^2}{2\sigma^2}}\frac{e^{\ii \Omega_{\bm k'}  t }}{\sqrt{2\omega_{\bm k'}}}\right)
    \\
    &\:\:\:\:\:\:\:\:\:\:\:\:\:\:\:\:\:\:\:\:\:\:\:\:\:\:\:\:\:\:\:\:\:\:\:\:\:\:\:\:\:\:\:\:\:\:\:\:\:\:\:\:\:\:\:\:\:\:\:\:\:\:\:\:\:\:\:\:\:\:\:\:\:\:\:\:\:\:\:\:\:+\frac{1}{(\pi \sigma^2)^{\frac{n}{2}}}\left(\int \dd^n \bm{k} e^{-\frac{|\bm k-\bm k_0|^2}{2\sigma^2}}\frac{e^{\ii \Omega_{\bm k}  t' }}{\sqrt{2\omega_{\bm k}}}\right)\left(\int \dd^n\bm k' e^{-\frac{|\bm k'-\bm k_0|^2}{2\sigma^2}}\frac{e^{-\ii \Omega_{\bm k'}  t }}{\sqrt{2\omega_{\bm k'}}}\right)\Bigg).\nonumber
    \end{align}
Now we notice that the time integrals can be recast as Dirac deltas, according to
\begin{align}
    \int \dd t e^{\ii (\Omega \pm \omega_{\bm k})t} &= 2\pi \delta(\Omega \pm \omega_{\bm k}) = 2\pi \delta(\Omega \pm \sqrt{\bm k^2+m^2}) = 2\pi \frac{\sqrt{\bm k^2+m^2}}{|\bm k|}\left(\delta\left(|\bm k|-\sqrt{\Omega^2 - m^2}\right)+(\delta\left(|\bm k|+\sqrt{\Omega^2 - m^2}\right)\right)\nonumber\\
    &=  2\pi \frac{\sqrt{\bm k^2+m^2}}{|\bm k|}\delta\left(|\bm k|-\sqrt{\Omega^2 - m^2}\right) = 2\pi \frac{\Omega}{\sqrt{\Omega^2 - m^2}}\delta\left(|\bm k|-\sqrt{\Omega^2 - m^2}\right),
\end{align}
where we used the fact that $|\bm k|$ is always positive.

With these results we see that the terms above that combine time exponents of the form $(\Omega + \omega_{\bm k})t$ vanish, because the argument of the delta will never yield zero. Note the technicality that the vacuum contribution yield zero in the adiabatic limit (i.e., the long time limit is thought of as the limit of a sequence of smooth enough switching functions of larger and larger support ~\cite{Satz_2007,LoukoCurvedSpacetimes}). That is, the vacuum contribution and counter-rotating term do not contribute in the infinitely long time. We are left with:
\begin{align}
    p_{g\rightarrow e} &= \frac{1}{(2\pi)^{n}}
    \frac{\lambda^2}{(\pi \sigma^2)^{\frac{n}{2}}}\left(2\pi\int \dd^n\bm{k} e^{-\frac{|\bm k-\bm k_0|^2}{2\sigma^2}}\frac{\delta\left(\Omega - \sqrt{\bm k^2 + m^2}\right)}{\sqrt{2\omega_{\bm k}}}\right) \left(2\pi\int \dd^n\bm k' e^{-\frac{|\bm k'-\bm k_0|^2}{2\sigma^2}}\frac{\delta\left(\Omega - \sqrt{{\bm k'}^2 + m^2}\right)}{\sqrt{2\omega_{\bm k'}}}\right)\nonumber\\
    &= \frac{(2\pi)^2\lambda^2}{(2\pi)^{n}(\pi \sigma^2)^{\frac{n}{2}}}\left|\int 
    \dd|\bm k| \dd \Omega_{n-1} |\bm k|^{n-1}e^{-\frac{|\bm k-\bm k_0|^2}{2\sigma^2}}\frac{\delta\left(\Omega - \sqrt{\bm k^2 + m^2}\right)}{\sqrt{2\omega_{\bm k}}} \right|^2\nonumber\\
    &= \frac{(2\pi)^2\lambda^2e^{-\frac{|\bm k_0|^2}{\sigma^2}}}{2(2\pi)^{n}(\pi \sigma^2)^{\frac{n}{2}}}\left|\int \dd|\bm k| \dd \Omega_{n-1} |\bm k|^{n-2}e^{-\frac{|\bm k|^2}{2\sigma^2}}e^{-\frac{|\bm k||\bm k_0|\cos\theta}{\sigma^2}}\delta\left(|\bm k|-\sqrt{\Omega^2 - m^2}\right)(\bm k^2+m^2)^{\frac{1}{4}} \right|^2\nonumber\\
    &= \frac{(2\pi)^2\lambda^2e^{-\frac{|\bm k_0|^2+(\Omega^2-m^2)}{\sigma^2}}\Omega(\Omega^2-m^2)^{n-2}}{2(2\pi)^{n}(\pi \sigma^2)^{\frac{n}{2}}}\left|\int \dd \Omega_{n-1} e^{-\frac{\sqrt{\Omega^2-m^2}|\bm k_0|\cos\theta}{\sigma^2}} \right|^2\\
    &= \frac{(2\pi)^2\lambda^2e^{-\frac{|\bm k_0|^2+(\Omega^2-m^2)}{\sigma^2}}\Omega(\Omega^2-m^2)^{n-2}}{2(2\pi)^{n}(\pi \sigma^2)^{\frac{n}{2}}} \left|2 \pi^{\frac{n}{2}} \left(\frac{|\bm k_0|\sqrt{\Omega^2-m^2}}{2 \sigma^{2}}\right)^{1-\frac{n}{2}}I_{\frac{n-2}{2}}\left(\frac{|\bm k_0|\sqrt{\Omega^2-m^2}}{ \sigma^{2}}\right)\right|^2\nonumber\\
    &= \frac{(2\pi)^2\lambda^2e^{-\frac{|\bm k_0|^2+(\Omega^2-m^2)}{\sigma^2}}\Omega(\Omega^2-m^2)^{n-2}}{2(2\pi)^{n}(\pi \sigma^2)^{\frac{n}{2}}}\left(\frac{|\bm k_0|\sqrt{\Omega^2-m^2}}{2 \sigma^{2}}\right)^{2-n} 4 \pi^{n} I_{\frac{n-2}{2}}\left(\frac{|\bm k_0|\sqrt{\Omega^2-m^2}}{ \sigma^{2}}\right)^2\nonumber\\
    &= 2\lambda^2 \pi^{2-\frac{n}{2}}e^{-\frac{|\bm k_0|^2+(\Omega^2-m^2)}{\sigma^2}}\frac{\Omega(\Omega^2-m^2)^{\frac{n}{2}-1}}{\sigma^{4-n}|\bm k_0|^{n-2}} I_{\frac{n-2}{2}}\left(\frac{|\bm k_0|\sqrt{\Omega^2-m^2}}{ \sigma^{2}}\right)^2\nonumber.
\end{align}
Notice that this result recovers the one obtained in \cite{erickson} once one sets the mass to zero. Notice that our computations assumed the detector gap $\Omega$ to be larger than the mass of the field, otherwise the \tb{leading order}  excitation probability in this limit would yield zero. This is compatible with the known fact that in order for a particle detector to absorb a particle, it must have a gap larger than its mass.

\section{The Real Vector Field Detecting a Particle}\label{ap:vector}

In this appendix we provide the details for the \tb{leading order}  excitation probability when a detector probes a one-particle excitation of a real massless vector field, carried out in Eq.~\eqref{eq:probVector}. We assume to have a quantized real vector field with two polarizations expanded in terms of creation and annihilation operators as in Eq.~\eqref{eq:fieldEM}. The interaction of the field with a two-level system that undergoes a trajectory $\mf{z}(\tau)$ is prescribed in terms of the interaction Hamiltonian density from Eq.~\eqref{eq:hIvector} and we consider the initial pure state for the field, $\ket{\varphi}$, given by Eq.~\eqref{eq:psiVector}, where we have $\norm{f_1}_2^2+\norm{f_2}_2^2=1$ due to the normalization of the state. The probability for the detector to transition from the ground to excited state is given by
\begin{equation}
    p_{g\rightarrow e} = \lambda^2\int \dd V \dd V' \Lambda_\nu^*(\mf x')\Lambda_\mu(\mf x)e^{\ii\Omega(\tau-\tau')}\bra{\varphi}\hat{E}^\nu(\mf x')\hat{E}^\mu(\mf x)\ket{\varphi},
\end{equation}
and it can be obtained in an analogous way to what was done in the real scalar case. It is then enough to compute the two-point function of the state $\ket{\varphi}$. In order to compute it, let us first state the following preliminary results,
\begin{equation}
\begin{aligned}
    \bra{0}\hat{a}_{\bm k',l'}\hat{a}^\dagger_{\bm p',s'}\hat{a}_{\bm p,s} \hat{a}^\dagger_{\bm k,l}\ket{0} &= \delta_{s'l'}\delta_{sl}\delta(\bm p'-\bm k')\delta(\bm p-\bm k).\\
    \bra{0}\hat{a}_{\bm k',l'}\hat{a}_{\bm p',s'}\hat{a}_{\bm p,s}^\dagger \hat{a}^\dagger_{\bm k,l}\ket{0} &= \delta_{ss'}\delta_{ll'}\delta(\bm p-\bm p')\delta(\bm k-\bm k')+\delta_{ls'}\delta_{sl'}\delta(\bm k-\bm p')\delta(\bm k'-\bm p).
\end{aligned}
\end{equation}
With these, we have
\begin{align}
    \bra{\varphi}\hat{E}^\nu(\mf x')&\hat{E}^\mu(\mf x)\ket{\varphi} = \sum_{ss'll'}\int \dd^3\bm k \dd^3 \bm k'\dd^3 \bm p \dd^3 \bm p' f_{l'}^*(\bm k')f_l(\bm k)\nonumber\\
    &\:\:\:\:\:\:\:\:\:\:\:\:\:\:\times\bra{0}\hat{a}_{\bm k',l'}\left(u_{\bm p',s'}(\mf x') \hat{a}_{\bm p',s'}+u_{\bm p',s'}^*(\mf x') \hat{a}^\dagger_{\bm p',s'}\right)\epsilon^\nu(\bm p',s')\epsilon^\mu(\bm p,s)\left(u_{\bm p,s}(\mf x) \hat{a}_{\bm p,s}+u_{\bm p,s}^*(\mf x) \hat{a}^\dagger_{\bm p,s}\right)\hat{a}^\dagger_{\bm k,l} \ket{0}\nonumber\\
    &= \sum_{ss'll'}\int \dd^3\bm k \dd^3 \bm k'\dd^3 \bm p \dd^3 \bm p' f_{l'}^*(\bm k')f_l(\bm k)\epsilon^\nu(\bm p',s')\epsilon^\mu(\bm p,s)\nonumber\\
    &\:\:\:\:\:\:\:\:\:\:\:\:\:\:\left(u_{\bm p',s'}(x') \bra{0}\hat{a}_{\bm k',l'}\hat{a}_{\bm p',s'}+u_{\bm p',s'}^*(\mf x')\bra{0}\hat{a}_{\bm k',l'} \hat{a}^\dagger_{\bm p',s'}\right)\left(u_{\bm p,s}(\mf x) \hat{a}_{\bm p,s}\hat{a}^\dagger_{\bm k,l} \ket{0}+u_{\bm p,s}^*(\mf x) \hat{a}^\dagger_{\bm p,s}\hat{a}^\dagger_{\bm k,l} \ket{0}\right)\nonumber\\
    &= \sum_{ss'll'}\int \dd^3\bm k \dd^3 \bm k'\dd^3 \bm p \dd^3 \bm p' f_{l'}^*(\bm k')f_l(\bm k)\epsilon^\nu(\bm p',s')\epsilon^\mu(\bm p,s)\nonumber\\
    &\:\:\:\:\:\:\:\:\:\:\:\:\:\:\Bigg(u_{\bm p',s'}(\mf x')u_{\bm p,s}(\mf x) \bra{0}\hat{a}_{\bm k',l'}\hat{a}_{\bm p',s'} \hat{a}_{\bm p,s}\hat{a}^\dagger_{\bm k,l} \ket{0}+u_{\bm p',s'}^*(\mf x')u_{\bm p,s}(\mf x)\bra{0}\hat{a}_{\bm k',l'} \hat{a}^\dagger_{\bm p',s'}\hat{a}_{\bm p,s}\hat{a}^\dagger_{\bm k,l}\ket{0}\nonumber\\
    &\:\:\:\:\:\:\:\:\:\:\:\:\:\:\:\:\:+u_{\bm p,s}^*(\mf x)u_{\bm p',s'}(\mf x') \bra{0}\hat{a}_{\bm k',l'}\hat{a}_{\bm p',s'} \hat{a}^\dagger_{\bm p,s}\hat{a}^\dagger_{\bm k,l} \ket{0}+u_{\bm p',s'}^*(\mf x')u_{\bm p,s}^*(\mf x)\bra{0}\hat{a}_{\bm k',l'} \hat{a}^\dagger_{\bm p',s'} \hat{a}^\dagger_{\bm p,s}\hat{a}^\dagger_{\bm k,l} \ket{0}\Bigg)\nonumber\\
    &= \sum_{ss'll'}\int \dd^3\bm k \dd^3 \bm k'\dd^3\bm  p \dd^3\bm  p' f_{l'}^*(\bm k')f_l(\bm k)\epsilon^\nu(\bm p',s')\epsilon^\mu(\bm p,s)\nonumber\\*
    &\:\:\:\:\:\:\:\:\:\:\:\:\:\:\Bigg(u_{\bm p',s'}^*(\mf x')u_{\bm p,s}(\mf x)\bra{0}\hat{a}_{\bm k',l'} \hat{a}^\dagger_{\bm p',s'}\hat{a}_{\bm p,s}\hat{a}^\dagger_{\bm k,l}\ket{0}+u_{\bm p,s}^*(x)u_{\bm p',s'}(\mf x') \bra{0}\hat{a}_{\bm k',l'}\hat{a}_{\bm p',s'} \hat{a}^\dagger_{\bm p,s}\hat{a}^\dagger_{\bm k,l} \ket{0}\Bigg)\nonumber\\
    &= \sum_{ss'}\int \dd^n\bm k \dd^3 \bm k'\Bigg(\epsilon^\nu(\bm k',s')\epsilon^\mu(\bm k,s)f_{s'}^*(\bm k')f_s(\bm k)u_{\bm k',s'}^*(x')u_{\bm k,s}(x)\nonumber
    \\*
    &\:\:\:\:\:\:\:\:\:\:\:\:\:\:\:\:\:\:\:\:\:\:\:\:\:\:\:\:\:\:\:\:\:\:\:\:\:\:\:\:\:\:\:\:\:\:\:\:\:\:\:\:\:\:\:\:\:\:\:\:+f_{s}^*(\bm k)f_{s}(\bm k)\epsilon^\nu(\bm k',s')\epsilon^\mu(\bm k',s')u_{\bm k',s'}^*(\mf x)u_{\bm k',s'}(\mf x')\nonumber\\*
    &\:\:\:\:\:\:\:\:\:\:\:\:\:\:\:\:\:\:\:\:\:\:\:\:\:\:\:\:\:\:\:\:\:\:\:\:\:\:\:\:\:\:\:\:\:\:\:\:\:\:\:\:\:\:\:\:\:\:\:\:\:\:\:\:\:\:\:\:\:\:\:\:\:\:\:\:\:\:\:\:\:\:\:\:\:+f_{s}^*(\bm k)f_{s'}(\bm k')\epsilon^\nu(\bm k',s')\epsilon^\mu(\bm k,s)u_{k,s}^*(\mf x)u_{\bm k',s'}(\mf x')\Bigg).
\end{align}
Therefore the transition probability can be cast as a sum of three terms:
\begin{equation}
\begin{aligned}
    p_{g\rightarrow e} &= \lambda^2\int \dd V \dd V' \Lambda_\nu^*(\mf x')\Lambda_\mu(\mf x) e^{\ii\Omega(\tau-\tau')} \bra{\varphi}\hat{E}^\nu(\mf x')\hat{E}^\mu(\mf x)\ket{\varphi}\\
    &= \lambda^2\sum_{ss'}\int \dd V \dd V'\int \dd^3 \bm k \dd^3 \bm k' \Lambda_\nu^*(\mf x')\Lambda_\mu(\mf x)e^{\ii\Omega(\tau-\tau')} \epsilon^\nu(\bm k',s')\epsilon^\mu(\bm k,s)f_{s'}^*\bm (k')f_s(\bm k)u_{\bm k',s'}^*(\mf x')u_{\bm k,s}(\mf x)\\
    &\:\:\:\:\:+\lambda^2\sum_{ss'}\int \dd V \dd V'\int \dd^3\bm k \dd^3 \bm k' \Lambda_\nu^*(\mf x')\Lambda_\mu(\mf x)e^{\ii\Omega(\tau-\tau')} f_{s}^*(\bm k)f_{s}(\bm k)\epsilon^\nu(\bm k',s')\epsilon^\mu(\bm k',s')u_{\bm k',s'}^*(\mf x)u_{\bm k',s'}(\mf x')\\
    &\:\:\:\:\:\:\:\:\:\:+\lambda^2\sum_{ss'}\int \dd V \dd V'\int \dd^3\bm k \dd^3 \bm k' \Lambda_\nu^*(\mf x')\Lambda_\mu(\mf x)e^{\ii\Omega(\tau-\tau')} f_{s}^*(\bm k)f_{s'}(\bm k')\epsilon^\nu(\bm k',s')\epsilon^\mu(\bm k,s)u_{\bm k,s}^*(\mf x)u_{\bm k',s'}(\mf x')\\
    &= \lambda^2\sum_{ss'}\int \dd V \dd V' \Lambda_\nu^*(\mf x')\Lambda_\mu(\mf x)e^{\ii\Omega(\tau-\tau')}  \int d^3 \bm k' f_{s'}^*(\bm k')\epsilon^\nu(\bm k',s') u_{\bm k',s'}^*(\mf x')\int \dd^3 \bm k f_s(\bm k)\epsilon^\mu(\bm k,s)u_{\bm k,s}(\mf x)\\
    &\:\:\:\:\:+\lambda^2(\norm{f_1}^2+\norm{f_2}^2)\int \dd V \dd V'\sum_{s'}\int \dd^3 \bm k' \Lambda_\nu^*(\mf x')\Lambda_\mu(\mf x)e^{\ii\Omega(\tau-\tau')} \epsilon^\nu(\bm k',s')\epsilon^\mu(\bm k',s')u_{\bm k',s'}^*(\mf x)u_{\bm k',s'}(\mf x')\\
    &\:\:\:\:\:\:\:\:\:\:+\lambda^2\sum_{ss'}\int \dd V \dd V' \Lambda_\nu^*(\mf x')\Lambda_\mu(\mf x)e^{\ii\Omega(\tau-\tau')}\int \dd^3 \bm k' f_{s'}(\bm k')\epsilon^\nu(\bm k',s')u_{\bm k',s'}(x')\int  \dd^3\bm k f_{s}^*(\bm k)\epsilon^\mu(\bm k,s) u_{\bm k,s}^*(\mf x).
\end{aligned}
\end{equation}

By using the functions defined in Eq.~\eqref{eq:FEM}, we can recast the \tb{leading order}  excitation probability as
\begin{equation}
\begin{aligned}
     p_{g\rightarrow e} &= \lambda^2\int \dd V \dd V' \Lambda_\nu^*(\mf x')\Lambda_\mu(\mf x) e^{\ii\Omega(\tau-\tau')}\left(W^{\nu\mu}(\mf x',\mf x)+F^{\nu}(\mf x')F^{\mu*}(\mf x) + F^{\nu*}(\mf x')F^{\mu}(\mf x)\right)\\
     &= p_{g\rightarrow e}^{vac} + p_{g\rightarrow e}^{particle}.
\end{aligned}
\end{equation}

For the case of pointlike detectors, we must prescribe a switching function that takes into account the vector-like character of the coupling, that is $\chi_\mu(\tau)$. This is the most general choice for switching function. For simplicity it is somewhat common  to simplify the switching, by assuming that $\chi_\mu(\tau) = \chi(\tau) X_\mu(\tau)$, where $\chi(\tau)$ is a scalar switching function and $X_\mu(\tau)$ is a vector field whose components are constant in a frame adapted to the detector's trajectory. The transition probability can then be obtained by considering the following spacetime smearing vector field 
\begin{equation}
    \Lambda_\mu(\mf x) = \chi_\mu(\tau)\frac{\delta^{(3)}(\bm x - \bm z(\tau))}{u^0(\tau) \sqrt{-g}},
\end{equation}
where ${\mf u}(\tau)$ is the four-velocity of the trajectory. The probability for a pointlike detector can then be written as
\begin{equation}
    \begin{aligned}
    p_{g\rightarrow e} = \lambda^2\int \dd \tau \dd \tau' \chi_\nu^*(\tau')\chi_\mu(\tau) e^{\ii\Omega(\tau-\tau')}\left(W^{\nu\mu}(\tau',\tau)+F^{\nu}(\tau')F^{\mu*}(\tau) + F^{\nu*}(\tau')F^{\mu}(\tau)\right).
    \end{aligned}
\end{equation}

\subsection{Probing the electric field}

    The \tb{electric} field seen by a given inertial observer in Minkowski spacetime can be written as
    \begin{equation}
        \hat{E}^\mu(\mf x) = \ii \sum_{s=1}^2 \int\frac{ \dd^3 \bm k}{(2\pi)^{\frac{3}{2}}} \sqrt{\frac{{\color{black}|\bm k|}}{2}}\left(\hat{a}_{\bm k,s}^\dagger e^{-\ii \mf k\cdot \mf x}- \hat{a}_{\bm k,s}e^{\ii \mf k\cdot \mf x}\right)\epsilon^\mu(\bm k,s).
    \end{equation}
    To proceed with the computations, we must choose polarization vectors $\epsilon^\mu(\bm k,s)$, with $s = 1,2$, that are orthogonal to the spacelike portion of $k^\mu$. We parametrize these according to
    \begin{equation}
    \begin{aligned}
        &\bm k = |\bm k| (\sin\theta \cos\phi \: \bm e_x + \sin \theta \sin \phi \: \bm e_y + \cos \theta \bm e_z ),\\
        &\bm \epsilon (\bm k,1) =  \cos\theta \cos\phi \: \bm e_x + \cos \theta \sin \phi \: \bm e_y - \sin \theta \bm e_z ,\\
        &\epsilon (\bm k,2) = -\sin\phi \: \bm e_x + \cos \phi \: \bm e_y.
    \end{aligned}
    \end{equation}
    The effective basis of solutions to the K.G. equation in this case is given by
    \begin{equation}
        u_{\bm k,s}(\mf x) = -\ii\sqrt{\frac{|\bm k|}{2}}\frac{e^{\ii \mf k\cdot \mf x}}{(2\pi)^\frac{3}{2}},
    \end{equation}
    which is independent of the spin $s$.
    
    Under the assumptions of an inertial pointlike detector (Eq. \eqref{eq:pointEM}) probing the one particle state defined by the momentum spectral functions in Eqs. \eqref{eq:f1EM} and \eqref{eq:f2EM}, and choosing coordinates such that the dipole moment of the detector can be written as $X = {\color{black}\Delta^2}\bm e_z$ \tb{(where $\Delta$ has units of length)}, we obtain:
    \begin{align}
        p_{g\rightarrow e} \!=\! \lambda^2{\color{black}\Delta^2}\!\!\int \dd t'\dd t e^{\ii\Omega(t-t')} \!\Bigg(\!&\int \frac{\dd^3 \bm k}{(2\pi)^3} \sin^2 \theta \frac{|\bm k|}{2}e^{i|\bm
        k|(t-t')}\\
        &+|\alpha_1|^2\!\!\int \dd^3 \bm k \dd^3 \bm k'\!\!\left(\frac{e^{-\frac{|\bm k-\bm k_0|^2}{2\sigma^2}}}{(\pi\sigma^2)^{\frac{3}{4}}}(-\sin\theta)\sqrt{\frac{|\bm k|}{2}}\frac{e^{-\ii|\bm k|t'}}{(2\pi)^{\frac{3}{2}}}\right)\!\!\!\left(\frac{e^{-\frac{|\bm k'-\bm k_0|^2}{2\sigma^2}}}{(\pi\sigma^2)^{\frac{3}{4}}}(-\sin\theta')\sqrt{\frac{|\bm k'|}{2}}\frac{e^{\ii|\bm k'|t}}{(2\pi)^{\frac{3}{2}}}\right)\!\!\Bigg)\nonumber\\
        &+|\alpha_1|^2\!\!\int \dd^3 \bm k \dd^3 \bm k'\!\!\left(\frac{e^{-\frac{|\bm k-\bm k_0|^2}{2\sigma^2}}}{(\pi\sigma^2)^{\frac{3}{4}}}(-\sin\theta)\sqrt{\frac{|\bm k|}{2}}\frac{e^{\ii|\bm k|t'}}{(2\pi)^{\frac{3}{2}}}\right)\!\!\!\left(\frac{e^{-\frac{|\bm k'-\bm k_0|^2}{2\sigma^2}}}{(\pi\sigma^2)^{\frac{3}{4}}}(-\sin\theta')\sqrt{\frac{|\bm k'|}{2}}\frac{e^{-\ii|\bm k'|t}}{(2\pi)^{\frac{3}{2}}}\right)\!\!\Bigg).\nonumber
    \end{align}
    Using the fact that 
    \begin{equation}
        \int \dd t \: e^{\ii(\Omega \pm |\bm k|)t} = 2\pi \delta(\Omega \pm |\bm k|),
    \end{equation}
    we can set the first two terms to zero in the adiabatic long-time limit, while the last term yields:
    \begin{equation}
        p_{g\rightarrow e} = \lambda^2{\color{black}\Delta^2}\frac{|\alpha_1|^2\Omega^5  e^{-\frac{(|\bm k_0|^2+\Omega^2)}{\sigma^2}}}{2(\pi \sigma^2)^{\frac{3}{2}}(2\pi)^{2}}\abs{\int \dd\theta \dd\phi\sin^2\theta  e^{\frac{\Omega(k_{0x}\sin\theta\cos\phi+k_{0y}\sin\theta\sin\phi+k_{0z}\cos\theta)}{\sigma^2}}}^2.
    \end{equation}
    The integral over $\phi$ can be solved in terms of the modified Bessel function of the first kind,
    \begin{equation}
        \int \dd\phi \: \exp\left(\frac{\Omega(k_{0x}\sin\theta\cos\phi+k_{0y}\sin\theta\sin\phi+k_{0z}\cos\theta)}{\sigma^2}\right) = 2 \pi e^{\frac{\Omega k_{0z}}{\sigma^2} \cos\theta}I_0\left(\Omega\frac{\sqrt{k_{0x}^2+k_{0y}^2}}{\sigma^2} \sin\theta\right).
    \end{equation}
    The result of the integral over $\theta$ yields a closed-form expression for the \tb{leading order}  excitation probability for a general $\bm k_0$,
    \begin{align}
        p_{g\rightarrow e}&=\lambda^2{\color{black}\Delta^2} |\alpha_1|^2\frac{\Omega ^3 e^{-\frac{{|\bm k_0|}^2+\Omega ^2}{\sigma ^2}}}{16 \sqrt{\pi } {|\bm k_0|}^2 \sigma ^3}
        \\&\:\:\:\:\:\: \Bigg\{\!2  I_0\!\!\left(\frac{{|\bm k_0|\Omega\sin ^2\left(\frac{\vartheta}{2} \right)} }{\sigma ^2}\right)\!\cos ^2\!\!\left(\frac{\vartheta }{2}\right) \! \left[I_1\!\left(\frac{{|\bm k_0|} \Omega  \cos ^2\left(\frac{\vartheta }{2}\right)}{\sigma ^2}\right)\!\sigma ^2 \!\cos\vartheta -2  I_0\!\left(\frac{{|\bm k_0|} \Omega  \cos ^2\left(\frac{\vartheta }{2}\right)}{\sigma ^2}\right)\!{|\bm k_0|} \Omega \sin ^2\!\left(\frac{\vartheta }{2}\right)\!\right]\nonumber\\
        &\:\:\:\:\:\:\:\:\:\:\:\:\:\:+I_1\!\left(\frac{{|\bm k_0|} \Omega  \sin^2\left(\frac{\vartheta }{2}\right)}{\sigma ^2}\right) \!\!\left[  I_1\!\left(\frac{{|\bm k_0|} \Omega \cos ^2\left(\frac{\vartheta }{2}\right)}{\sigma^2} \right)\!{|\bm k_0|} \Omega  \sin^2\vartheta-2 I_0\!\left(\frac{{|\bm k_0|} \Omega  \cos ^2\left(\frac{\vartheta }{2}\right)}{\sigma
       ^2}\right)\!\sigma ^2 \!\sin^2\!\left(\frac{\vartheta }{2}\right)\! \cos\vartheta \right]\!\Bigg\}^2\!\!,\nonumber
    \end{align}
    where $\vartheta$ denotes the relative angle between $\bm k_0$ and $\bm X$. In the particular cases where the polarization vector is a) parallel, and b) orthogonal to the center of the momentum distribution of the one-particle state we obtain the following simplified results:
    \begin{equation}
        p_{g\rightarrow e}^{\parallel} = \lambda^2{\color{black}\Delta^2}|\alpha_1|^2 e^{-\frac{|\bm k-\bm k_0|^2}{\sigma^2}}\frac{\sigma \Omega^3}{4 \sqrt{\pi}|\bm k_0|^2}I_1\left(\frac{\Omega|\bm k_0|}{\sigma^2}\right)^2,
    \end{equation}
    and    
    \begin{equation}
        p_{g\rightarrow e}^{\perp} =  \lambda^2\frac{{\color{black}\Delta^2}|\alpha_1|^2 \Omega^5}{16 \sqrt{\pi}\sigma^3}e^{-\frac{|\bm k-\bm k_0|^2}{\sigma^2}}\left(I_1\left(\frac{\Omega|\bm k_0|}{2\sigma^2}\right)^2+I_0\left(\frac{\Omega|\bm k_0|}{2\sigma^2}\right)^2\right)^2,
    \end{equation}
    respectively. 

\section{The Spinor Field Detecting a Particle}\label{ap:spinor}

    In this appendix we compute the transition probabilities associated with the fermionic particle detector presented in Section \ref{sec:Fermion}.
    A spin $1/2$ fermionic field in curved spacetimes can be expanded in terms of a basis of solutions to Dirac's equation according to Eq. \eqref{eq:fermionField}. We consider the interaction of such field with a two-level system that undergoes a trajectory $\mf z(\tau)$ with free Hamiltonian given by Eq. \eqref{eq:detectorH}. The interaction between the detector and the field is prescribed in terms of the interaction Hamiltonian weight \eqref{hIfermion}, where we remark that now the spacetime smearing function is a spinor field.
    
    We consider the initial pure state for the field $\ket{\varphi}$ given by Eq. \eqref{eq:psiFermion}. The probability amplitude for the detector to transition from the ground to excited state while the filed transitions from $\ket{\varphi}$ to an arbitrary state $\ket{\text{out}}$ at first order in $\lambda$ can then be obtained by
    \begin{equation}
        \bra{e,\text{out}}\hat{U}_I\ket{g,\psi} = \lambda\int \dd V e^{\ii\Omega\tau} \bra{\text{out}}\hat{\bar{\psi}}(\mf x)\slashed{j}(\mf x)\Lambda(\mf x)\ket{\varphi}+\mathcal{O}(\lambda^2).
    \end{equation}
    The transition probability regardless of the final state of the field will them be given by
    \begin{equation}
        p_{g\rightarrow e} = \lambda^2\int \dd V \dd V'  e^{\ii\Omega(\tau'-\tau)} \bra{\varphi}\bar{\Lambda}(\mf x)\slashed{j}(\mf x)\hat{\psi}(\mf x)\hat{\bar{\psi}}(\mf x')\slashed{j}(\mf x')\Lambda(\mf x')\ket{\varphi}.
    \end{equation}
    Let us then calculate the contracted two-point function above,
    \begin{align}
        \bra{\varphi}\bar{\Lambda}(\mf x)\slashed{j}(\mf x)&\hat{\psi}(\mf x)\hat{\bar{\psi}}(\mf x')\slashed{j}(\mf x')\Lambda(\mf x')\ket{\varphi}\nonumber\\
        &= \sum_{s,s',r,r'=1}^2 \int \dd^3 \bm k \dd^3 \bm k' \dd^3 \bm p\, \dd^3 \bm p'\nonumber\\
        &\:\:\:\:\:\:\:\:\:\:\:\:\:\:\:\:\:\:\:\:\:\:\:\:\:\:\:\:\:\times\bra{0}\bar{\Lambda}(\mf x)\slashed{j}(\mf x)\Big(\hat{a}_{\bm k',r'}f_{r'}^*(\bm k') +\hat{b}_{\bm k',r'}g_{r'}(\bm k')\Big)\left( u_{\bm p,s}(\mf x)\hat{a}_{\bm p,s} + v_{\bm p,s}(\mf x) \hat{b}^\dagger_{\bm p,s}\right)\nonumber\\
        &\:\:\:\:\:\:\:\:\:\:\:\:\:\:\:\:\:\:\:\:\:\:\:\:\:\:\:\:\:\:\:\:\:\:\:\:\:\:\:\:\:\:\:\:\:\:\:\:\:\:\:\:\:\:\:\:\:\:\times\left( \bar{u}_{\bm p',s'}(x')\hat{a}^\dagger_{\bm p',s'} + \bar{v}_{\bm p',s'}(\mf x') \hat{b}_{\bm p',s'}\right)\Big(f_r(\bm k)\hat{a}^\dagger_{\bm k,r} +g_r^*(\bm k)\hat{b}^\dagger_{\bm k,r}\Big)\slashed{j}(\mf x')\Lambda(\mf x')\ket{0}\nonumber\\
        &= \sum_{s,s',r,r'=1}^2 \int \dd^3 \bm k \dd^3 \bm k' \dd^3 \bm p\, \dd^3\bm p'\\
        &\:\:\:\:\:\:\:\:\:\:\:\:\:\:\:\:\:\:\:\:\:\times\bar{\Lambda}(\mf x)\slashed{j}(\mf x)\Bigg(g_{r'}(\bm k')v_{\bm p,s}(\mf x) \bar{v}_{\bm p',s'}(\mf x') g_r^*(\bm k)\bra{0}\hat{b}_{\bm k',r'}\hat{b}^\dagger_{\bm p,s}\hat{b}_{\bm p',s'}\hat{b}^\dagger_{\bm k,r}\ket{0}\nonumber\\
        &\:\:\:\:\:\:\:\:\:\:\:\:\:\:\:\:\:\:\:\:\:\:\:\:\:\:\:\:\:\:\:\:\:\:\:\:\:\:\:\:\:\:\:\:\:\:\:\:\:\:\:\:\:\:\:\:\:\:+
        g_{r'}(\bm k')u_{\bm p,s}(\mf x)\bar{u}_{\bm p',s'}(\mf x')g_r^*(\bm k) \bra{0}\hat{b}_{\bm k',r'}\hat{a}_{\bm p,s}\hat{a}^\dagger_{\bm p',s'}\hat{b}^\dagger_{\bm k,r}\ket{0}\nonumber\\
        &\:\:\:\:\:\:\:\:\:\:\:\:\:\:\:\:\:\:\:\:\:\:\:\:\:\:\:\:\:\:\:\:\:\:\:\:\:\:\:\:\:\:\:\:\:\:\:\:\:\:\:\:\:\:\:\:\:\:\:\:\:\:\:\:\:\:\:\:\:\:\:\:\:\:\:+f^*_{r'}(\bm k')u_{\bm p,s}(\mf x)\bar{u}_{\bm p',s'}(\mf x')f_r(\bm k)\bra{0}\hat{a}_{\bm k',r'} \hat{a}_{\bm p,s}\hat{a}^\dagger_{\bm p',s'}\hat{a}^\dagger_{\bm k,r}\ket{0}\Bigg)\slashed{j}(\mf x')\Lambda(\mf x')\nonumber
    \end{align}
    
    It is then enough to compute the following matrix elements:
    \begin{align}
        \bra{0}\hat{a}_{\bm k',r'}\hat{a}_{\bm p,s}\hat{a}^\dagger_{\bm p',s'}\hat{a}^\dagger_{\bm k,r}\ket{0}& = \delta_{s's}\delta_{rr'}\delta(\bm p-\bm p')\delta(\bm k-\bm k') - \delta_{s'r'}\delta_{sr}\delta(\bm p-\bm k)\delta(\bm k'-\bm p') \nonumber\\
        \bra{0}\hat{b}_{\bm k',r'}\hat{a}_{\bm p,s}\hat{a}^\dagger_{\bm p',s'}\hat{b}^\dagger_{\bm k,r}\ket{0}& = \delta_{ss'}\delta_{rr'}\delta(\bm p-\bm p')\delta(\bm k-\bm k')\\
        \bra{0}\hat{b}_{\bm k',r'}\hat{b}^\dagger_{\bm p,s}\hat{b}_{\bm p',s'}\hat{b}^\dagger_{\bm k,r}\ket{0}& = \delta_{rs'}\delta_{r's} \delta(\bm p'-\bm k)\delta(\bm p-\bm k')\nonumber
    \end{align}
    We can now calculate the contraction of the detector smearing functions with the two-point function:
    \begin{align}
        \bra{\varphi}\bar{\Lambda}(\mf x)\slashed{j}(\mf x)&\hat{\psi}(\mf x)\hat{\bar{\psi}}(\mf x')\slashed{j}(\mf x')\Lambda(\mf x')\ket{\varphi}\nonumber \\
        &= \sum_{s,s',r,r'=1}^2 \int \dd^3 \bm k \dd^3 \bm k' \dd^3 \bm p \dd^3\bm p'\nonumber\\
        &\:\:\:\:\:\:\times\bar{\Lambda}(\mf x)\slashed{j}(\mf x)\Bigg(g_{r'}(\bm k')v_{\bm p,s}(\mf x) \bar{v}_{\bm p',s'}(\mf x') g_r^*(\bm k)\delta_{rs'}\delta_{r's} \delta(\bm p'-\bm k)\delta(\bm p-\bm k')\nonumber\\
        &\:\:\:\:\:\:\:\:\:\:\:\:\:\:\:\:\:\:\:\:\:\:\:\:\:\:\:\:\:\:\:\:\:\:\:\:+
       g_{r'}(\bm k')u_{\bm p,s}(\mf x)\bar{u}_{\bm p',s'}(\mf x')g_r^*(\bm k) \delta_{ss'}\delta_{rr'}\delta(\bm p-\bm p')\delta(\bm k-\bm k')\nonumber\\[11pt]
       &\:\:\:\:\:\:\:\:\:\:\:\:\:\:\:\:\:\:\:\:\:\:\:\:\:\:\:\:\:\:\:\:\:\:\:\:\:\:\:+f^*_{r'}(\bm k')u_{\bm p,s}(\mf x)\bar{u}_{\bm p',s'}(\mf x')f_r(\bm k)\delta_{s's}\delta_{rr'}\delta(\bm p-\bm p')\delta(\bm k-\bm k')\nonumber\\
       &\:\:\:\:\:\:\:\:\:\:\:\:\:\:\:\:\:\:\:\:\:\:\:\:\:\:\:\:\:\:\:\:\:\:\:\:\:\:\:\:\:\:- f^*_{r'}(\bm k')u_{\bm p,s}(\mf x)\bar{u}_{\bm p',s'}(\mf x')f_r(\bm k)\delta_{s'r'}\delta_{sr}\delta(\bm p-\bm k)\delta(\bm k'-\bm p')\Bigg)\slashed{j}(\mf x')\Lambda(\mf x')\nonumber\\
       &= \sum_{s,r=1}^2\int \dd^3 \bm k \dd^3 \bm p \bar{\Lambda}(\mf x)\slashed{j}(\mf x)\Bigg(g_{s}(\bm p)v_{\bm p,s}(\mf x) \bar{v}_{\bm k,r}(\mf x') g_r^*(\bm k)+g_{r}(\bm k)u_{\bm p,s}(\mf x)\bar{u}_{\bm p,s}(\mf x')g_r^*(\bm k)\nonumber\\
       &\quad\quad\quad\quad\quad\quad\quad\quad\quad\quad\quad\quad\quad+ f^*_{r}(\bm k)u_{\bm p,s}(\mf x)\bar{u}_{\bm p,s}(\mf x')f_r(\bm k) - f^*_{r}(\bm p)u_{\bm k,s}(\mf x)\bar{u}_{\bm p,r}(\mf x')f_s(\bm k)\Bigg)\slashed{j}(\mf x')\Lambda(\mf x')\nonumber\\
       &= \bar{\Lambda}(\mf x)\slashed{j}(\mf x)\left(\sum_{r=1}^2(\norm{g_r}^2+\norm{f_r}^2)W_0(\mf x,\mf x')+ G(\mf x)\bar{G}(\mf x') - F(\mf x)\bar{F}(\mf x') \right)\slashed{j}(\mf x')\Lambda(\mf x')\nonumber\\
       &= \bar{\Lambda}(\mf x)\slashed{j}(\mf x)\left(W_0(\mf x,\mf x') +G(\mf x)\bar{G}(\mf x') - F(\mf x)\bar{F}(\mf x') \right)\slashed{j}(\mf x')\Lambda(\mf x')
       \end{align}
    where we have used the spinors $F(\mf x)$ and $G(\mf x)$ and the matrix Wightman function of the vacuum that were defined in Eq. \eqref{eq:Fspinor}. The final \tb{leading order}  excitation probability of the detector will then be given at leading order by
    \begin{equation}\label{eq:refMe}
        p_{g\rightarrow e} = \lambda^2\int \dd V \dd V'  e^{-\ii\Omega(\tau-\tau')} \bar{\Lambda}(\mf x)\slashed{j}(\mf x)\left(W_0(\mf x,\mf x') +G(\mf x)\bar{G}(\mf x') - F(\mf x)\bar{F}(\mf x')\right)\slashed{j}(\mf x')\Lambda(\mf x').
    \end{equation}
    
    The case of pointlike detectors with a switching function $\chi(\tau)$ can be obtained by considering the following four-current density
    \begin{equation}
        j^\mu(\mf x) = \frac{\delta^{(3)}(\bm x - \bm z(\tau))}{u^0(\tau) \sqrt{-g}}u^\mu(\tau),
    \end{equation}
    where ${\mf u}(\tau)$ is the four-velocity of the trajectory. Notice that the Dirac delta in the current above washes out all position dependence in the smearing spinor field $\Lambda(\mf x)$, so that it only depends on the proper time of the trajectory. We can then write it as $\Lambda(\mf x) = \chi(\tau)$. With these, the (leading order) excitation probability for a pointlike fermionic detector can then be written as
    \begin{equation}\label{eq:pointlikeSpinor}
        p_{g\rightarrow e} = \lambda^2\int \dd \tau \dd \tau' e^{\ii\Omega(\tau-\tau')}\bar{\chi}(\tau)\slashed{u}(\tau)\left(W_0(\tau,\tau')+G(\tau)\bar{G}(\tau') - F(\tau)\bar{F}(\tau')\right)\slashed{u}(\tau)\chi(\tau').
    \end{equation}
    
    We now compute the \tb{leading order} deexcitation probability of the detector. In this case, we obtain the following probability amplitude for the field to go from the state $\ket{\varphi}$ to an arbitrary final state:
    \begin{equation}
        \bra{g,\text{out}}\hat{U}_I\ket{e,\psi} \approx \lambda\int \dd V e^{-\ii\Omega\tau} \bra{\text{out}}\bar{\Lambda}(\mf x)\slashed{j}(\mf x)\hat{\psi}(\mf x)\ket{\varphi}.
    \end{equation}
    The \tb{leading order}  deexcitation probability will then be given by the sum over all possible final states of the field,
    \begin{equation}
        p_{g\rightarrow e} = \lambda^2\int \dd V \dd V'  e^{-\ii\Omega(\tau-\tau')} \bra{\varphi}\hat{\bar{\psi}}(\mf x')\slashed{j}(\mf x')\Lambda(\mf x')\bar{\Lambda}(\mf x)\slashed{j}(\mf x)\hat{\psi}(\mf x)\ket{\varphi}.
    \end{equation}
    Let us then calculate the sandwiched two-point function that shows up above,
    \begin{align}
        \bra{\varphi}\hat{\bar{\psi}}(\mf x')\slashed{j}(\mf x')&\Lambda(\mf x')\bar{\Lambda}(\mf x)\slashed{j}(\mf x)\hat{\psi}(\mf x)\ket{\varphi} \nonumber\\
        &= \sum_{s,s',r,r'=1}^2 \int \dd^3 \bm k \dd^3 \bm k' \dd^3 {\bm p} \dd^3{\bm p}'\nonumber\\
        &\:\:\:\:\:\:\:\:\:\:\: \times \Bigg( \bra{0}(\hat{a}_{\bm k',r'}f_{r'}^*(\bm k') +\hat{b}_{\bm k',r'}g_{r'}(\bm k'))\left( \bar{u}_{{\bm p}',s'}(\mf x')\hat{a}^\dagger_{{\bm p}',s'} + \bar{v}_{{\bm p}',s'}(\mf x') \hat{b}_{{\bm p}',s'}\right)\slashed{j}(\mf x')\Lambda(\mf x')\nonumber\\[-10pt]
        &\:\:\:\:\:\:\:\:\:\:\:\:\:\:\:\:\:\:\:\:\:\:\:\:\:\:\:\:\:\:\:\:\:\:\:\:\:\:\:\:\:\:\:\:\:\:\:\:\:\:\:\:\:\:\:\:\:\:\:\:\:\:\:\:\:\:\:\:\:\:\:\times\bar{\Lambda}(\mf x)\slashed{j}(\mf x)\left( u_{{\bm p},s}(\mf x)\hat{a}_{{\bm p},s} + v_{{\bm p},s}(\mf x) \hat{b}^\dagger_{{\bm p},s}\right)(f_r(\bm k)\hat{a}^\dagger_{\bm k,r} +g_r^*(\bm k)\hat{b}^\dagger_{\bm k,r})\ket{0}\Bigg)\nonumber\\
        &= \sum_{s,s',r,r'=1}^2 \int \dd^3 \bm k \dd^3 \bm k' \dd^3 {\bm p} \dd^3{\bm p}'\\
        &\quad\quad\times\Bigg(f_{r'}^*(\bm k')f_r(\bm k)\bar{u}_{{\bm p}',s'}(\mf x')\slashed{j}(\mf x')\Lambda(\mf x')\bar{\Lambda}(\mf x)\slashed{j}(\mf x)u_{{\bm p},s}(\mf x)\bra{0}\hat{a}_{\bm k',r'}\hat{a}^\dagger_{{\bm p}',s'}\hat{a}_{{\bm p},s}\hat{a}^\dagger_{\bm k,r}\ket{0}\nonumber\\
        &\quad\quad\quad\quad\quad\quad+f_{r'}^*(\bm k')f_r(\bm k)\bar{v}_{{\bm p}',s'}(\mf x')\slashed{j}(\mf x') \Lambda(\mf x')\bar{\Lambda}(\mf x)\slashed{j}(\mf x)v_{{\bm p},s}(\mf x)\bra{0}\hat{a}_{\bm k',r'}\hat{b}_{{\bm p}',s'}\hat{b}^\dagger_{{\bm p},s}\hat{a}^\dagger_{\bm k,r}\ket{0}\nonumber\\
        &\quad\quad\quad\quad\quad\quad\quad\quad\quad\quad+g_{r'}(\bm k')g_r^*(\bm k)\bar{v}_{{\bm p}',s'}(\mf x')\slashed{j}(\mf x') \Lambda(\mf x')\bar{\Lambda}(\mf x)\slashed{j}(\mf x)v_{{\bm p},s}(\mf x)\bra{0}\hat{b}_{\bm k',r'}\hat{b}_{{\bm p}',s'}\hat{b}^\dagger_{{\bm p},s}\hat{b}^\dagger_{\bm k,r}\ket{0}\Bigg).\nonumber
    \end{align}
    It is then enough to calculate the following matrix elements:
    \begin{align}
        \bra{0}\hat{b}_{\bm k',r'}\hat{b}_{{\bm p}',s'}\hat{b}^\dagger_{{\bm p},s}\hat{b}^\dagger_{\bm k,r}\ket{0}& = \delta_{s's}\delta_{rr'}\delta({\bm p}-{\bm p}')\delta(\bm k-\bm k') - \delta_{s'r}\delta_{sr'}\delta({\bm p}'-\bm k)\delta(\bm k'-{\bm p}) \nonumber\\
        \bra{0}\hat{a}_{\bm k',r'}\hat{b}_{{\bm p}',s'}\hat{b}^\dagger_{{\bm p},s}\hat{a}^\dagger_{\bm k,r}\ket{0}& = \delta_{ss'}\delta_{rr'}\delta({\bm p}-{\bm p}')\delta(\bm k-\bm k')\\
        \bra{0}\hat{a}_{\bm k',r'}\hat{a}^\dagger_{{\bm p}',s'}\hat{a}_{{\bm p},s}\hat{a}^\dagger_{\bm k,r}\ket{0}& = \delta_{rs}\delta_{r's'} \delta({\bm p}-\bm k)\delta({\bm p}'-\bm k')\nonumber
    \end{align}
    We now compute the contraction of the two-point  function with the detector's spinor smearing function:
    \begin{align}
        \bra{\varphi}\hat{\bar{\psi}}(\mf x')\slashed{j}(\mf x')&\Lambda(\mf x')\bar{\Lambda}(\mf x)\slashed{j}(\mf x)\hat{\psi}(\mf x)\ket{\varphi}\nonumber\\
        &= \sum_{s,s',r,r'=1}^2 \int \dd^3 \bm k \dd^3 \bm k' \dd^3 {\bm p} \dd^3{ p}'\nonumber\\
        &\:\:\:\:\:\:\:\:\:\:\:\:\times\Bigg(f_{r'}^*(\bm k')f_r(\bm k)\bar{u}_{{\bm p}',s'}(\mf x')\slashed{j}(\mf x')\Lambda(\mf x')\bar{\Lambda}(\mf x)\slashed{j}(\mf x)u_{{\bm p},s}(\mf x)\delta_{rs}\delta_{r's'} \delta({\bm p}-\bm k)\delta({\bm p}'-\bm k')\nonumber\\
        &\:\:\:\:\:\:\:\:\:\:\:\:\:\:\:\:\:\:\:\:\:\:+f_{r'}^*(\bm k')f_r(\bm k)\bar{v}_{{\bm p}',s'}(\mf x')\slashed{j}(\mf x') \Lambda(\mf x')\bar{\Lambda}(\mf x)\slashed{j}(\mf x)v_{{\bm p},s}(\mf x)\delta_{ss'}\delta_{rr'}\delta({\bm p}-{\bm p}')\delta(\bm k-\bm k')\nonumber\\[11pt]
        &\:\:\:\:\:\:\:\:\:\:\:\:\:\:\:\:\:\:\:\:\:\:\:\:\:\:+g_{r'}(\bm k')g_r^*(\bm k)\bar{v}_{{\bm p}',s'}(\mf x')\slashed{j}(\mf x') \Lambda(\mf x')\bar{\Lambda}(\mf x)\slashed{j}(\mf x)v_{{\bm p},s}(\mf x)\delta_{s's}\delta_{rr'}\delta({\bm p}-{\bm p}')\delta(\bm k-\bm k')\nonumber
        \\
        &\:\:\:\:\:\:\:\:\:\:\:\:\:\:\:\:\:\:\:\:\:\:\:\:\:\:\:\:\:\:- g_{r'}(\bm k')g_r^*(\bm k)\bar{v}_{{\bm p}',s'}(\mf x')\slashed{j}(\mf x') \Lambda(\mf x')\bar{\Lambda}(\mf x)\slashed{j}(\mf x)v_{{\bm p},s}(\mf x)\delta_{s'r}\delta_{sr'}\delta({\bm p}'-\bm k)\delta(\bm k'-{\bm p})\Bigg)\nonumber\\
        &= \sum_{s,s'=1}^2 \int \dd^3 \bm k \dd^3 \bm k'\bar{\Lambda}(\mf x)\slashed{j}(\mf x) \\&\:\:\:\:\:\:\:\:\times\Bigg(f_{s'}^*(\bm k')f_s(\bm k)u_{\bm k,s}(\mf x)\bar{u}_{\bm k',s'}(\mf x')+f_{s'}^*(\bm k')f_{s'}(\bm k)\left(\int \dd^3 {\bm p}\:v_{{\bm p},s}(\mf x)\bar{v}_{{\bm p},s}(\mf x')\right) \delta(\bm k-\bm k')\nonumber\\
        &\:\:\:\:\:\:\:\:\:\:\:\:\:\:\:\:+g_{s'}(\bm k')g^*_{s'}(\bm k)\left(\int\dd^3 {\bm p} v_{{\bm p},s}(\mf x)\bar{v}_{{\bm p},s}(\mf x')\right)\delta(\bm k-\bm k')-g_{s'}(\bm k')g_s^*(\bm k)v_{\bm k',s'}(\mf x)\bar{v}_{\bm k,s}(\mf x')\Bigg)\slashed{j}(\mf x')\Lambda(\mf x')\nonumber\\
        &= \bar{\Lambda}(\mf x)\slashed{j}(\mf x)\Bigg( F(\mf x)\bar{F}(\mf x')- G(\mf x')\bar{G}(\mf x)+\sum_{s=1}^2(\norm{f_{s}}^2+\norm{g_s}^2)\overline{W}_0(\mf x,\mf x')\Bigg)\slashed{j}(\mf x')\Lambda(\mf x')\nonumber
    \end{align}
    where we have defined the spinors $F(\mf x)$ and $G(\mf x)$ as in Eq. \eqref{eq:Fspinor} and the Wightman matrix $\overline{W}_0(\mf x,\mf x')$ as in \eqref{eq:Woconjugate}. With these, the deexcitation  probability of the detector (at leading order) can be written as
    \begin{equation}
        p_{e\rightarrow g} = \lambda^2\int \dd V \dd V' e^{-\ii\Omega(\tau-\tau')} \bar{\Lambda}(\mf x)\slashed{j}(\mf x)\Bigg(F(\mf x)\bar{F}(\mf x')- G(\mf x')\bar{G}(\mf x)+\bar{W}_0(\mf x',\mf x)\Bigg)\slashed{j}(\mf x')\Lambda(\mf x').
    \end{equation}
    The reduction to the pointlike case is analogous to Eq.~\eqref{eq:pointlikeSpinor}

\subsection{A Particular case: vacuum excitation of an inertial detector in flat spacetime}

    We now consider a inertial pointlike detector in Minkowski spacetime. We will consider a fermionic field quantized in some inertial quantization frame $(t,\bm x)$. In the usual plane wave basis the mode solutions are given by
    \begin{align}
        u_{\bm p,1}(\mf x)&=\frac{1}{(2\pi)^{\frac{3}{2}}}\sqrt{\frac{\omega_{\bm p}+m }{2 \omega_{\bm p} }}\left(\begin{array}{c}
        1 \\
        0 \\
        \frac{p_{z} }{\omega_{\bm p}+m } \\
        \frac{p_{x}+i p_{y}}{\omega_{\bm p}+m }
        \end{array}\right)e^{\ii \mf p\cdot \mf x},
        &&&  u_{\bm p,2}(x)&=\frac{1}{(2\pi)^{\frac{3}{2}}}\sqrt{\frac{\omega_{\bm p}+m }{2 \omega_{\bm p} }}\left(\begin{array}{c}
        0\\
        1\\
        \frac{p_{x}-i p_{y}}{\omega_{\bm p}+m }\\
        \frac{-p_z}{\omega_{\bm p}+m }  
        \end{array}\right)e^{\ii \mf p\cdot \mf x},
        \\
        v_{\bm p,1}(x)&=\frac{1}{(2\pi)^{\frac{3}{2}}}\sqrt{\frac{\omega_{\bm p}+m }{2 \omega_{\bm p} }}\left(\begin{array}{c}
        \frac{p_{z} }{\omega_{\bm p}+m } \\
        \frac{p_{x}+i p_{y} }{\omega_{\bm  p}+m } \\
        1 \\
        0
        \end{array}\right)e^{-\ii \mf p\cdot \mf x},
        &&& v_{\bm p,2}(x) &=\frac{1}{(2\pi)^{\frac{3}{2}}}\sqrt{\frac{\omega_{\bm p}+m }{2 \omega_{\bm p} }}\left(\begin{array}{c}
        \frac{p_{x}-i p_{y} }{\omega_{\bm p}+m } \\
        \frac{-p_{z} }{\omega_{\bm p}+m } \\
        0 \\
        1
        \end{array}\right)e^{-\ii \mf p\cdot \mf x}.
    \end{align}
    To provide the description of the fermionic particle detector, we must specify the current $j^\mu(\mf x)$ and the spinor $\Lambda(\mf x)$. We particularize to to the choices in Eq.~\eqref{eq:spinorChoices}. Finally we fix the  the field state to be $\ket{\varphi}$ given in Eq. \eqref{eq:psiFermion} with the Gaussian choices of momentum distribution functions of Eq. \eqref{eq:fSpinor}.
    
    Same as before, in the long time adiabatic limit only the co-rotating term contributes. It is then enough to compute the $G(\mf x)$ spinor. We split it in terms of its spin components, $G_1(\mf x)$ and $G_2(\mf x)$. By choosing the Dirac basis associated with the frame such that the $z$ component is aligned with the $\bm p_0$ vector, we have
    \begin{align}
        G_1(t) &= \frac{2\pi}{(2\pi)^{3/2}}\int \dd |\bm p| \dd\theta\: |\bm p|^2 \sin\theta \sqrt{\frac{\omega_{\bm p}+m}{2\omega_{\bm p}}}\begin{pmatrix}
            \frac{|\bm p|\cos\theta}{\omega_{\bm p} +m}\\ 0 \\ 1 \\ 0
        \end{pmatrix}
        e^{\ii\Omega_{\bm p} t}  \frac{\beta_1}{(\pi\sigma^2)^{3/4}} e^{-\frac{|\bm p|^2 + |\bm p_0|^2}{2\sigma^2}} e^{-\frac{|\bm p| |\bm p_0| \cos\theta}{\sigma^2}}\\
        &= \frac{1}{\sqrt{2\pi}}\frac{\beta_1}{(\pi\sigma^2)^{3/4}}\int \dd |\bm p| \: |\bm p|^2 \sqrt{\frac{\omega_{\bm p}+m}{2\omega_{\bm p}}}\begin{pmatrix}
            \frac{|\bm p|}{\omega_{\bm p} +m}\frac{2\sigma^2}{|\bm p| |\bm p_0|}\left(\cosh\left(\frac{|\bm p| |\bm p_0|}{\sigma^2}\right)-\frac{\sigma^2}{|\bm p| |\bm p_0|}\sinh\left(\frac{|\bm p| |\bm p_0|}{\sigma^2}\right)\right)\\ 0 \\ \frac{2\sigma^2}{|\bm p||\bm p_0|}\sinh\left(\frac{|\bm p||\bm p_0|}{\sigma^2}\right) \\ 0
        \end{pmatrix}
        e^{\ii\Omega_{\bm p} t}   e^{-\frac{|\bm p|^2 + |\bm p_0|^2}{2\sigma^2}},\nonumber
    \end{align}
    and
    \begin{align}
        G_2(t) &= \frac{2\pi}{(2\pi)^{3/2}}\int \dd |\bm p| \dd\theta\: |\bm p|^2 \sin\theta \sqrt{\frac{\omega_{\bm p}+m}{2\omega_{\bm p}}}\begin{pmatrix}
            0\\-\frac{|\bm p|\cos\theta}{\omega_{\bm p} +m}\\ 0 \\ 1 
        \end{pmatrix}
        e^{\ii\Omega_{\bm p} t}  \frac{\beta_2}{(\pi\sigma^2)^{3/4}} e^{-\frac{|\bm p|^2 + |\bm p_0|^2}{2\sigma^2}} e^{-\frac{|\bm p||\bm p_0| \cos\theta}{\sigma^2}}\\
        &= \frac{1}{\sqrt{2\pi}}\frac{\beta_2}{(\pi\sigma^2)^{3/4}}\int \dd |\bm p| \: |\bm p|^2 \sqrt{\frac{\omega_{\bm p}+m}{2\omega_{\bm p}}}\begin{pmatrix}
            0\\-\frac{|\bm p|}{\omega_{\bm p} +m}\frac{2\sigma^2}{|\bm p| |\bm p_0|}\left(\cosh\left(\frac{|\bm p| |\bm p_0|}{\sigma^2}\right)-\frac{\sigma^2}{|\bm p| |\bm p_0|}\sinh\left(\frac{|\bm p| |\bm p_0|}{\sigma^2}\right)\right)\\ 0 \\ \frac{2\sigma^2}{|\bm p||\bm p_0|}\sinh\left(\frac{|\bm p| |\bm p_0|}{\sigma^2}\right)
        \end{pmatrix}
        e^{\ii\Omega_{\bm p} t}   e^{-\frac{|\bm p|^2 + |\bm p_0|^2}{2\sigma^2}}\!.\nonumber
    \end{align}
    
    For this case then, the probability from Eq. \eqref{eq:pointlikeSpinor} reads
    \begin{equation}
        p_{g\rightarrow e} = \tm{\lambda^2}\sum_{s = 1}^2\int \dd t \dd t' e^{-\ii\Omega (t-t')}\Lambda^\dagger \bar{G}_s(t) G_s(t')\Lambda.
    \end{equation}
    All that is left is to compute
    \begin{align}
        \int \dd t & e^{-\ii\Omega t}\Lambda^\dagger G_1(t) = \frac{1}{\sqrt{2\pi}}\frac{\beta_1}{(\pi\sigma^2)^{3/4}}\frac{2\pi\Omega}{\sqrt{\Omega^2-m^2}}(\Omega^2 - m^2) \sqrt{\frac{\Omega+m}{2\Omega}}e^{-\frac{|\bm p_0|^2 + \Omega^2 - m^2}{2\sigma^2}}\frac{2 \sigma^2}{|\bm p_0|\sqrt{\Omega^2 - m^2}}\nonumber\\&\Bigg(\!B_1^*\sinh\!\!\left(\frac{|\bm p_0|\sqrt{\Omega^2 - m^2}}{\sigma^2}\right)+\frac{\sqrt{\Omega^2-m^2}}{\Omega + m}A_1^*\left(\!\cosh\!\!\left(\frac{|\bm p_0|\sqrt{\Omega^2 - m^2}}{\sigma^2}\right)\!-\!\frac{2\sigma^2}{|\bm p_0|\sqrt{\Omega^2 - m^2}}\sinh\!\!\left(\frac{|\bm p_0|\sqrt{\Omega^2 - m^2}}{\sigma^2}\right)\!\right)
       \!\! \Bigg)\nonumber\\
        &= \frac{2\beta_1 \sqrt{\sigma\Omega(\Omega+m)}}{\pi^{1/4}|\bm p_0|}e^{-\frac{|\bm p_0|^2 + \Sigma_m^2}{2\sigma^2}}\!\left(\!B_1^*\sinh\left(\frac{|\bm p_0|\Sigma_m}{\sigma^2}\right)+\frac{\Sigma_m}{\Omega + m}A_1^*\left(\cosh\left(\frac{|\bm p_0|\Sigma_m}{\sigma^2}\right)\!-\!\frac{\sigma^2}{|\bm p_0|\Sigma_m}\sinh\left(\frac{|\bm p_0|\Sigma_m}{\sigma^2}\!\right)\!\right)\!
        \right),
    \end{align}
     where we have defined $\Sigma_m = \sqrt{\Omega^2 - m^2}$ and used the following representation of the Dirac delta function
     \begin{equation}
        \int \dd t e^{\ii(\Omega \pm \omega_{\bm p})t} = 2 \pi \delta(\Omega\pm \omega_{\bm p}) =  2\pi \frac{\sqrt{{\bm p}^2+m^2}}{|{\bm p}|}\delta\left(|{\bm p}|-\sqrt{\Omega^2 - m^2}\right) =  \frac{2\pi\Omega}{\Omega^2-m^2}\delta\left(|{\bm p}|-\sqrt{\Omega^2 - m^2}\right).
    \end{equation}
    An analogous expression holds for the term involving the time integral of the contraction of $G_2(t)$ and $\Lambda$. We then obtain the following expression for the \tb{leading order}  excitation probability of the fermionic detector,
    \begin{align}
        p_{g\rightarrow e}\! =\! \frac{4\lambda^2 \sigma\Omega(\Omega+m)}{\sqrt{\pi}|\bm p_0|^2}e^{-\frac{p_0^2 + \Sigma_m^2}{\sigma^2}}\!\Bigg|&\beta_1B_1^*\sinh\!\left(\frac{|\bm p_0|\Sigma_m}{\sigma^2}\right)\!\!\!+\!\frac{\Sigma_m}{\Omega + m}\beta_1A_1^*\!\left(\!\cosh\!\left(\frac{|\bm p_0|\Sigma_m}{\sigma^2}\right)\!-\!\frac{\sigma^2}{|\bm p_0|\Sigma_m}\sinh\!\left(\frac{|\bm p_0|\Sigma_m}{\sigma^2}\right)\!\right)\\&+\beta_2B_2^*\sinh\!\left(\frac{|\bm p_0|\Sigma_m}{\sigma^2}\right)\!-\!\frac{\Sigma_m}{\Omega + m}\beta_2A_2^*\left(\!\cosh\!\left(\frac{|\bm p_0|\Sigma_m}{\sigma^2}\right)\!-\!\frac{\sigma^2}{|\bm p_0|\Sigma_m}\!\sinh\!\left(\frac{|\bm p_0|\Sigma_m}{\sigma^2}\right)\!\right)
        \Bigg|^2.\nonumber
    \end{align}
    Finally, we factor out the $\sinh$ term and obtain the expression
    \begin{align}
        p_{g\rightarrow e} \!=\! \frac{4\lambda^2 \sigma\Omega(\Omega+m)}{\sqrt{\pi} |\bm p_0|^2}e^{-\frac{ |\bm p_0|^2 + \Sigma_m^2}{\sigma^2}}\!\sinh^2\!\left(\frac{ |\bm p_0|\Sigma_m}{\sigma^2}\right)\!\!\Bigg|&\beta_1B_1^*\!+\!\beta_2B_2^*\!+\!(\beta_1A_1^*\!-\!\beta_2A_2^*)\frac{\Sigma_m}{\Omega + m}\!\left(\!\text{coth}\!\left(\frac{ |\bm p_0|\Sigma_m}{\sigma^2}\right)\!-\!\frac{\sigma^2}{ |\bm p_0|\Sigma_m}\right)\!
        \Bigg|^2\!.
    \end{align}
\section{The Complex Scalar Field Detecting a Particle}\label{ap:complex}

In this section we compute the \tb{leading order}  excitation probability for the complex scalar field particle detector model proposed in Section \ref{sec:Complex}. A complex scalar field can be expanded in terms of creation and annihilation operators of particles and antiparticles according to Eq. \eqref{eq:fieldComplex}, where we assume the creation and annihilation operators to satisfy canonical commutation relations (or anti-commutation in the case of the Grassmann scalar), given by Eq. \eqref{eq:generalComm}.
We consider the interaction of such field with a two-level system that undergoes a trajectory $\mf z(\tau)$ and whose free Hamiltonian is assumed to be given by Eq. \eqref{eq:detectorH}. The interaction between the detector and field is prescribed in terms of the interaction Hamiltonian weight of Eq. \eqref{hIcomplex}. Notice that now the spacetime smearing function is a complex scalar function that transforms properly according to $U(1)$ transformations.

Consider an initial pure state for the field $\ket{\varphi}$ given by Eq. \eqref{eq:psiComplex}. The probability amplitude for the detector to transition from the ground to excited state can then be computed by
\begin{equation}
    \bra{e,\text{out}}\hat{U}_I\ket{g,\psi} \approx \lambda\int \dd V \Lambda(\mf x) e^{\ii\Omega\tau} \bra{\text{out}}\hat{\psi}^\dagger(\mf x)\ket{\varphi}.
\end{equation}
The probability itself will then be given by
\begin{equation}
    p_{g\rightarrow e} = \lambda^2\int \dd V \dd V' \Lambda^*(\mf x')\Lambda(\mf x) e^{\ii\Omega(\tau-\tau')} \bra{\varphi}\hat{\psi}(\mf x')\hat{\psi}^\dagger(\mf x)\ket{\varphi}.
\end{equation}
To evaluate it we need to calculate the two-point function of the field in the state $\ket{\varphi}$. In order to do so, we advance  the following preliminary results:
\begin{equation}
\begin{aligned}
    \bra{0}\hat{a}_{{\bm k}'}\hat{a}_{{\bm p}'}\hat{a}_{\bm p}^\dagger \hat{a}^\dagger_{\bm k}\ket{0} &= \bra{0}\hat{a}_{{\bm k}'}\hat{a}_{{\bm p}'}\ket{{\bm p},{\bm k}}  = \delta({\bm p}-{\bm p}')\delta({\bm k}-{\bm k}')\mp\delta({\bm k}-{\bm p}')\delta({\bm k}'-{\bm p}),\\
    \bra{0}\hat{b}_{{\bm k}'}\hat{b}^\dagger_{{\bm p}'}\hat{b}_{\bm p} \hat{b}^\dagger_{\bm k}\ket{0} &= \bra{0}\hat{b}_{{\bm k}'}\hat{b}^\dagger_{{\bm p}'}\ket{0}\delta({\bm p}-{\bm k}) = \delta({\bm p}'-{\bm k}')\delta({\bm p}-{\bm k}),\\
   \bra{0}\hat{b}_{{\bm k}'}\hat{a}_{{\bm p}'}\hat{a}^\dagger_{\bm p} \hat{b}^\dagger_{\bm k}\ket{0} &=\delta({\bm p}'-{\bm p})\delta({\bm k}'-{\bm k}).
\end{aligned}
\end{equation}
With these,
\begin{align}
    \bra{\varphi}\hat{\psi}(\mf x')&\hat{\psi}^\dagger(\mf x)\ket{\varphi}\nonumber \\&= \int \dd^n{\bm k} \dd^n {\bm k}'\dd^n {\bm p}\, \dd^n {\bm p}' \nonumber\\
    &\times\bra{0}(f^*({\bm k}')\hat{a}_{{\bm k}'}+g({\bm k}')\hat{b}_{{\bm k}'}) \left(u_{{\bm p}'}(\mf x') \hat{a}_{{\bm p}'}+u^*_{{\bm p}'}(\mf x') \hat{b}^{\dagger}_{{\bm p}'}\right)\left(u_{\bm p}^*(\mf x) \hat{a}^\dagger_{\bm p}+u_{\bm p}(\mf x) \hat{b}_{\bm p}\right)(f({\bm k})\hat{a}^\dagger_{\bm k} +g^*({\bm k})\hat{b}^\dagger_{\bm k}) \ket{0}\nonumber\\
    &= \int \dd^n{\bm k} \dd^n {\bm k}'\dd^n {\bm p}\, \dd^n {\bm p}' \Bigg(g^*({\bm k})u_{\bm p}(\mf x)u^*_{{\bm p}'}(\mf x')g({\bm k}')\bra{0} \hat{b}_{{\bm k}'} \hat{b}^{\dagger}_{{\bm p}'} \hat{b}_{\bm p}\hat{b}^\dagger_{\bm k}\ket{0}\nonumber\\
    &\:\:\:\:\:\:\:\:\:\:\:\:\:\:\:\:\:\:\:\:\:\:\:\:\:\:\:\:\:\:\:\:\:\:\:\:\:\:\:\:\:\:\:\:\:\:\:\:\:\:\:\:\:\:\:\:\:\:\:\:\:\:\:\:\:\:\:\:+f({\bm k})u_{\bm p}^*(\mf x)u_{{\bm p}'}(\mf x') f^*({\bm k}')\bra{0}\hat{a}_{{\bm k}'}\hat{a}_{{\bm p}'} \hat{a}^\dagger_{\bm p}\hat{a}^\dagger_{\bm k}\ket{0}\nonumber\\
    &\:\:\:\:\:\:\:\:\:\:\:\:\:\:\:\:\:\:\:\:\:\:\:\:\:\:\:\:\:\:\:\:\:\:\:\:\:\:\:\:\:\:\:\:\:\:\:\:\:\:\:\:\:\:\:\:\:\:\:\:\:\:\:\:\:\:\:\:\:\:\:\:\:\:\:\:\:\:\:\:\:\:\:\:\:\:\:\:\:\:\:\:\:\:\:\:\:\:+g^*({\bm k})u_{\bm p}^*(\mf x) u_{{\bm p}'}(\mf x') g({\bm k}')\bra{0} \hat{b}_{{\bm k}'}\hat{a}_{{\bm p}'}\hat{a}^\dagger_{\bm p}\hat{b}^\dagger_{\bm k}\ket{0}\Bigg)\nonumber\\
    &= \int \dd^n{\bm k} \dd^n {\bm k}'\dd^n {\bm p} \,\dd^n {\bm p}'\Bigg(g^*({\bm k})u_{\bm p}(\mf x)u^*_{{\bm p}'}(\mf x')g({\bm k}')\delta({\bm k}-{\bm p})\delta({\bm k}'-{\bm p}')\\
    &\:\:\:\:\:\:\:\:\:\:\:\:\:\:\:\:\:\:\:\:\:\:\:\:\:\:\:\:\:\:\:\:\:\:\:\:\:\:\:\:\:\:\:\:\:\:\:\:\:\:\:\:\:\:\:\:\:\:\:+f({\bm k})u_{\bm p}^*(\mf x)u_{{\bm p}'}(\mf x') f^*({\bm k}')\delta({\bm p}-{\bm p}')\delta({\bm k}-{\bm k}')\nonumber\\[11pt]
    &\:\:\:\:\:\:\:\:\:\:\:\:\:\:\:\:\:\:\:\:\:\:\:\:\:\:\:\:\:\:\:\:\:\:\:\:\:\:\:\:\:\:\:\:\:\:\:\:\:\:\:\:\:\:\:\:\:\:\:\:\:\:\:\:\:\:\:\:\:\:\:\:\:\:\:\mp f({\bm k})u_{\bm p}^*(\mf x)u_{{\bm p}'}(\mf x') f^*({\bm k}')\delta({\bm k}-{\bm p}')\delta({\bm k}'-{\bm p})\nonumber\\
    &\:\:\:\:\:\:\:\:\:\:\:\:\:\:\:\:\:\:\:\:\:\:\:\:\:\:\:\:\:\:\:\:\:\:\:\:\:\:\:\:\:\:\:\:\:\:\:\:\:\:\:\:\:\:\:\:\:\:\:\:\:\:\:\:\:\:\:\:\:\:\:\:\:\:\:\:\:\:\:\:\:\:\:\:\:\:\:\:\:\:\:\:\:+g^*({\bm k})u_{\bm p}^*(\mf x) u_{{\bm p}'}(\mf x') g({\bm k}')\delta({\bm k}-{\bm k}')\delta({\bm p}-{\bm p}')\Bigg)\nonumber\\
    &= \int \dd^n{\bm k} \dd^n {\bm p}  \Bigg(g^*({\bm k})u_{\bm k}(\mf x)u^*_{{\bm p}}(\mf x')g({\bm p})+f({\bm k})u_{\bm p}^*(\mf x)u_{{\bm p}}(\mf x') f^*({\bm k})\nonumber\\[-10pt]
    &\:\:\:\:\:\:\:\:\:\:\:\:\:\:\:\:\:\:\:\:\:\:\:\:\:\:\:\:\:\:\:\:\:\:\:\:\:\:\:\:\:\:\:\:\:\:\:\:\:\:\:\:\:\:\:\:\:\:\:\:\:\:\:\:\:\:\:\:\:\:\:\:\:\:\:\:\:\:\:\:\:\:\:\:\:\:\:\:\:\:\:\:\:\mp f({\bm k})u_{\bm p}^*(\mf x)u_{{\bm k}}(\mf x') f^*({\bm p})+g^*({\bm k})u_{\bm p}^*(\mf x) u_{{\bm p}}(\mf x') g({\bm k})\Bigg)\nonumber\\
    &= \underbrace{G^*(\mf x')G(\mf x)+\norm{g}^2W_0(\mf x',\mf x)}_{\text{anti-particle}} +\underbrace{
    \norm{f}^2W_0(\mf x',\mf x) \mp F(\mf x')F^*(\mf x)}_{\text{particle}}\nonumber,
\end{align}
where we have used the definitions form the Equations  \eqref{eq:Fcomplex} and the signs are given by the choices of commutation/anticommutation made in Eq.~\eqref{eq:generalComm}.

Therefore, using that $\ket{\varphi}$ is a normalized state, we obtain $\norm{f}^2+\norm{g}^2 =1$ and the transition probability can be cast as
\begin{align}\label{eq:ending}
     p_{g\rightarrow e} &= \lambda^2\int \dd V \dd V' \Lambda^*(\mf x')\Lambda(\mf x) e^{\ii\Omega(\tau-\tau')}\left(G^*(\mf x')G(\mf x)+ W_0(\mf x',\mf x) \mp F(\mf x')F^*(\mf x)\right).
\end{align}

The case of a pointlike detector with a complex switching function $\chi(\tau)$ can then be obtained by using the following complex spacetime smearing function
\begin{equation}
    \Lambda(\mf x) = \chi(\tau)\frac{\delta^{(n)}(\bm x - \bm z(\tau))}{u^0(\tau)\sqrt{-g}},
\end{equation}
where $\mf{u}(\tau)$ denotes the four-velocity of the detector's trajectory and the above expression can be used in any coordinate system $(t,\bm x)$. The \tb{leading order}  excitation probability for a point-like detector then reads
\begin{equation}
    p_{g\rightarrow e} = \tm{\lambda^2}\int \dd \tau \dd \tau' \chi^*(\tau')\chi(\tau) e^{\ii\Omega(\tau-\tau')}\left(G^*(\tau')G(\tau)+W(\tau',\tau)\mp F(\tau')F^*(\tau) \right).
\end{equation}

\section{\tb{Proof that the probabilities for fields with fermionic statistics are always positive}}\label{ap:harmlessLabel}

    In this Appendix we show that the transition probability for particle detectors linearly coupled to fields that satisfy satisfy Fermi-Dirac statistics is always positive. Although there are first principle arguments for the probabilities in Eqs. \eqref{eq:pgeFermion} and \eqref{eq:probComplexGeneral} to be positive numbers, when looking at the expressions themselves, it might not be entirely clear whether the result is positive for any field state. In this appendix we will focus in the case of complex scalar detectors, but the arguments naturally carry to the fermionic case.
    
    Equation \eqref{eq:probComplexGeneral} shows the \tb{leading order}  excitation probability of the linear complex detector detector interacting with a \mbox{one-particle} state. The contribution due to the antiparticle content is easily seen to be positive, while the particle content contributes negatively. For our argument, it is then enough to consider the case where the field state contains only particle content, that is, $g(\bm k) = 0$. The \tb{leading order} probability then reads
    \begin{align}
        p_{g\rightarrow e} &= \lambda^2\int \dd V \dd V' \Lambda^*(\mf x')\Lambda(\mf x) e^{\ii\Omega(\tau-\tau')}\left(W(\mf x',\mf x)- F(\mf x')F^*(\mf x)\right),
    \end{align}
    where $F(\mf x)$ is defined in Eq. \eqref{eq:Fcomplex}. Using the expression for $F$ and for the Wightman function, it is possible to rewrite the excitation probability in terms of an integral over momentum space ${\bm k}$. In order to simplify notation, we define
    \begin{equation}
        U({\bm k}) = \tm{\lambda}\int \dd V \Lambda^*(\mf x) e^{-\ii\Omega \tau} u_{\bm k}(\mf x).
    \end{equation}
    With this, we can write the \tb{leading order} excitation probability as
    \begin{equation}
        p_{g\rightarrow e} = \int \dd^n {\bm k}\, U^*({\bm k}) U({\bm k}) - \int \dd^n {\bm k}\,f({\bm k})U({\bm k})\int \dd^n {\bm k}'\, f^*({\bm k}')U^*({\bm k}') = U\cdot U - (U \cdot f^*)\:\:(f^*\cdot U ),
    \end{equation}
    where $\cdot$ denotes the $L^2$ inner product in momentum space. Then the Cauchy-Schwarz inequality can be used to show that the probability above is indeed always positive. We have
    \begin{equation}
        |U \cdot f^*|^2 \leq (U \cdot U) (f^* \cdot f^*) = (U \cdot U),
    \end{equation}
    where we have used the fact that $f$ is normalized with respect to this inner product and therefore so is $f^*$. \tb{The result above in particular shows that the counter-rotating contribution to the excitation probability ($|U \cdot f^*|^2$) is always smaller than the vacuum contribution ($U \cdot U$), and this result also holds for the case of the real scalar field theory.} It is then easy see that the particle content contributes less than the vacuum contribution to the excitation probability and decreases the likelihood for the excitation of the detector to occur.
    
    \tb{Notice that although we have considered the case of linearly coupled complex particle detector models in the analysis above, it can trivially be generalized to the case of a fermionic particle detector if one replaces the space of functions $L^2$ with the space of spinors, with its natural inner product, $(\psi,\phi) = \bar{\psi}\phi$. In fact, the excitation probability of the fermionic particle detector model can be written solely in terms of the inner product of the spinors $\Lambda(\mf x) \slashed{j}(\mf x)$ with the basis of functions for this space, associated with the field's mode decomposition. With this generalization, the argument above also proves that, in the fermionic model, the particle contribution to the excitation probability is always less than the vacuum contribution and thus, the probability is always positive.}
    
    

  \twocolumngrid

\bibliography{references}
    
\end{document}